%% file: prop2.tex
\documentclass[12pt]{article}
\usepackage{graphicx}
\include{header}
\begin{document}
\pagenumbering{roman}
\include{intro}
\newpage
\tableofcontents
\include{executive}
\pagenumbering{arabic}
\setcounter{page}{1}
\include{motivation}
\include{inner}

\include{highvoltage}
\include{outerveto}

\include{electronics}
\include{slow}
\include{dchooz_laser}
\include{calibration}

\include{clean}
\include{software}
\include{costs}

\include{bib}
\include{appendix}
\end{document}

%% file: header.tex
\newcommand{\nue}{$\nu_{e} \:$}
\newcommand{\numu}{$\nu_{\mu} \:$}
\newcommand{\nuebar}{\overline{\nu}_{e}}

\newcommand{\sstc}{$\sin ^{2} 2\theta_{13} \:$}
\newcommand{\tha}{$\theta_{12} \:$}
\newcommand{\thb}{$\theta_{23} \:$}
\newcommand{\thc}{$\theta_{13} \:$}
\def\adm2{\Delta{{m}^2_{{atm}}}}
\def\sdm2{\Delta{{m}^2_{{sol}}}}
\def\be{\begin{equation}}
\def\ee{\end{equation}}
\def\Dm2{\Delta{m}^2}
\def\t13{\theta_{{13}}}
\def\tsol{\theta_{{sol}}}
\def\simgt{\lower.5ex\hbox{$\; \buildrel > \over \sim \;$}}
\newcommand{\quq}{\theta_{13}}

\newcommand{\mxang}{\sin^2(2\theta)}
\newcommand{\mxangg}{\sin^2(2\theta_{13})}
\newcommand{\dmsqat}{\Delta m^2_{atm}}

%% file: intro.tex
\parindent=12pt
\title{\bf Proposal for U.S. participation 
in  \\
Double-CHOOZ: \\ 
A New $\theta_{13}$ Experiment at the Chooz Reactor
\\
}
\author{
S.~Berridge$^g$,
W.~Bugg$^g$, 
J.~Busenitz$^a$, 
S.~Dazeley$^e$, 
\\
G.~Drake$^b$,
Y.Efremenko$^g$, 
M.~Goodman$^{b*}$,
J.~Grudzinski$^b$,
\\
V.~Guarino$^b$,
G. Horton-Smith$^d$,
Y.~Kamyshkov$^g$, 
T.~Kutter$^e$
\\
C.~Lane$^c$,   
J.~LoSecco$^f$,
R.~McNeil$^e$,
W.~Metcalf$^e$, 
\\
D.~Reyna$^b$,
I.~Stancu$^a$, 
R.~Svoboda$^{e*}$, 
R.~Talaga$^b$}

\date{October 14, 2004}
\maketitle

\begin{center} $^a$ University of Alabama,
$^b$ Argonne National Laboratory,
$^c$ Drexel University,\\
$^d$ Kansas State University, 
$^e$ Louisiana State University,\\
$^f$ University of Notre Dame,
$^g$ University of Tennessee\\
{* US Contacts: \verb=phsvob@lsu.edu=, \verb=maury.goodman@anl.gov=}
\end{center}

\vspace*{0.2in}
\begin{center}
{\bf Abstract}
\end{center}

\par It has recently been widely recognized that a reactor 
anti-neutrino
disappearance experiment with two or more detectors is one of the most
cost-effective ways to extend our reach in sensitivity for the neutrino
mixing angle $\theta_{13}$ without ambiguities from CP violation and matter
effects\cite{R:WhitePaper}.  The physics capabilities of a new
reactor experiment together with superbeams and
neutrino factories have also been studied~\cite{R:Huber2003, R:Minakata2003}
but these latter
are considered by many to be more ambitious projects due to their
higher costs, and hence to be farther in the future.
\par
We propose to contribute to an international collaboration
to modify the existing neutrino physics facility at the Chooz-B Nuclear
Power Station in France. The experiment, known as Double-CHOOZ, is expected
to reach a sensitivity of \sstc $> 0.03$ over a three
year run, 2008-2011. 
This would cover roughly 85\% of the remaining allowed region.  
The costs and time to first results for this critical parameter can be
minimized since our project takes advantage of an existing
infrastructure.

%% file: executive.tex
{\noindent\bf\Large Executive Summary}
\\

{\noindent\bf\large\underline{Opportunity}}\\
There has been superb progress in understanding the neutrino sector
of elementary
particle physics in the past few years.  It is now widely recognized that
the possibility exists for a rich program of measuring CP violation and
matter effects in future accelerator $\nu$ experiments, using
superbeams,
off-axis detectors, neutrino factories and beta beams.  However,
this possibility can be fulfilled only
if the value of the neutrino mixing parameter
$\quq$ is such that $\mxangg \simgt 0.01$.
A new experiment at CHOOZ will
be sensitive over most of that range, $\mxangg > 0.03$
and has a great opportunity for an exciting and important discovery,
a non-zero value to $\quq$.  It would also serve as a crucial stop
toward a more 
sensitive new experiment.
In comparison to other possible sites, the far detector lab at
CHOOZ already exists.  It is possible, for a modest total cost, and a
U.S. contribution of around \$5M, to rapidly deploy 
Double-CHOOZ and start taking data by 2008.  The sensitivity of the
original CHOOZ experiment would be surpassed in only 4 months.  The
full sensitivity of Double-CHOOZ would be reached in 3 years.
Double-CHOOZ presents an ideal opportunity to leverage the US 
experience in neutrino detection with the European resources of nuclear
reactor and neutrino laboratory as well as their expertise in low background
detectors.

\vspace*{6pt}

{{\noindent\bf\large\underline{Purpose of the Experiment}}\\
In the presently accepted paradigm to describe the neutrino sector, there are
three mixing angles.  One is measured by solar neutrinos and the KamLAND
experiment,  one by atmospheric neutrinos and the long-baseline
accelerator projects.  Both angles are large.  
The third angle, $\quq$, has not yet been measured to
be nonzero but
has been constrained to be small by CHOOZ.

\vspace*{6pt}

{{\noindent\bf\large\underline{From CHOOZ to Double-CHOOZ}}\\
The basic feature of reactor $\quq$ experiments 
is to search for energy dependent $\bar{\nu}_e$
disappearance.
CHOOZ used a single five-ton detector when the reactors were new.
We propose to use two significantly-improved detectors (hence the
name ``Double-CHOOZ") with increased luminosity.
The best current limit on $\quq$ comes from the CHOOZ experiment and is a function
of $\dmsqat$, which has been measured using atmospheric neutrinos
by Super-Kamiokande.  The latest reported value of $\dmsqat$ 
is
$1.2~<~\dmsqat~<~3.0~\times~10^{-3}$eV$^2$ 
with a best fit reported at 2.0.  The
CHOOZ limits for $\dmsqat$ of 2.6 and 2.0 $\times~10^{-3}$eV$^2$
are $\mxangg~<~$
0.14 and 0.19.  Global fits using the solar data limit the value for small
$\dmsqat$ to less than 0.12.  
The dominant systematic errors, such as cross-sections, flux uncertainties,
and the absolute target volume, will be eliminated in a relative 
measurement with two identical detectors.  
Increased statistics will be achieved by running longer and 
using a larger detector, and with both CHOOZ reactors
running at full power.  In three years, Double-CHOOZ will
achieve 250~t~GW~y (ton-Gigawatt-years)
while the 
CHOOZ value was 12~t~GW~y.
This will permit a mixing angle sensitivity of
$\mxangg~>~0.03$.  

\vspace*{6pt}

{{\noindent\bf\large\underline{Organization of the Proposal}}\\
The design of the Double-CHOOZ experiment is described in
Section~1.  Many parts of the Section draw on the previous 
European LOI, which was written with substantial input by some members
of the US collaboration.  Proposed US  systems 
are described in Section 2.  
These are the
inner detector photomultiplier tubes, the high voltage
system, electronics, slow monitoring, a laser system, calibration
deployment and radiopurity maintenance.
In Section 3, we present the
costs and schedule for the experiment, with emphasis on the U.S. components
which are presented in this proposal.  The management structure for the
U.S. collaboration within Double-CHOOZ is included.

%% file: motivation.tex
\section{Description of the Double-CHOOZ Experiment}
\subsection{Introduction}

\par A group of European scientists is proposing a new two-detector 
experiment known as Double-CHOOZ\cite{bib:choozloi}
 at the Chooz-B Nuclear Power Station,
 the site of a previous
reactor neutrino oscillation experiment, in order to search for a non-zero
value of the mixing angle \thc.  The collaboration, which currently consists
of 52 physicists from 14 institutions in Europe, has received preliminary
approval in France, a critical first step. Information about the physics
opportunity, detector design and simulation, liquid scintillator and buffer
liquids, calibration, backgrounds, systematic errors, and sensitivity are
provided in the document ``Letter of Intent for Double-CHOOZ: A Search for
the Mixing Angle \thc''\cite{bib:choozloi}. The authors of this
proposal are from U.S. institutions which have joined and are seeking
funding.  We review the motivation
for such an experiment and its sensitivity in Section 1.
The systems that the U.S. groups are proposing to work on are
described in Section 2.  
The cost and schedule are provided in Section 3.
A list of collaborators and a brief background of U.S.
participants are given in an Appendix.

\par The members of this proposal have participated in the International Working Group
with the goal of carrying through an experimental reactor neutrino program
sensitive to \sstc $> 0.01$. Reaching such a sensitivity at the level 
of better
than 1\% is a
difficult task as it requires total systematic uncertainties at the 1\%
level, something which has never been achieved in a reactor experiment.
Double-CHOOZ, with an expected sensitivity of \sstc $>0.03$, is an important
step in this program for two reasons: (1) by exploring the region
$0.03 < $ \sstc $< 0.2$ there is an excellent chance for a new discovery, 
and (2) it is a realistic setting to learn more about backgrounds and
systematic errors crucial for mounting a more sensitive next-generation
experiment.
Concerning the last point, a number of initiatives are being
considered that could potentially improve the Double-CHOOZ sensitivity
by running longer with
larger detectors located at greater depths\cite{R:Angra, R:Braidwood,
R:DayaBay, R:DiabloCanyon, R:Kaska}. These
more sensitive experiments have greater cost, longer time scale, and
are a more challenging extrapolation from previous experiments. In contrast,
Double-CHOOZ can be done at modest cost and on a relatively short time scale
using an existing facility.
Double-CHOOZ will provide a crucial bridge to more
ambitious and more sensitive experiments in the future.

\subsection{Physics Motivation}

\par In the presently accepted paradigm to describe the neutrino sector, there
are three mixing angles (\tha, \thb, \thc) that quantify the mixing of
the neutrino mass and flavor eigenstates.  \tha has been measured by solar
neutrino experiments and the KamLAND reactor experiment. \thb has been 
measured by atmospheric neutrino experiments and the K2K long-baseline
experiment. The third angle, \thc, has not yet been measured but has been
constrained to be small by the CHOOZ reactor
experiment\cite{R:CHOOZ}.  

\par 
The possibility exists for a rich program of measuring CP violation and matter
effects in future accelerator neutrino experiments, which has led to an
intense worldwide effort to develop neutrino superbeams, off-axis detectors,
neutrino factories, and beta beams\cite{bib:aps}. However, the possibility of measuring
CP violation in the foreseeable future
can be fulfilled only if the value of the neutrino mixing
parameter \thc is such that \sstc $\simgt 0.01$. A timely new reactor experiment
sensitive to \thc in this range has excellent discovery potential for
finding a non-zero value of this important parameter. In addition,
a short time scale for a measurement or improved limit on \thc will help
long-baseline accelerator experiments better exploit their full potential.
\subsubsection{$\nuebar$ detection principle}

\par
Reactor antineutrinos are detected through their
interaction by inverse neutron decay (threshold of 1.806~MeV) 
\begin{equation}
\bar\nu_e + p \rightarrow e^+ + n~.
\end{equation}
If we use an averaged
fuel composition typical during a reactor cycle corresponding to  
$^{235}$U (55.6\%), $^{239}$Pu (32.6\%), $^{238}$U  (7.1\%) and 
$^{241}$Pu (4.7\%), the mean energy release per fission  W is  
203.87~MeV and the energy weighted cross section amounts to
$<\sigma>_{\rm fission} = 5.825 \times 10^{-43}~{\rm cm}^2 \, \, {\rm per} \,
\, {\rm fission}~. $
For the purpose of simple scaling, a reactor with a power of
1~GW$_{th}$ induces a rate of $\sim$450 events per year in a detector
 containing $10^{29}$~protons, at a distance of 1~km. \\

Experimentally one takes advantage of the coincidence signal of the
prompt positron followed in space and time by the delayed neutron
capture. This very clear signature allows us to reject most of
the accidental
backgrounds. The energy of the incident antineutrino is then related to the
energy of the positron by the relation
\begin{equation}
\label{eq:energy}
E_{\nuebar}=E_{{e}^+} +(m_{n}-m_{p})+O(E_{\nuebar}/m_{n})~.
\end{equation}
Experimentally, the visible energy seen in the detector is given by  
$E_{vis}=E_{{e}^+} + 511~{keV}$, where $E_{{e}^+}$
is the sum of the rest mass and kinetic energy of the positron.
and the
additional $511\,\mathrm{keV}$ come from the annihilation of the
positron with an electron when it stops in the matter.  

\subsubsection{$\nuebar$ oscillations}

\par Reactor neutrino disappearance experiments measure the survival probability 
$P_{\nuebar \rightarrow \nuebar}$ of the electron antineutrinos 
emitted from the nuclear power plant.
A disappearance experiment does not measure
the $\delta$-CP phase. Furthermore, because of the low energy as well
as the short baseline considered, matter effects are 
negligible\cite{minakatareactor2002}. 
Assuming a ``normal'' mass hierarchy scenario
 the $\nuebar$ survival probability 
can be written\cite{CHOOZU13,PetcovNHIH}

\par 
\begin{eqnarray}
\label{3nuSP}
P_{\nuebar\to\nuebar} & = &
1-2\sin^2\theta_{13}\cos^2\theta_{13}\sin^2\left(\frac{\Delta{m}^2_{31}L}{4E}\right)
\\ \nonumber
& - & 
\frac{1}{2}\cos^4\theta_{13}\sin^2(2\theta_{12})\sin^2\left(\frac{\Delta{m}^2_{21}L}{4E}\right)\nonumber\\ 
& + & 
2\sin^2\theta_{13}\cos^2\theta_{13}\sin^2\theta_{12}\left(\cos\left(\frac{\Delta{m}^2_{31}L}{2E}-\frac{\Delta{m}^2_{21}L}{2E}\right)-\cos\left(\frac{\Delta{m}^2_{31}L}{2E}\right)\right) \nonumber
\end{eqnarray}

The first two terms in Equation~\ref{3nuSP} contain respectively the
atmospheric driven ($\Dm2_{31} = \adm2$) and solar driven 
($\Dm2_{21} = \sdm2$, $\theta_{12} \sim \tsol$)
contributions, while the third term
is an interference between solar and
atmospheric driven oscillations whose amplitude is a function of
$\t13$.
Reactor experiments provide a clean measurement of the
mixing angle $\t13$, free from any contamination coming from matter
effects and other parameter correlations or 
degeneracies\cite{minakatareactor2002, huberreactor2003}. 

\subsection{Configuration of the Detectors for Double-CHOOZ}
\subsubsection{The CHOOZ nuclear reactors}

\par
\begin{figure}[h]
\begin{center}
\includegraphics[width=\textwidth]{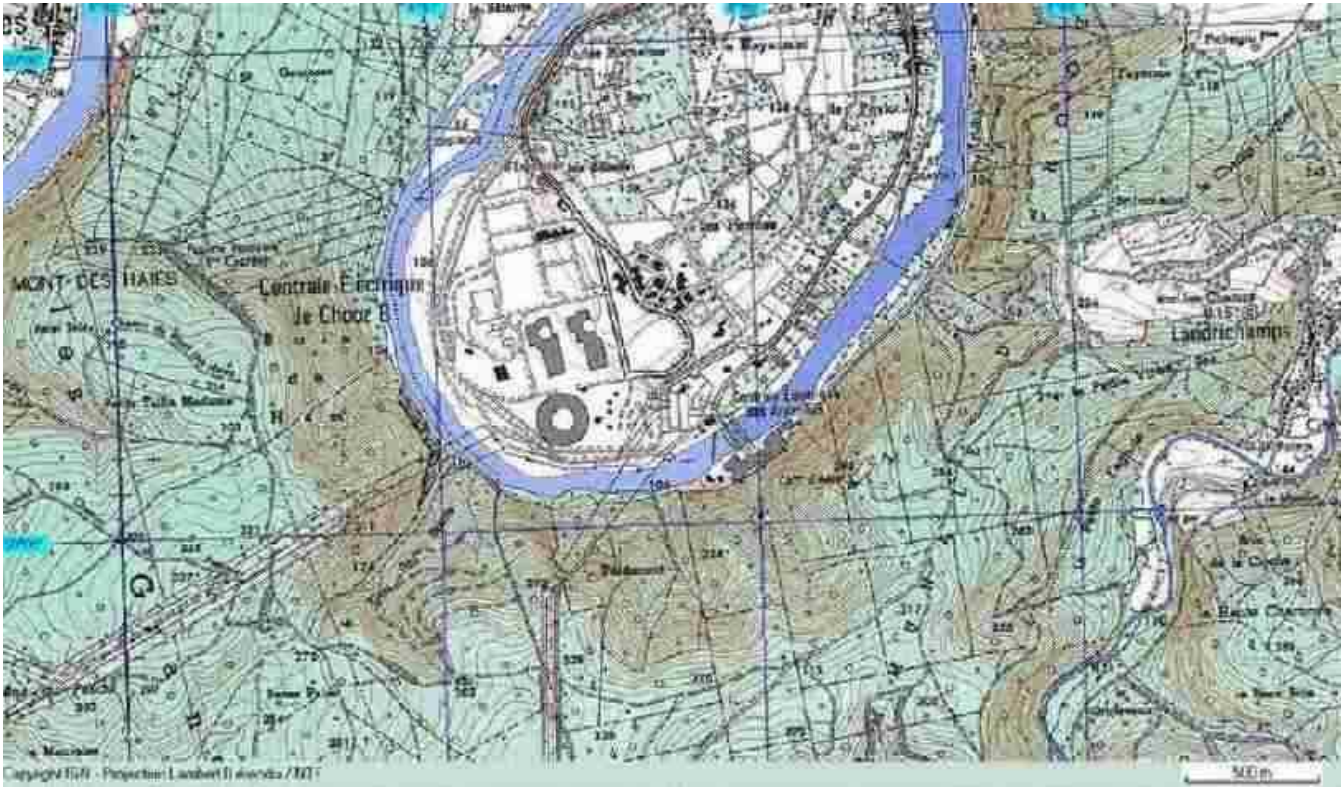}
\caption[Map of the experiment site]{Map of the experiment site. The
  two cores are separated by a distance of 100~meters. The far
  detector site is located at 1.0 and 1.1~km from the two cores.}
\label{fig:map}
\end{center}
\end{figure}

The antineutrinos are produced by the pair
of reactors located at the CHOOZ-B nuclear power station operated by
the French  company Electricit\'e de France (EDF) in partnership
with the Belgian utilities Electrabel S.A./N.V. and Soci\'et\'e Publique
d'Electricit\'e.  They are located in the Ardennes region, northeast of
France, very close to the Belgian border, in a loop of the Meuse
river (See Figure~\ref{fig:map}.
Both reactors are of
the  most recent  N4 type (4 steam generators) with a thermal power of
4.27~GW$_{{th}}$, and 1.5~GW$_{{e}}$.  They operate with an 
80\% load factor.  
These are Pressurized Water Reactors (PWR) and are fed with
UOx fuel. 
205 fuel assemblies are contained within each reactor core.
 The entire reactor vessel is a cylinder of 13.65~meters high and 
4.65~meters in diameter.
The first reactor started operating at full power in May 1997,
and the second one in September 1997.

\par
\begin{figure}[htbp]
\begin{center}
\includegraphics[width=\textwidth]{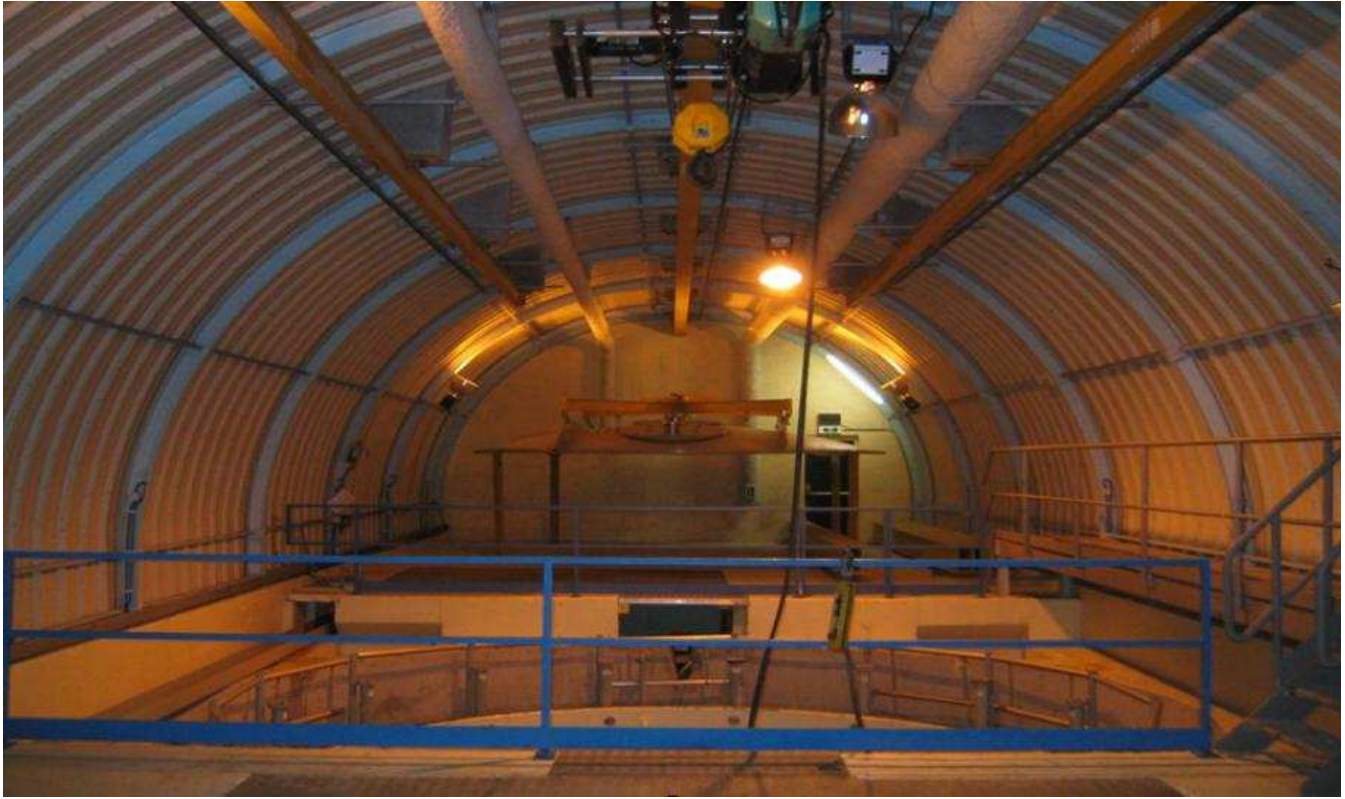}
\caption[Picture of the CHOOZ far detector site]
{Picture of the CHOOZ far detector site taken in September
 2003. The original CHOOZ laboratory hall constructed by EDF, 
 located close the the old CHOOZ-A underground power plant, is still in
 perfect condition and will be re-used.}
\label{fig:choozfarfoto}
\end{center}
\end{figure}

 
The Double-CHOOZ experiment will run two identical detectors of medium size,
containing 12.7 cubic meters of liquid scintillator target doped with
0.1\% of Gadolinium.
The neutrino laboratory of the CHOOZ experiment, located 1.0 and
1.1~km respectively 
from the two cores of the CHOOZ nuclear plant will be used again
(see Figure \ref{fig:choozfarfoto}). 

\par 
A sketch of  the Double-CHOOZ far detector is 
shown in Figure~\ref{fig:choozfar}. 
The CHOOZ far site is shielded by about 300~m.w.e. of 2.8~${g/cm}^3$ rocks. 
An artificial overburden of a few tens of meters
height has to be built for the CHOOZ-near detector.  
The required overburden ranges from 53 to
80 m.w.e. depending on the near detector location, between 100 and 200
meters away from the cores.
A sketch of this detector is shown in Figure~\ref{fig:chooznear}. 
An initial study has been commissioned by the French electricity
power company EDF to determine the best combination of
location-overburden and to optimize the cost of the project.
\subsubsection{Detector design}

\begin{figure}[htbp]
\begin{center}
\includegraphics[width=0.6\textwidth]{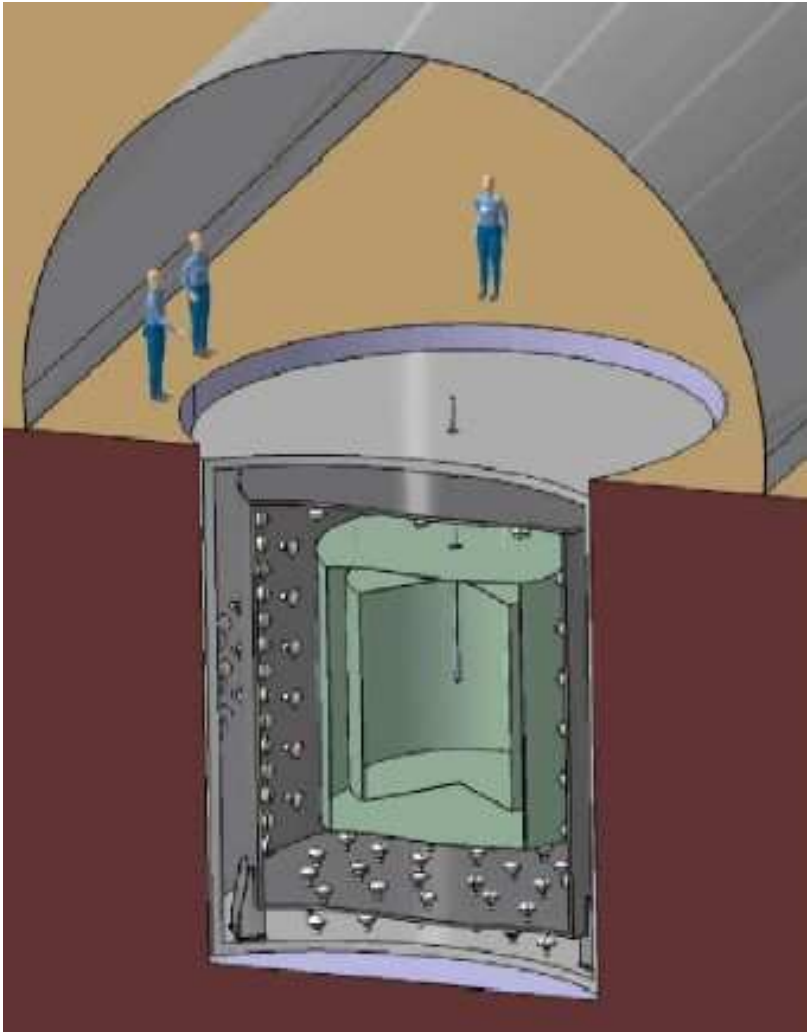}
\caption[Double-CHOOZ far detector]
{The Double-CHOOZ far detector, at the CHOOZ underground site. 
The detector is located in the tank used for the CHOOZ experiment (7 meters high and 7
meters in diameter) that is still available. About 12.7~${m}^3$
of a dodecane+PXE based
liquid scintillator doped with gadolinium will be contained in a 
transparent acrylic cylinder surrounded by the $\gamma$-catcher 
region and the buffer. 
}
\label{fig:choozfar}
\end{center}
\end{figure}

\par The Double-CHOOZ far detector will consist of
a target cylinder of 120~cm radius 
and 280~cm height, providing a volume of 12.7$~{m}^3$.
The near and far detectors will be identical inside the PMTs supporting
structure. 
This will allow a relative normalization systematic error
 of  $\sim$0.6\%.
However, due to the different overburdens (60-80 to 300 m.w.e.), 
the outer shielding will not be identical since the
cosmic ray background varies between Double-CHOOZ near and Double-CHOOZ far. 
The overburden of the near detector has been chosen in order to keep
the signal to background ratio above 100.
In this case, a knowledge of the backgrounds within a
factor two keeps the associated systematic error well below one percent.

\begin{figure}[htbp]
\begin{center}
\includegraphics[width=0.9\textwidth]{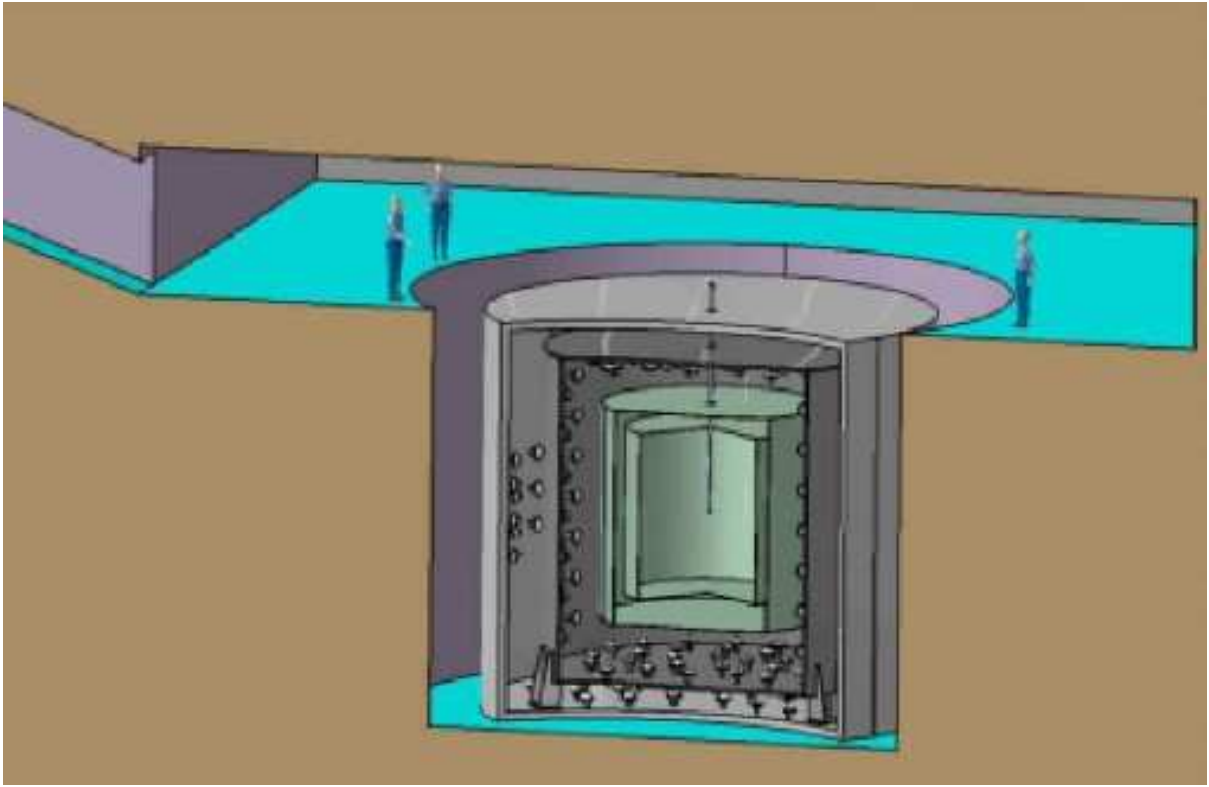}
\caption[CHOOZ-near detector]
{The CHOOZ-near detector at the new underground
site, close to the reactor cores. 
This detector is identical to the Double-CHOOZ far detector up to and
including  the PMT  surface. 
The veto region will be enlarged to better reject the cosmic 
muon induced backgrounds (see Section~\ref{sec:background}).}
\label{fig:chooznear}
\end{center}
\end{figure}

\begin{figure}[htbp]
\begin{center}
\includegraphics[width=0.8\textwidth]{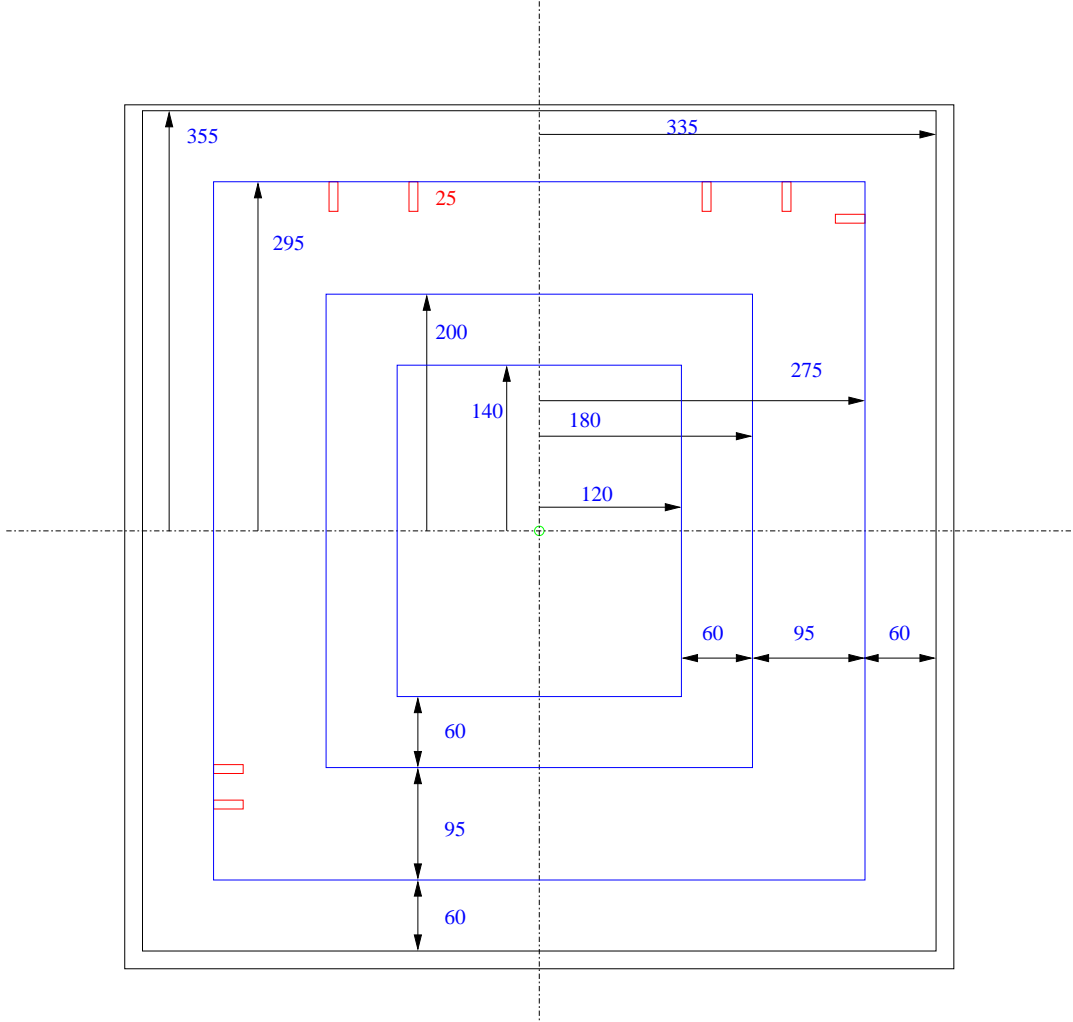}
\caption[Dimensions of the Double-CHOOZ far detector]
{Dimensions of the Double-CHOOZ far detector (in cm). Starting from the
  center we have: the neutrino target region composed of Gd doped liquid
  scintillator (12.7~m$^3$), the $\gamma$-catcher region composed of
  unloaded liquid scintillator (28.1~m$^3$), 
 the non scintillating buffer region (100.0~m$^3$),
  and the veto (110.0~m$^3$). The CHOOZ-near detector is
  identical up to and including the PMT support structure; however, 
  its external muon veto is slightly larger to better reject the
  cosmic muon induced backgrounds.}
\label{fig:detectorsize}
\end{center}
\end{figure}

Starting from the center of the target the detector elements are as
follows (see Figures~\ref{fig:choozfar}, ~\ref{fig:chooznear} and 
~\ref{fig:detectorsize}).

\begin{itemize}
\item{\bf $\nuebar$ target} (12.7 m$^3$)\\
A 120 cm radius, 280 cm height, 6-10~mm width acrylic cylinder,
filled with 0.1\% Gd loaded liquid scintillator target
(see Section~\ref{sec:scintillator}).
\item{\bf $\gamma$-catcher} (28.1 m$^3$)\\
A 60 cm buffer of non-loaded liquid scintillator with the same
optical properties as the $\nuebar$ target 
(light yield, attenuation length).
This scintillating buffer around the target is necessary to
measure the gammas from the neutron capture on Gd, to measure the
positron annihilation, and to reject the background from fast neutrons.

\item{\bf Non Scintillating Buffer} (100 m$^3$)\\
A 95 cm thick cylindrical buffer of non scintillating liquid, to decrease the
level of accidental background (mainly the contribution from
 photomultiplier tubes radioactivity).
\item{\bf PMT supporting structure}
\item{\bf Inner Veto system}(100 m$^3$) \\
A 60 cm thick cylindrical 
veto region filled with liquid scintillator for the far
detector, and a slightly larger one (about 100~cm) for the near detector.
\item{\bf Outer Veto system}(5000 tubes) \\ A four-layer
proportional tube system will identify and locate throughgoing
muons with 98\% efficiency.
\end{itemize}

\par Table~\ref{choozto2chooz} summarizes the control of the
systematic uncertainties that was achieved in the CHOOZ experiment
as well as the goal for Double-CHOOZ. The main
uncertainties at CHOOZ came from the uncertainty in the
knowledge of the
antineutrino flux coming from the reactor. 
This systematic error vanishes by adding a near detector.
The non-scintillating buffer will reduce the singles rates in each
detector by two orders of magnitude with respect to CHOOZ, which
had no such buffer. 
The positron detection threshold will be about 500~keV, well below the
1.022~MeV physical threshold of the inverse beta decay reaction.
Such a low threshold has three advantages:
\begin{itemize}
\item{The systematic error due to this threshold is suppressed.
It was an 0.8\% source of systematic error in 
CHOOZ\cite{choozlast}.}
\item{Backgrounds below 1~MeV reactor neutrino
threshold can be measured.}
\item{The onset of the positron spectrum provides an additional
  calibration point between the near and far detectors.}
\end{itemize}
This reduction of the singles events relaxes or even suppresses
the localization cuts that were used in CHOOZ\cite{choozlast}
such as the distance of an event to the PMT
surface and the distance between the positron and the neutron.  
These cuts were difficult to calibrate.  Factors affecting
other cuts 
will be 
carefully calibrated between the two detectors.
Most important will be the calibration of the energy selection of the delayed neutron
after its capture on a Gd nucleus (with a mean energy release of 8 MeV
in gammas). The requirement is $\sim$100~keV on the precision of this cut
between both detectors, which is feasible with standard techniques using
radioactive sources (energy calibration) and lasers (optical
calibrations) at different positions throughout the detector active
volume. The sensitivity of a reactor
experiment of Double-CHOOZ scale ($\sim$300~GW$_{{th}}$.ton.year) is mostly
given by the total number of events detected in the far detector. The requirement on the
positron energy scale is then less stringent since the weight of the
spectrum distortion is low in the analysis.   

\par A summary of key detector parameters is given in Table~\ref{tab:detsum}.
\begin{table}[htbp]
\begin{center}
\begin{tabular}{|lrr|}
\hline
 & \multicolumn{1}{c}{CHOOZ} & \multicolumn{1}{c|}{Double-CHOOZ}  \\
\hline
 Reactor fuel cross sections & 1.9\% & ---   \\
 Number of protons     & 0.8\% & 0.2\%  \\
 Detection efficiency   & 1.5\% & 0.5\%  \\
 Reactor power         & 0.7\% & ---    \\
 Energy per fission    & 0.6\% & ---    \\
\hline
\end{tabular}
\caption[Systematic errors in CHOOZ 
and Double-CHOOZ goals]{Summary of the systematic errors
  in CHOOZ and the goals for Double-CHOOZ. 
}
\label{choozto2chooz}
\end{center}
\end{table}
\begin{table}[htbp]
\begin{center}
\begin{tabular}{|rrc|}
\hline
$\nuebar$ far detector events & 60,000 & \\
$\nuebar$ near detector events & 3 Million & \\
Far Detector Distance & 1.05 km & \\
Far Detector Overburden & 300 m.w.e. & \\
$\nuebar$ target volume & 12.67 m$^3$ & Gd loaded LS (0.1\%) \\
$\nuebar$ target dimensions  & 120 cm $\times$ 280 cm & radius $\times$ height \\
$\gamma$ catcher volume & 28.1 m$^3$ & unloaded LS \\
$\gamma$ catcher thickness & 60 cm & \\
Buffer volume & 100 m$^3$ & non-scintillating organic liquid\\
Veto volume & 110 m$^3$ & low scintillating organic liquid \\

Near Detector Distance & 100-200 m & \\
Near Detector Overburden & 50-80 mwe & \\
Signal to Noise Goal & 100:1 & \\
Thermal Power & 4.25 Gw & each of 2 cores \\
Electric Power & 1.3 GWe & each of 2 cores \\
Running time at full power & 3 years & \\
Number of Phototubes & 512 8" &  per detector \\
Phototubes Coverage & 12.9\%  &   \\
$\mxang$ goal in 3 years & 0.03 & (90\% CL) \\
\hline
\end{tabular}
\caption[Detectors] {Summary of the some parameters  of the
proposed Double-CHOOZ experiment.}
\label{tab:detsum}
\end{center}
\end{table}

\subsection{Scintillator}
\label{sec:scintillator}
\subsubsection{Liquid inventory}

\par The Double-CHOOZ detector design requires 
different liquids in the separate detector volumes as shown in 
Figures~\ref{fig:choozfar} and \ref{fig:chooznear}.
The inner most volume of 12.7~m$^3$, the $\nuebar$-target, 
contains a proton rich liquid 
scintillator mixture loaded with gadolinium (Gd-LS) at a concentration 
of approximately 1~g/liter. The neutron capture time in 0.1\%
Gd loaded scintillator is 30$\mu$s.
The adjacent volume, the $\gamma$-catcher,
has a volume of 28~m$^3$ and is filled with an unloaded liquid 
scintillator. The photomultipliers are immersed in a non-scintillating 
buffer in order to shield the active volume from the gamma rays
emitted by them. The volume of the buffer liquid is  
approximately 100~m$^3$. Last, an 
instrumented optically isolated
volume of approximately
110~m$^3$ encloses the whole setup serving as a shield against
external radiation and as a muon veto system. Table~\ref{tab:detsum}
includes a summary of the liquid inventory of a single detector system.

\par The selection of the organic liquids are guided by physical and 
technical requirements, as well as by safety considerations. 
In particular, the solvent mixtures or their 
components have a high flash point 
(e.g. phenyl-xylylethane (PXE): flash point (fp) 145~$^o$C, 
dodecane: fp 74~$^o$C, mineral oil: fp 110~$^o$C).  
The $\nuebar$-target and $\gamma$-catcher have as solvent a  
mixture of 80\% dodecane and 20\% PXE, or alternatively
trimethyl-benzene (PC). Mineral oil is under study 
as an alternative to dodecane. 
A similar solvent mixture matching the density of the 
$\gamma$-catcher and $\nuebar$-target, will be used as the buffer 
liquid, however with the addition of a scintillation light quencher 
(e.g. DMP). The veto volume contains low-scintillating organic liquid and will be 
equipped with PMTs. 

\par Research with gadolinium loaded scintillator at MPIK and LNGS/INR 
indicates that suitable gadolinium loaded scintillators can be 
produced using the chemistry of beta-diketone complexes as well
as using a single carboxylic acid stabilized by careful control 
of pH. 
Several prototype detectors filled with 
different scintillator samples are continuously
measured in the LENS low-background facility at Gran Sasso
since October 2003 to study the stability of the scintillator 
as well as backgrounds. No change in light yield nor in attenuation
length has been observed.
Furthermore, research is being carried out to achieve stability 
with respect to interaction with detector container materials, through
the adjustment of inert solvent components of the 
scintillator while simultaneously retaining high scintillation yields.
A variety of tests have been carried out to date that show satisfactory
optical properties using  either  Beta-diketonate (BDK) Gd-LS
or Carboxylate (CBX) Gd-LS \cite{bib:choozloi}.

From the results of the laboratory research, we now have two 
working Gd-LS formulations and we expect that
both the BDK and CBX systems will comply with the design goals 
of Double-CHOOZ. The designation of the default and backup LS formulation 
will take place in 2005.  A further outcome is the detailed engineering
of the Gd-LS production scheme. This will be a critical input
for the finalization of the scintillator fluid systems discussed in 
the next subsection. The final selection of the buffer and veto liquids will 
be done contingent upon the mechanical design of the containment 
vessels and the definition of the Gd-LS formulation.

\subsubsection{Scintillator fluid systems}

\par
The scintillator fluid systems (SFSs) include the {\bf off-site SFS} for 
production, purification and storage of the Gd-LS, as well as the 
$\gamma$-catcher LS. 
The {\bf on-site SFS} will be on the reactor area,  
close to the experimental location.

The SFSs scheme envisions the production and storage of the 
complete Gd-LS for both the near and 
far detector, in order to assure identical {\it proton
per volume} concentrations.
The off-site SFS will include ISO-containers for storage and subsequent
transport to the experimental site. Moreover, it will include
a purification column, a nitrogen purging unit, a mixing chamber,
nitrogen blankets 
and auxiliary systems. A similar system, known as Module-0\cite{mod0}, 
has been constructed by groups associated with
Borexino. Since the specifications for Module-0 are more demanding
than required for Double-CHOOZ, no problems are anticipated.

The on-site SFSs will consist of an 
area above ground close to the detector sites for the transport tanks 
which will be connected to the detector by a tubing system.
The purpose of the on-site SFS is to transfer the different liquids
from their transport container into the detector volumes in a safe and 
clean way. The different detector volumes will be filled 
simultaneously and kept at equal hydrostatic
pressures.

\subsection{Calibration}
\label{sec:calibration}
The main goal of the calibration effort is to reach maximum
sensitivity to neutrino oscillations by comparing the positron
energy spectra measured by the Double-CHOOZ far and Double-CHOOZ 
near detectors.
Calculations show that a relative difference both in geometry
(construction) and in response of detectors slightly distorts the
ratio of the spectra in both detectors. Therefore, appropriate
calibration coefficients and concise measurements of the systematic
uncertainties need to be measured in situ.
The calibration sources (See Table~\ref{t:techniques})
must be deployed regularly throughout
the detector active volume to monitor the detector
response to positrons, neutron captures, gammas and
the backgrounds in the Double-CHOOZ experiment. This requires
a dedicated mechanical system in order to introduce calibration
sources into the different regions of the detector.
\begin{table}[htpb]
\begin{center}
 \begin{tabular}{|ll|}
 \hline
 Technique & Calibrations \\
 \hline 
 Optical Fibers, Diffusive Laser ball & Timing and Charge Slopes
 and Pedestals, \\
                   & attenuation length of detector components \\
 \hline
 Neutron Sources: Am-Be, $^{252}$Cf & Neutron response, relative
 and \\
                                    & absolute efficiency, capture time \\
 \hline
 Positron Sources: $^{22}$Na, $^{68}$Ge & $e^{+}$ response,
 energy scale, trigger thresh. \\
 \hline
 Gamma Sources: & Energy linearity, stability, resolution, \\
                & spatial and temporal variations. \\
 $^{137}$Cs & $\beta^{-}$, 0.662 MeV \\
 $^{22}$Na & $\beta^{+}$, 1.275 MeV + annih \\
 $^{54}$Mn & EC, 0.835 MeV \\
 $^{65}$Zn & 1.35 MeV \\
 $^{60}$Co & EC, 1.173, 1.33 MeV \\
 $^{68}$Ge & EC, $\beta^{+}$ 1.899 MeV + annih \\
 $^{88}$Y  & EC, 0.898, 1.836 MeV \\
 H neutron capture & 2.223 MeV \\
 $^{241}$Am-$^{9}$Be & ($\alpha$,n) 4.44 MeV ($^{12}$C)\\
 Gd neutron capture & Spectrum in 8 MeV window \\
$^{12}C$ neutron capture & 4.97 MeV \\
\hline
 $^{228}$Th & 2.615 MeV \\
 $^{40}$K & EC, $\beta^{+},\beta^{-}$, 11\% 1.46 MeV \\
\hline
 \end{tabular}
 \caption[Techniques available to calibrate the Double-CHOOZ 
experiment]{Table showing the different techniques that are 
available to calibrate the Double-CHOOZ experiment.}
 \label{t:techniques}
\end{center}
\end{table}

\par
There are a number of specific tasks for a successful calibration of
the detectors. These include optical calibrations (single
photoelectron (PE) response, multiple PE response, detector
component optical constants), electronic calibrations (trigger
threshold, timing and charge slopes and pedestals, dead time), 
energy (energy scale and resolution), and neutron and positron detection
efficiency and response. In addition, detector calibrations 
test the Monte-Carlo and analysis code to verify the accuracy of
the simulations, throughout the detector (spatially), and during
the lifetime of the experiment.

\par The optical calibrations are based on the experience with CHOOZ
 and the CTF-Borexino experiments. In CTF-Borexino the
optical calibration consists of a UV pulsed-laser (jitter less than 1~ns)
coupled to an optical fiber illuminating separately each PMT. This
allows the single PE response to be measured since the amplitude
of the pulse is tuned to approximately a single PE. This technique
allows the gain, timing slope, charge slope and pedestals to be
determined relative to individual PMTs and to the triggers. In
addition to the optical fiber calibration, the light attenuation
in the liquid scintillator will be monitored using a diffusive laser
ball source, as has been successfully used by many experiments
over the last twenty years\cite{Ahmad:2002jz}.
This source illuminates all the PMTs isotropically and allows the
attenuation length of the detector components and the PMT angular
response to be measured as a function of photon wavelength.

\par We will calibrate the detector energy
response to gammas from 1~MeV to about 10~MeV corresponding to the
endpoint of the fission product beta decays. 
It is necessary to also know the energy
scale in the window 6-10~MeV to be able to identify
the delayed second trigger as a neutron. The required 
accuracy is 100~keV.
This will be accomplished by deploying various higher
energy gamma calibration sources (see Table~\ref{t:techniques})
and by detailed Monte-Carlo simulations in the energy region where
there are no calibration sources.

The overall energy scale can be determined from the position of
the 0.662~MeV peak of the $^{137}$Cs source, and then verified by
calibration with several gamma sources (see
Table~\ref{t:techniques}) in different energy ranges: $^{54}$Mn
(0.835~MeV),  $^{65}$Zn (1.351~MeV),
$^{60}$Co, and $^{228}$Th (2.614~MeV). 
The capture of
neutrons from an Am-Be source scintillator (to be discussed later)
can also be used as a high energy gamma source as it produces
prompt 4.4~MeV gammas. We will also use the natural sources from
radioactive impurities of the detector materials ($^{40}$K,
$^{208}$Tl …) and neutrons produced by cosmic muons for energy
calibration. Since these sources are present permanently, they are
useful for monitoring the stability of the energy response. 

\par Positron detection can be calibrated with a
$^{22}$Na source. It emits a 1.275~MeV primary
gamma accompanied by a low energy positron which annihilates
inside the source container. The primary and annihilation gammas
from the source mimic the positron annihilation resulting from an
antineutrino event inside the detector. Another 
positron
source is $^{68}$Ge, which produces positrons with higher
energies, and therefore calibrates higher energy positrons.
$^{68}$Ge decays by EC to $^{68}$Ga and $\beta^{+}$-decays to
stable $^{68}$Zn with an endpoint of 1.9~MeV. This isotope also
has the advantage that it produces only low energy gammas in
coincidence with the nuclear decay, and the $\beta^{+}$ has an
endpoint of 1.889~MeV 89\% of the time. A second purpose of this
source (if a source is constructed so that the beta is absorbed by
the shielding surrounding the source) is to tune the trigger
threshold to be sensitive to annihilation gammas and to monitor
its stability. A $^{68}$Ge source was successfully used in
the Palo Verde reactor neutrino experiment \cite{paloverdenim1}.

\par There are two suitable and accessible neutron sources for neutron
calibration: the Am-Be source and $^{252}$Cf spontaneous fission
source. These sources emit neutrons with different energy spectra
from what is expected from inverse beta decay, and thus the
importance of these differences needs to be quantified. To
decrease the background during neutron source deployment, neutrons
from Am-Be should be tagged by the 4.4~MeV gamma emitted in
coincidence with the neutron. This will allow the neutron capture
detection efficiency to be determined independent of knowing the
precise rate of the neutron source, because every time a 4.44~MeV
gamma is detected a neutron is released \cite{SCroft}.   It should
be noted that the use of this gamma involves a significant correction
for the n-p elastic scattering which often takes place in coincidence, thus the 
uncertainty is typically larger than for single gamma sources.
The absolute neutron detection efficiency will be determined 
with a $^{252}$Cf source by
using the known neutron multiplicity (known to 0.3\%). For the
source placed in the center, the size of the Gd region will be larger
than the neutron capture mean free path, so that the neutron
capture will be studied independent of the presence of the acrylic
vessel. In order to tag the neutron events, a small fission
chamber will be used to detect the fission products.
Therefore, neutron source calibrations will provide us with the
relevant data to calibrate the detector response to neutrons. In
particular, neutron sources will allow us to measure the absolute
neutron efficiency, to determine and monitor the appropriate
thresholds of neutron detection, and to measure the neutron
capture time for both the far and near detectors.
\subsection{Background}
\label{sec:background}

\par The signature for a neutrino 
event is a prompt signal with a minimum energy of about 
1~MeV and a delayed 8~MeV signal after neutron capture in gadolinium. 
This may be mimicked by background events which 
can be divided into two classes: 
accidental and correlated events. The former occur when a neutron like event 
by chance falls into the 
time window (typically few 100~$\mu$s)  after an event in the 
scintillator with an energy of more than one MeV. 
The latter is formed by neutrons which slow down by scattering 
in the scintillator, deposit  $> 1$~MeV visible energy  and are captured in 
the Gd~region. 
The sources and rates of various backgrounds are used in this
subsection to determine the necessary overburden of the near 
detector and the purity limits for detector components.\\

\subsubsection{Intrinsic  backgrounds}

\par Background due to beta and gamma events 
above  about 1~MeV can take place
 in the scintillator or in the acrylic vessels which
contain the liquid. 
The contribution from the Uranium and Thorium chains is reduced to a few elements, 
as all alpha events show quenching with visible energies well below 1~MeV. 
Furthermore the short delayed Bi-Po coincidences in both chains can be
detected event by event, 
and hence rejected. The decays of $^{234}$Pa (beta decay, $Q = 2.2$~MeV), 
$^{228}$Ac (beta decay, $Q = 2.13$~MeV) and $^{208}$Tl (beta decay, $Q = 4.99$~MeV) 
need to be considered. 
Assuming radioactive equilibrium the beta/gamma background rate due to both chains can be 
estimated by 
$b_1 \simeq M_{U} \cdot 6\cdot 10^3~{\mathrm s^{-1}} +M_{Th} \cdot 4\cdot 10^3~{\mathrm s^{-1}}$,
where the total mass of U and Th is given in grams. Taking into account the total 
  scintillator mass of the neutrino target plus the $\gamma$-catcher, this rate can be expressed by  
 $b_1\simeq 3~{\mathrm s^{-1}} (c_{U,Th}/10^{-11})$ , where $c_{U,Th}$ is the 
  mass concentration of Uranium and Thorium in the liquid. 

\par The contribution from $^{40}$K can be expressed by
$b_2\simeq 1~{\mathrm s^{-1}} (c_{K}/10^{-9})$, where $c_{K}$ is the mass concentration of natural 
   K in the liquid. 
The background contribution due to U, Th and K in the acrylic vessels can be written as 
   $b_3 \simeq 2~{\rm s^{-1}} (a_{K}/10^{-7}) + 5~{\rm s^{-1}} (a_{U,Th}/10^{-9})$,
   where $a_{K}$ and $a_{U,Th}$ describe the mass concentrations of K, U and Th in the acrylic. 
Measured values of a in the CTF of Borexino and by SNO show that
in principle the beta/gamma rate in the detector due to intrinsic 
 radioactive elements can be kept at levels well below 1~s$^{-1}$. 

\par The dominant contribution to 
the external gamma background is expected to come from the photomultipliers (PMTs) and 
structure material. 
Again contributions from U, Th and K have to be considered. 
However, because of the shielding of the buffer region only the 2.6~MeV gamma 
emission from $^{208}$Tl has to be taken into account. 
The shielding factor $S$ due to the buffer liquid is calculated to be $S \sim 10^{-2}$. 
The resulting gamma background in the neutrino target plus the $\gamma$-catcher 
can be written as $b_{ext} \simeq 2~{\rm s^{-1}} (N_{{PMT}}/500)$, 
where $N_{PMT}$ is the number of PMTs.  Section \ref{sec:pmt} contains
a more detailed study of this background.

Neutrons inside the target may be produced by spontaneous fission of heavy elements and 
by ($\alpha$,n)-reactions. For the rate of both contributions the concentrations of U and Th in 
the liquid are the relevant parameters. 
The neutron rate in the target region can be written as  
$n_{int}\simeq 0.4~{\rm s^{-1}} (c_{U,Th}/10^{-6})$. 
For the aimed concentration values the intrinsic contribution to 
the neutron background is negligible.

\subsubsection{External background sources}

Several sources contribute to the external neutron background. 
We first discuss external cosmic muons which produce neutrons in the target region via spallation and muon capture. 
Those muons intersect the detector and should be identified by the veto
systems. 
However, some neutrons may be captured after the veto time window. 
Therefore we estimate the  rate of neutrons, which are 
generated by spallation processes of through going muons and by stopped negative muons which 
are captured by nuclei.\\

The first contribution is estimated by calculating the muon 
flux for different shielding values 
and taking into account an $E^{0.75}$ dependence for 
the cross section of neutron production, 
where $E$ is the depth dependent mean energy of the total muon flux. 
The absolute neutron flux is finally obtained by 
considering measured values in several experiments 
(LVD\cite{lvd}, MACRO\cite{Ambrosio:1998wu}, CTF\cite{CTF}) 
in the Gran Sasso underground laboratory
and extrapolating these results by comparing muon fluxes and mean energies for the different shielding factors. 
Table~\ref{t2} gives the expected neutron rate depending on the shielding.
\begin{table}[htbp]
\begin{center}

\begin{tabular}{|lrrrrr|}
\hline
\multicolumn{1}{|c}{Overburden}  & \multicolumn{1}{c}{$\mu$ rate}  
& \multicolumn{1}{c}{$<E_\mu>$}  & \multicolumn{1}{c}{Neutrons}  
& \multicolumn{1}{c}{$\mu$ stopping rate}   &  \multicolumn{1}{c|}{Neutrons}   \\
& & & through going $\mu$ & & stopping $\mu$ \\
(m.w.e.) & (s$^{-1}$) & (GeV) & (s$^{-1}$) & (s$^{-1}$) & (s$^{-1}$) \\
\hline
40      &       $1.1 \cdot 10^{3}$    &  14  & 2        & 
$5 \cdot 10^{-1}$     &  0.7       \\
60    & $5.7 \cdot 10^{2}$    &  19  & 1.4       & 
$3 \cdot 10^{-1}$     &  0.4       \\
80    & $3.5 \cdot 10^{2}$    &  23  & 1        & 
$1.2 \cdot 10^{-1}$     &  0.2       \\
100   & $2.4 \cdot 10^{2}$    &  26  & 0.7  & 
$6 \cdot 10^{-2}$     &  0.08       \\
300   & $2.4 \cdot 10^{1}$    &  63  & 0.15     & 
$2.5 \cdot 10^{-3}$     &  0.003       \\
\hline
\end{tabular}
\caption[Estimated neutron rate in the active detector region due to
    through going and stopped cosmic muons.]
{\label{t2} Estimated neutron rate in the active detector
  region due to through going cosmic muons.}
\end{center}
\end{table}

Negative muons which are stopped in the target region can be captured by nuclei where a  
neutron is released afterward. 
The rate can be estimated quite accurately by 
calculating the rate of stopped muons as a 
function of the depth of shielding and 
taking into account the ratio between the $\mu$-life time and $\mu$-capture times.
As the capture time in Carbon is known to be around 25 $\mu$s
($\approx$1~ms in H) 
only about 10\% of captured muons may create a neutron.
Since the concentration in Gd is so low, its effect can be neglected here.
The estimated results are shown in Table~\ref{t2}.
The neutron generation due to through going muons dominates.

\subsubsection{Beta-neutron cascades}

\par Muon spallation on $^{12}$C nuclei in the organic liquid scintillator 
may generate
$^8$He, $^9$Li, and $^{11}$Li which may undergo beta decay with a 
neutron emission.
Those background events show the same signature as a 
neutrino event.
For small overburdens the muon flux is too high to allow tagging 
by the muon veto,
as the lifetimes of these isotopes are between 0.1~s and 1~s.
The cross sections for the production of $^8$He, $^9$Li have been 
measured by a group of TUM at the SPS at CERN
with muon energies of 190~GeV (NA54 experiment\cite{NA54}). In this 
experiment only the combined production  $^8$He + $^9$Li were obtained 
without the ability to separate them.  


A conservative estimate of 2 events per day in the target region can be
estimated for 300~m.w.e. shielding by assuming $E^{0.75}$ scaling as we
did in calculating the neutron flux.  An alternative scaling has been
suggested whereby the number of $^{9}$Li/$^{8}$He-producing interactions
varies in proportion to the flux of muons over $500~GeV$\cite{horthonsmith}, 
leading to a lower event rate of 0.4 per day.

In Table~\ref{tab:na54choozfar} all radioactive $^{12}$C-spallation products including the 
beta-neutron cascades are
shown with the estimated event rates in both detectors. \\
 
\par The Q-values of the beta-neutron cascade decays
is 8.6~MeV, 11.9~MeV, 20.1~MeV for $^8$He, 
$^9$Li, and $^{11}$Li, respectively.
In the experiment the $^8$He production rate might be measured if we 
set a dedicated trigger after a muon event in the target region looking 
for the double cascade of energetic betas 
($^8$He $\rightarrow$ $^8$Li $\rightarrow$ $^8$Be) occurring 
in 50\% of all $^8$He decays.  
Nothing similar exists in the case of $^9$Li, but the beta endpoint 
here is above the positron endpoint induced by reactor
antineutrinos.  
KamLAND has measured that
the production of $^9$Li (lifetime 838~ms) dominates the production of 
$^8$He (lifetime 119~ms) by at
least a factor of 8.  
Nevertheless, from the results of the 
NA54 experiment\cite{NA54}, the total 
cross section of  $^8$He + $^9$Li is known, and if the $^8$He is
evaluated separately,  some redundancy on the total $\beta$-neutron cascade will be available. 
The neutrons emitted in the $^8$He decays are typically around 
1~MeV.   With an overburden more than 50 mwe, these backgrounds
are acceptable.

\begin{table}[htbp]
\begin{center}
\begin{tabular}{|c|rr|rr|}
\noalign{\bigskip}
\hline
           & \multicolumn{2}{|c}{Near detector} & 
\multicolumn{2}{c|}{Far detector} \\
\hline
  Isotopes & $R_{\mu}$ &  $R_{\mu}$  & $R_{\mu}$  &  $R_{\mu}$   \\
           & ($E^{0.75}$ scaling) & ($E>500$~GeV) & ($E^{0.75}$
  scaling) &  ($E>500$~GeV)\\             & \multicolumn{4}{c|}{per day} \\ 
\hline 
$^{12}$B & \multicolumn{4}{c|}{not measured} \\
$^{11}$Be & $<18$  &  $<3.8$ &  $<2.0$  &  $<0.45$ \\
$^{11}$Li & \multicolumn{4}{c|}{not measured} \\
$^{9}$Li  & $17 \pm 3$ & $3.6$ &  $1.7 \pm 0.3$ & $0.36$ \\
$^{8}$Li  & $31 \pm 12$ & $6.6$ &  $3.3 \pm 1.2$ & $0.7$\\ 
$^{8}$He  & \multicolumn{4}{c|}{$^{8}$He \& $^{9}$Li measured together}\\
$^{6}$He  & $126 \pm 12$ & $26.8$ &  $13.2 \pm 1.3$ & $2.8$\\
$^{11}$C & $7100\pm455$ & $1510$  & $749\pm48$ & $159.3$ \\
$^{10}$C  & $904\pm114$ & $192$ & $95\pm12$ & $20.2$ \\
$^{9}$C   & $38\pm12$ & $8.1$   & $4.0\pm1.2$ & $0.85$\\
$^{8}$B   & $60\pm11$ & $12.7$  & $5.9\pm1.2$ & $1.25$\\
$^{7}$Be  & $1800\pm180$ & $382.9$ & $190\pm19$ & $40.4$\\
\noalign{\smallskip} 
\hline
\end{tabular}
\caption[Radioactive isotopes induced by muons in liquid scintillator
  targets at the Double-CHOOZ near and far detectors.]
{Radioactive isotopes produced
  by muons and their secondary shower particles in liquid scintillator
  targets at the Double-CHOOZ near and far detectors. 
The rates $R_{\mu}$ (events/d) are given for a target of 
$4.4 \times 10^{29}$ $^{12}$C (For a mixture of 80\% Dodecane and 20\%
PXE, 12.7~m$^3$)  at a depth of 60~m.w.e. for the near detector 
and 300~m.w.e. for the far detector. 
Columns 3 and 5 correspond to an estimate of the
number of events assuming that the isotopes are produced only by high
energy muon showers $E>500$~GeV\cite{horthonsmith}. A neutrino signal
rate of 85 events per day is expected at Double-CHOOZ far, without
oscillation effect (for a power plant running at nominal power, both
  dead  time and detector efficiency are not taken into account here).}
\label{tab:na54choozfar}
\end{center}
\end{table}
\subsubsection{External neutrons and correlated events}

\par Very fast neutrons, generated by cosmic muons 
outside the detector, may penetrate into the target region.
As the neutrons are slowed down through scattering, recoil 
protons may give rise to a visible signal 
in the detector. This is  followed by a delayed neutron capture event.
Therefore, this type of background signal gives 
the right time correlation and can mimic a neutrino event.
Pulse shape discrimination in order to distinguish 
between $\beta$ events and recoil protons is in principle possible, but 
will only be used as a consistency check since the errors are not small.

\par A Monte-Carlo program has been written to estimate the 
correlated background rate for 
a shielding depth of 100~m.w.e. and a flat topology.
In order to test the code the correlated background for the Chooz experiment 
(different detector dimensions, 300~m.w.e. shielding)
has been calculated with the program.
The most probable background rate was determined to be 0.8 counts per day.
A background rate higher than 1.6 events per day is excluded by 90\%~C.L.
This has to be compared with the measured rate of 1.1 events per day.
We conclude that the Monte-Carlo program reproduces 
the real correlated background value  
roughly within a factor~2. \\

\par For Double-Chooz we calculated the correlated background rate 
for 100~m.w.e. shielding and estimated
the rates for other shielding values by 
taking into account the different muon fluxes and assuming
a $E^{0.75}$ scaling law for the probability to produce neutrons.  
The neutron capture rate in the Gd~loaded scintillator for 
an overburden of 100~m.w.e. is about 300/h.
However, only 0.5\% of those neutrons create a 
signal in the scintillator within the neutrino window
(i.e. between 1~MeV and 8~MeV), because most deposit much more energy  
during the multiple scattering processes.
The quenching factors for recoil protons and carbon nuclei has been taken into account.
In addition around 75\% from those events generate a signal in the 
inner muon veto above 4~MeV (visible 
$\beta$ equivalent energy).
In total the correlated background rate is estimated to be about 3~counts 
per day for 100~m.w.e. shielding.  This can be measured to high precision
using events tagged by the inner and/or outer veto.
In Table~\ref{t4} the estimated correlated background rates are shown for different shielding 
depths.\\
\begin{table}[htbp]
\begin{center}

\begin{tabular}{|lrr|}
\hline
\multicolumn{1}{|c}{Overburden}  & \multicolumn{1}{c}{Total neutron rate}  
&  \multicolumn{1}{c|}{Correlated background rate} \\

    (m.w.e.)     &      in $\nu$-target (h$^{-1}$)        &   (d$^{-1}$) \\ 
\hline
40      & 829       &      8.4        \\
60      & 543       &      5.4    \\
80      & 400       &      4.2    \\
100     & 286       &      3.0    \\
300     &  57       &      0.5    \\
\hline
\end{tabular}
\caption[Limits on the estimated neutron rate and the correlated
  background rate due to fast neutrons]
{\label{t4} Estimated neutron rate in the target region and the correlated background rate due to fast neutrons generated outside the detector by cosmic muons.}
\end{center}
\end{table}

\par The correlated background rates can be compared with accidental rates,
where a neutron signal falls into the time window opened by a 
$\beta^+$-like~event.
The background contribution due to accidental delayed coincidences can
be determined {\it in situ} by measuring the single counting rates of neutron-like and $\beta^+$-like events.
Therefore the accidentals are not so dangerous as correlated background events.
Radioactive elements in the detector materials will be carefully controlled,
especially in the scintillator itself, so the beta-gamma rate above 1~MeV 
will be only a few counts per second.
For a time window for the delayed coincidence of $\sim$200~$\mu$s 
(this should allow a highly efficient neutron detection 
in Gd~loaded scintillators),
and a veto efficiency of 98\%, 
the accidental background rates are estimated  in Table~\ref{t5}.
The rate of neutrons which cannot be correlated to muons (``effective neutron rate'') 
is calculated by $ n_{{eff}} = n_{{tot}} \cdot (1 - \epsilon ) $, 
where $n_{tot}$ is the total neutron rate 
(sum of the numbers given in Table~\ref{t2}) 
and $\epsilon$ is the veto efficiency.
If the veto efficiency is 98\% or better, the accidental background for the far
detector is far below one event per day (see following Table~\ref{t5}).\\

\begin{table}[htbp]
\begin{center}

\begin{tabular}{|lrr|}
\hline
\multicolumn{1}{|c}{Overburden}  & \multicolumn{1}{c}{Effective neutron rate}  
& \multicolumn{1}{c|}{Accidental background rate} \\
            (m.w.e.)     &      ($\rm h^{-1}$)           &  ($\rm d^{-1}$)        \\
\hline
40    &  97   & 2.4            \\
60    &  65   & 1.6       \\
80    &  43   & 1.0           \\
100   &  28   & 0.7       \\
300   &   6   & 0.15           \\
\hline
\end{tabular}
\caption[Estimated accidental event rates for different
  shielding depths.] {\label{t5} Example of estimated accidental event
  rates for different shielding depths. The rates scale with the total
  beta-gamma rate above 1~MeV 
(here $b_{tot} = b_{ext} + b \approx 2.5~\rm s^{-1}$), the time window 
(here $\tau = 200 \, \mu $s) and the effective neutron background rate.
A muon veto efficiency of 98\% was assumed.}
\end{center}
\end{table}

\par We conclude that correlated events are the 
most severe background source for the experiment.
Two processes mainly contribute: $\beta$-neutron cascades 
and very fast external neutrons.
Both types of events are coming from spallation processes of high energy muons.
In total the background rates for the 
near detector will be between 9/d and 23/d if a shielding
of 60~m.w.e. is chosen.
For the far detector a total background rate between 1/d and 2/d is estimated.

\subsection{Errors}

\label{sec:errors}

\par In the CHOOZ experiment, the total systematic error was
2.7\%. The goal of Double-CHOOZ is to reduce the overall systematic 
uncertainty to 0.6\%. 
A summary of the CHOOZ systematic errors is given in 
Table \ref{choozto2chooz}\cite{choozlast}. The right column 
presents the new experiment goals. 

\subsubsection{Detector systematic uncertainties}

\par The distance from the CHOOZ detector to the cores of the nuclear plant
was measured to within $\pm 10$~cm. 
This translates into a systematic error of 0.15\% in Double-CHOOZ, because the
effect becomes relatively more important for the near detector located 
 100-200~meters away from the reactor.
Specific studies are currently ongoing to guarantee this 10~cm error. 
 Furthermore, the ``barycenter'' of the neutrino emission in the
 reactor core must be monitored with the same precision. 
In a previous experiment at Bugey\cite{Bugey}, a 5~cm change of
this barycenter was measured and monitored, using the instrumentation
of the nuclear power plant\cite{garciaz1992}. Our goal is to
confirm that this error could be kept below 0.2\%.
\subsubsection{Volume measurement}
\par In the CHOOZ experiment, the volume measurement was done with an 
absolute precision of  0.3\%\cite{choozlast}.
The goal is to reduce this uncertainty by a factor of two, but only on the 
relative volume measurement between the two inner acrylic vessels (the
other volumes do not constitute the $\nuebar$ target).
We plan to use the same mobile tank to fill both targets; a pH-based
measurement is being studied as well.  
A more accurate measurement could be performed by combining a
traditional flux measurement with a weight measurement of the quantity
of liquid entering the acrylic vessel.
Furthermore we plan to build both inner acrylic targets at
the manufacturer and to move each of them as a single unit to the detector
site.  Both inner vessels will undergo precise filling tests at the
manufacturer.
\subsubsection{Density}

\par The uncertainty of the density of the scintillator is $\sim$0.1\%.
The target liquid will be prepared in a large single batch,
so that they can be used for the two detector fillings. 
The same systematic effect will
then occur in both detectors and will not contribute to the
overall systematic error.
However, the measurement and control of the temperature will be
important to guarantee the stability of the density in both targets
(otherwise it would contribute to the relative uncertainty, 
see Section~\ref{sec:sensitivity}). 
The temperature control and
circulation of the liquid in the external veto will be used to keep both
$\nuebar$ targets at a constant temperature.

\subsubsection{Fraction of hydrogen atoms}

\par This quantity is very difficult to measure, and the error is of the
order of 1\%; however, the target liquid will be prepared in a large 
single batch (see above). 
This will guarantee that, even if the absolute value is not known to
a high precision, both detectors will have the same number of hydrogen 
atoms per
gram. This uncertainty, which originates in the presence of unknown 
chemical compounds in the liquid, does not change with time.
The thermal neutron is captured either on hydrogen or on Gadolinium
(other reactions such as Carbon captures can be neglected). 
\subsubsection{Gadolinium concentration}

\par Gd concentration can be extracted from a time capture measurement
done with a neutron source calibration.
A very high precision can be reached on the neutron detection
efficiency (0.3\%)
by measuring the detected neutron multiplicity from a
Californium source (Cf).
This number is based on the precision quoted in Reference \cite{choozlast}, but
taking away the Monte-Carlo uncertainty, since we work with
two-identical detectors.
We can increase our sensitivity to very small differences in the
response from both detectors by using the same calibration source for
the measurements.
The Californium source calibration can be made all along the z-axis
of the detector, and is thus sensitive to spatial effects due to the variation of 
Gd concentration (staying far enough from the boundary of the target,
and searching for a top/down asymmetry). A difference between the time capture of both
detectors could also be detected with a sensitivity slightly less 
than 0.3\%. 
\subsubsection{Spatial effects}

\par A major advantage of the three volume design for Double-CHOOZ is the
absence of a fiducial volume cut.  The entire inner volume is the fiducial
volume, so the relative normalization depends on detector size and liquid
measurement, and does not depend on phototube properties and detailed
simulations.

\par The $\sim$1\% spill in/out effect oberved in the CHOOZ
experiment\cite{choozlast} cancels by using a set of two identical
detectors.  
An angle effect persists but is much smaller.
A 500~keV energy cut induces a positron inefficiency smaller than 0.1\%.
The relative uncertainties between both detectors lead thus  to an
even smaller systematic error.
\subsubsection{Selection cuts uncertainties}

\par The analysis cuts are potentially important 
sources of systematic errors. In
the CHOOZ experiment it was 1.5\%\cite{choozlast}. 
The goal of the new experiment is to reduce this error by a factor of three.
The CHOOZ experiment used 7 analysis cuts to select the $\nuebar$ 
(one of them had 3 cases, see Section~8.7 of\cite{choozlast}). 
In Double-CHOOZ we plan to reduce the number of selection cuts
 to 3 (one of them will be very loose, and may not even be used). 
This can be achieved because of reduction of the number of 
accidentals background events, only possible with the new detector
design.
To select $\nuebar$ events we have to identify the prompt positron 
followed by the delayed neutron (delayed in time and separated in space).
The trigger will require two local energy depositions of more than
500~keV in less than 200~$\mu$s. 

\par Since any $\nuebar$ interaction deposits at least 1~MeV (slightly less
due to the energy resolution effect) the energy cut at 500~keV does not 
reject any  $\nuebar$ events. 
As a consequence, there will not be any systematic error associated with
 the trigger. 
The only requirement is the  stability of the energy threshold, 
which is related to the energy calibration.

\par The energy spectrum of a neutron capture has two peaks, the first peak 
at 2.2~MeV tagging the neutron capture on hydrogen, 
and the second peak at around 8~MeV tagging the neutron capture on Gd.
The selection cut that identifies the neutron will be set at about 6~MeV,
which is above the energy of neutron capture on hydrogen and
all radioactive contamination.
At this energy of 6~MeV, an error of $\sim$100~keV on the selection
cut changes the number of neutrons by $\sim$0.2\%. 
This error on the relative calibration is achievable by using
the same Cf calibration source for both detectors.

\par The exact analytical behavior
 describing the neutron capture time on Gd is not known, 
so the absolute systematic error for a single detector
cannot be significantly improved with respect to
 CHOOZ\cite{choozlast}. However, the uncertainty originating from
the liquid properties disappears by comparing the near and far
 detector neutron time capture. The remaining effect deals with the
 control of the electronic time cuts. 
For completeness, a redundant system will be designed  in order to control
perfectly these selection cuts (for example
time tagging in a specialized unit and using Flash-ADC's).

\par The distance cut systematic error (distance between prompt and
delayed events) was published as 0.3\% in the CHOOZ
experiment\cite{choozlast}. This cut is difficult to calibrate, 
 since the rejected events are typically $\nuebar$ 
candidates badly reconstructed. In Double-CHOOZ, this cut
 will be either largely relaxed (two meters instead of one meter for instance) 
or totally suppressed, if the accidentals event rate is low enough, as
expected from current simulations).

\par
The Double-CHOOZ veto will consist of liquid scintillator and 
have a thickness of 60~cm at the far site, and
even larger at the near detector site. 
The veto inefficiency comes from the 
through going cables and the supporting structure material. This inefficiency was 
low enough in CHOOZ, and should be acceptable for the 
Double-CHOOZ far detector. 
However, it must be lowered for the near detector because the muon
flux is a factor 30 higher for a shallower overburden of 60~m.w.e..
A constant dead time will be applied in coincidence with each through
going muon. This has to be measured very carefully since the resulting dead time
will be very different for the two detectors: a few percent at the far
detector, and at around~30\% at the near detector. 
A  1\% precision on the knowledge of this dead time is required. 
This will be accomplished with several independent methods:
\begin{itemize}
\item{the use of a synchronous clock, to which the veto will be applied,}
\item{a measurement of the veto gate with a dedicated flash ADC,}
\item{the use of an asynchronous clock that randomly generates two
  signals mimicking the antineutrino tag 
  (with the time between them characteristic of the neutron capture on
  Gd). With this method, all dead times (originating from the veto as
  well as from the data acquisition system) will be measured simultaneously.
  A few thousand such events per day are needed,}
\item{the generation of sequences of veto-like test pulses 
(to compare the one predicted dead time to the actually measured).}
\end{itemize}

\par 
The trigger will be rather simple. It will use only the total analog sum
of  energy deposit in the detector. 
Two signals of more than 500~keV in 200~$\mu$s will be required.

\par
A summary of the systematic errors associated with $\nuebar$ event
selection cuts is given in Table \ref{tab:cuterrors}.
\begin{table}[htbp]
\begin{center}
\begin{tabular}{|lrrc|}
  \hline
  & \multicolumn{1}{c}{CHOOZ}          & \multicolumn{2}{c|}{Double-CHOOZ} \\
  \hline
  selection cut &   rel. error $(\%)$ & rel. error $(\%)$ & Comment\\
  \hline
  positron energy$^\ast$ &  $0.8$ & $0$ & not used \\
  positron-geode distance &  $0.1$ & $0$ & not used \\
  neutron capture &  $1.0$ &  $0.2$ &  Cf calibration\\
  capture energy containment & $0.4$ & 0.2 & Energy calibration \\
  neutron-geode distance &  $0.1$ & 0 & not used  \\
  neutron delay &  $0.4$ & 0.1 &  --- \\
  positron-neutron distance &  $0.3$ & $0-0.2$ & 0 if not used\\
  neutron multiplicity$^\ast$ &  $0.5$ & $0$ & not used\\
  combined$^\ast$ & $1.5$ & $0.2$-$0.3$ & ---\\
    \hline
    \multicolumn{3}{l}{$^\ast${\small average values}} 
\end{tabular}
\caption[Summary of the neutrino selection cut uncertainties]
{Summary of the neutrino selection cut uncertainties. CHOOZ
values have been taken from\cite{choozlast}.}
\label{tab:cuterrors}
\end{center}
\end{table}
We summarize in Table \ref{tab:syscancels} the systematic
uncertainties that largely cancel in the
Double-CHOOZ experiment.
\begin{table}[htbp]
\begin{center}
\begin{tabular}{|lrr|}
\hline
 & \multicolumn{1}{c}{CHOOZ} & \multicolumn{1}{c|}{Double-CHOOZ} \\
\hline
Reactor power             &         0.7\%   &   negligible \\
Energy per fission        &         0.6\%   &   negligible \\
$\nuebar$/fission         &         0.2\%   &   negligible \\
Neutrino cross section    &         0.1\%   &   negligible \\
Number of protons/$\mathrm{cm^3}$  &         0.8\%   &   0.2\% \\
Neutron time capture      &         0.4\%   &   negligible \\       
Neutron efficiency        &         0.85\%  &   0.2\% \\       
Neutron energy cut (E$_\gamma$ from Gd) & 0.4\%   &   0.2\%\\
\hline
\end{tabular}
\caption[ Summary of systematic errors
  that cancel or are significantly decreased in Double-CHOOZ]
{\label{tab:syscancels} Summary of systematic errors
  that cancel or are significantly decreased in Double-CHOOZ.} 
\end{center}
\end{table}
%
%
The error on the absolute knowledge of the chemical composition of the
Gd scintillator disappears. There  remains only the measurement error on the 
volume of target (relative between two detectors).
The error on the absolute knowledge of the gamma spectrum from a Gd
neutron capture disappears. However, there will be a calibration error 
on the difference between the 6~MeV energy cut in both detectors.   
Table \ref{systematics2} summarizes the identified systematic errors
that are currently being considered for the Double-CHOOZ experiment.
\begin{table}[htbp]
\begin{center}
\begin{tabular}{|lrr|}
\hline
 & \multicolumn{1}{c}{After CHOOZ} & \multicolumn{1}{c|}{Double-CHOOZ Goal} \\
\hline
 Solid angle & 0.2\%         & 0.2\% \\
 Volume      & 0.2\%             & 0.2\% \\
 Density     & 0.1\%             & 0.1\% \\
 Ratio H/C   & 0.1\%             & 0.1\%  \\
 Neutron efficiency & 0.2\%      & 0.1\%   \\
 Neutron energy & 0.2\%          & 0.2\% \\ 
 Spatial effects    & neglect  & neglect \\
 Time cut    & 0.1\%               & 0.1\% \\
 Dead time(veto) & 0.25\%        & $<$ 0.25\%  \\
 Acquisition & 0.1\%              &  0.1\% \\
 Distance cut  & 0.3\%      & $<$ 0.2\%  \\
\hline 
 Grand total & 0.6\%&  $<$ 0.6\% \\ 
\hline
\end{tabular}
\caption[Systematic errors that can
be achieved without improvement of the CHOOZ published  
systematic uncertainties]{\label{systematics2}
The column ``After CHOOZ'' lists the systematic errors that can
be achieved without improvement of the CHOOZ published systematic
uncertainties Reference \cite{choozlast}.
In Double-CHOOZ, we estimate the total systematic error on the 
normalization between the detectors to be less than 0.6\%. }
\end{center}
\end{table}
\subsubsection{Background subtraction error}

\par
The design of the detector will allow a Signal/Background (S/B)
ratio of about 100 to be achieved 
(compared to 25 at full reactor power 
in the first experiment\cite{choozlast}). 
The knowledge of the background at a level around 30-50\% will reduce
the background systematic uncertainties to an acceptable level.
In the Double-CHOOZ experiment, two background components have been
identified, uncorrelated and correlated.  Among those backgrounds, one has:
\begin{itemize}
\item{The accidental rate, that can be computed from the single event
  measurements, for each energy bin.}
\item{The fast neutrons creating recoil protons, and then a neutron
  capture. This background was dominant in CHOOZ\cite{choozlast}.  
The associated energy spectrum is relatively flat
  up to a few  tens of~MeV.}
\item{The cosmogenic muon induced events, such as $^9$Li and $^8$He, 
 that  have been studied and measured at the NA54 CERN 
experiment\cite{NA54} in a muon beam as well as 
in the KamLAND experiment\cite{Eguchi:2002dm}. 
 Their energy spectrum goes well above 8~MeV, and follows a well defined shape.}
\end{itemize}

\par The backgrounds that will be measured are:
\begin{itemize}
\item{Below 1~MeV (this was not possible in CHOOZ, due
  to the different detector design and the higher energy threshold)}
\item{Above 8~MeV (where there remains only 0.1\% of the neutrino
  signal).}
\item{By extrapolating from the various thermal power of the plant
     (refueling will result in two months per year at half power).}
\end{itemize}
From the measurement of the energy spectrum
in accidental events, and from
the extraction of the cosmogenic spectrum, the shape of the spectrum for
fast neutron events can be obtained with a precision greater than what
is required.
\subsubsection{Liquid scintillator stability and calibration}

\par From our simulations,  the calibration of the relative
energy scales at the 1\% level is
necessary. 
The specification of no more than 100~keV scale difference at 6~MeV is 
achieved if this 1\% level is obtained.
This level of calibration can be obtained with the program considered
in subsection \ref{sec:calibration}.  We will
 move the same calibration radioactive sources from
one detector to the other, and directly
compare the position of the well defined calibration peaks.  

\subsection{Sensitivity and Expectations}
\label{sec:sensitivity}

\par
\begin{figure}[ht]
\begin{center}
\includegraphics[angle=-90 , width=0.75\textwidth]{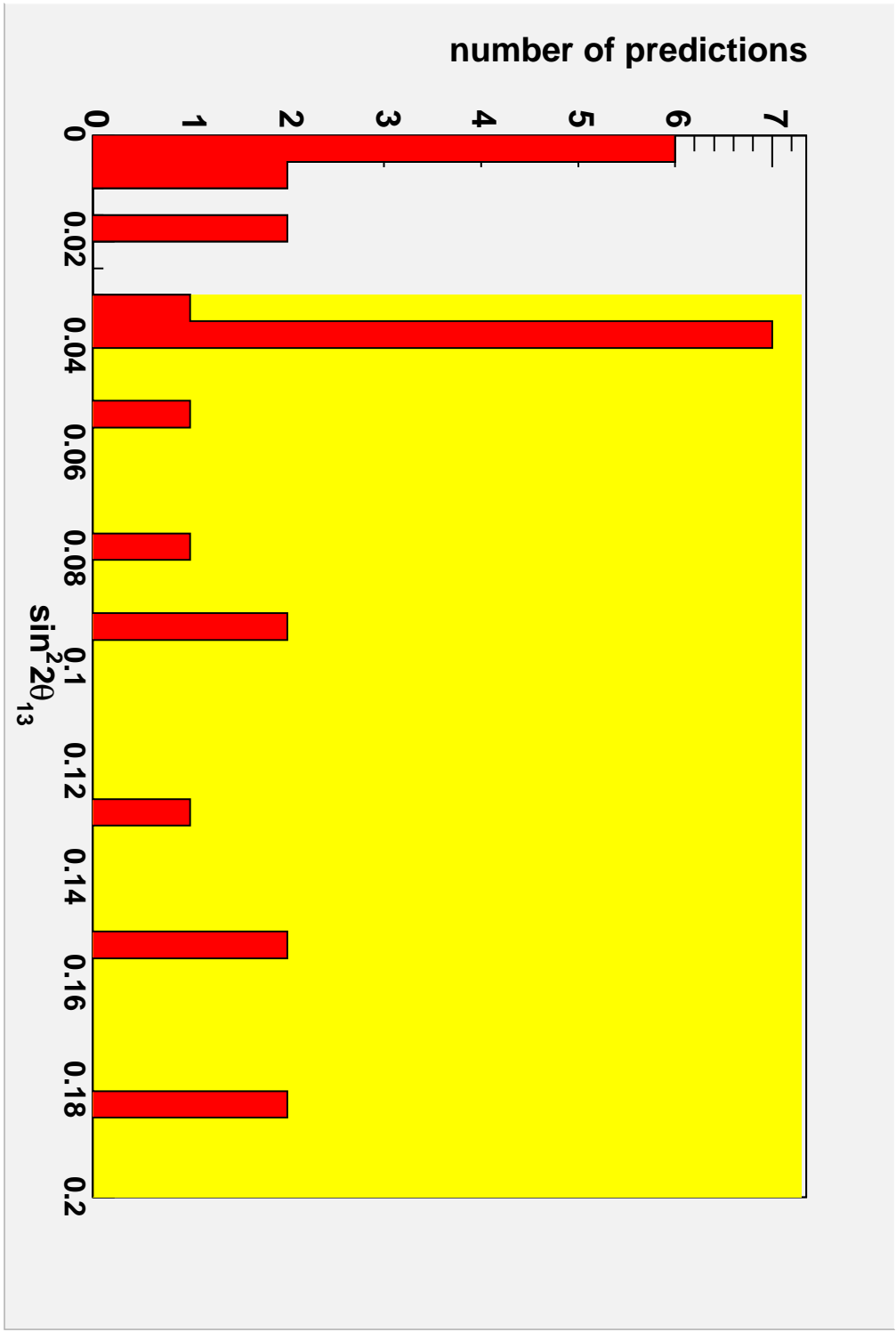}
  \caption[predictions]{Predictions for $\quq$.  
The sensitive region of Double-CHOOZ is shown in yellow.
}
  \label{fig:pred}
\end{center}
\end{figure}

 An overview of various predictions compiled for
Reference \cite{R:WhitePaper} is given in Table~\ref{tab:ThPredictions}.  
For more extensive reviews, see for example 
\cite{Barr:2000ka,Altarelli:2002hx,Barbieri:2003qd,Chen:2003zv}.
The conclusion from all these considerations about neutrino mass 
models is that a value of $\theta_{13}$ close to the CHOOZ bound 
would be quite natural, while smaller values become harder and harder 
to understand as the limit on $\theta_{13}$ is improved.
These predictions are graphically depicted in Figure \ref{fig:pred}.

\begin{table}[htb]
\begin{center}
\begin{tabular}{|lcc|} \hline\hline
Reference & $\sin\theta_{13}$ & $\sin^2 2\theta_{13}$ \\
\hline
\emph{SO(10)} & & \\
Goh, Mohapatra, Ng \cite{Goh:2003hf} & 0.18 & 0.13 \\
\hline
\emph{Orbifold SO(10)} & & \\
Asaka, Buchm\"uller, Covi \cite{Asaka:2003iy} & 0.1 & 0.04 \\
\hline
\emph{SO(10) + flavor symmetry} & & \\
Babu, Pati, Wilczek \cite{Babu:1998wi} & $5.5 \cdot 10^{-4}$ &
 $1.2 \cdot 10^{-6}$ \\
Blazek, Raby, Tobe \cite{Blazek:1999hz} & 0.05 & 0.01\\
Kitano, Mimura \cite{Kitano:2000xk} & 0.22 & 0.18 \\
Albright, Barr \cite{Albright:2001uh} & 0.014 & $7.8 \cdot 10^{-4}$ \\
Maekawa \cite{Maekawa:2001uk} & 0.22 & 0.18 \\
Ross, Velasco-Sevilla \cite{Ross:2002fb} & 0.07 & 0.02\\
Chen, Mahanthappa \cite{Chen:2002pa} & 0.15 & 0.09 \\
Raby \cite{Raby:2003ay} & 0.1 & 0.04 \\
\hline
\emph{SO(10) + texture} & & \\
Buchm\"uller, Wyler \cite{Buchmuller:2001dc} & 0.1 & 0.04 \\
Bando, Obara \cite{Bando:2003ei} & 0.01 .. 0.06 &
 $4 \cdot 10^{-4}$ .. 0.01 \\
\hline
\emph{Flavor symmetries} & & \\
Grimus, Lavoura \cite{Grimus:2001ex,Grimus:2003kq} & 0 & 0 \\
Grimus, Lavoura \cite{Grimus:2001ex} & 0.3 & 0.3 \\
Babu, Ma, Valle \cite{Babu:2002dz} & 0.14 & 0.08 \\
Kuchimanchi, Mohapatra \cite{Kuchimanchi:2002fi} & 0.08 .. 0.4 & 
 0.03 .. 0.5 \\
Ohlsson, Seidl \cite{Ohlsson:2002rb} & 0.07 .. 0.14 & 0.02 .. 0.08 \\
King, Ross \cite{King:2003rf} & 0.2 & 0.15 \\
\hline
\emph{Textures} & & \\
Honda, Kaneko, Tanimoto \cite{Honda:2003pg} & 0.08 .. 0.20 &
 0.03 .. 0.15 \\
Lebed, Martin \cite{Lebed:2003sj} & 0.1 & 0.04 \\
Bando, Kaneko, Obara, Tanimoto \cite{Bando:2003wb} & 0.01 .. 0.05 &
 $4 \cdot 10^{-4}$ .. 0.01 \\
Ibarra, Ross \cite{Ibarra:2003xp} & 0.2 & 0.15 \\
\hline
\emph{$3 \times 2$ see-saw} & & \\
Appelquist, Piai, Shrock \cite{bib:s1, bib:s2} & 0.05 & 0.01 \\
Frampton, Glashow, Yanagida \cite{Frampton:2002qc} & 0.1 & 0.04 \\
Mei, Xing \cite{Mei:2003gn} (normal hierarchy) & 0.07 & 0.02 \\
\hphantom{Mei, Xing \cite{Mei:2003gn}} (inverted hierarchy) & $>0.006$ &
 $> 1.6 \cdot 10^{-4}$ \\
\hline
\emph{Anarchy} & & \\
de Gouv\^{e}a, Murayama \cite{deGouvea:2003xe} & $>0.1$ & $>0.04$ \\
\hline
\emph{Renormalization group enhancement} & & \\
Mohapatra, Parida, Rajasekaran \cite{Mohapatra:2003tw} & 0.08 .. 0.1 &
 0.03 .. 0.04 \\
\hline
\end{tabular}
\caption{Incomplete selection of predictions for $\theta_{13}$.}
\label{tab:ThPredictions}
\end{center}
\end{table}

\par	The CHOOZ experiment\cite{R:CHOOZ}, which provides the most
sensitive limit on \thc to date, did not push the limit lower for two
related reasons: (1) It met its goal of showing that the atmospheric 
neutrino anomaly was not due to \numu $\rightarrow$ \nue oscillations
and (2) It was close to its systematic error limit due to uncertainties
in the reactor neutrino flux, 
and so stopped running. 
Double-CHOOZ will {\it cover} roughly 85\% of the available parameter
space in that graph and cover a similar percentage of those predictions.
We emphasize that a non-zero value for $\quq$ is expected, unlike the
situation in most searches for new physics!
\par
\begin{figure}[ht]
\begin{center}
\includegraphics[angle=-90 , width=0.75\textwidth]{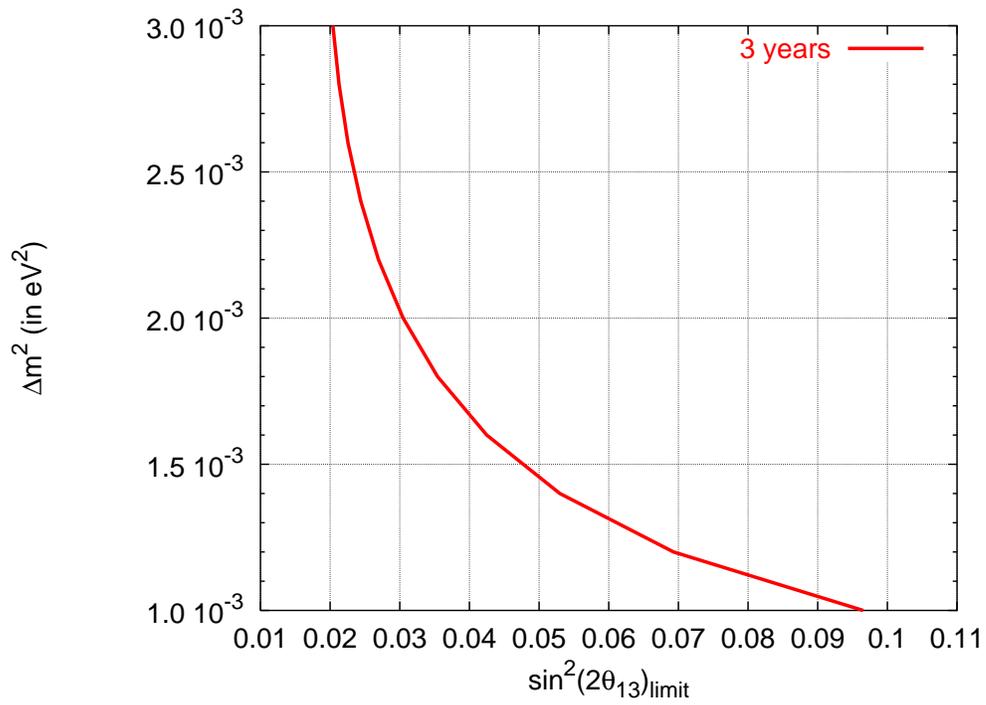}
  \caption[Double-CHOOZ sensitivity limit at 90\% C.L. (for 1
  d.o.f).]{Double-CHOOZ sensitivity limit at 90\% C.L. (for 1 d.o.f).}
  \label{F:DC_limits_A}
\end{center}
\end{figure}

The best present CHOOZ limit on \thc is a function of $\Delta m^{2}_{atm}$,
which has been measured using atmospheric neutrinos by Super-Kamiokande (SK)
and others. The latest reported value from SK\cite{R:SK_A}
is $1.5 < $ $\Delta m^{2}_{atm}$ $< 3.0\times 10^{-3} eV^{2}$ with a best
fit reported at $2.4\times10^{-3}\,$eV$^2$.
  The current CHOOZ limits for $\Delta m^{2}$ of 3.0,
2.0, and $1.5\times 10^{-3} eV^{2}$ are \sstc $<$ 0.12, 0.20, and 0.40
respectively. 
Figure~\ref{F:DC_limits_A} shows the limits expected from Double-CHOOZ after
3 years of operation as a function of $\Delta m^{2}_{atm}$, assuming
\thc is zero.  This depends strongly on the {\it relative}
 normalization systematic
error between the near and far detectors ($\sigma_{rel}$) which we estimate
will be ~0.5\%. 
Double-CHOOZ will significantly improve
the existing limits even if a positive value of \thc is not found.

%% file: inner.tex
\section{U.S. Systems}
\subsection{Inner Detector Photomultiplier Tubes}
\label{sec:pmt} 
Photomultiplier tubes (PMTs) are a major component of the Double-CHOOZ
detectors and will be a primary U.S. contribution to the experiment. We
are proposing the purchase of 512 photomultiplier tubes each for
the near and far Double-CHOOZ detectors, along with 16 spares for
a total of 1040.  There
are several reasons why this area is highly appropriate as a U.S. 
responsibility:\\

\begin{itemize}
\item The U.S. Double-CHOOZ group has extensive experience in selection, 
testing, and mounting of PMTs, having participated in CHOOZ, E645, IMB, 
KamLAND, LSND, mini-BooNE, SNO, and Super-Kamiokande.
\item It is a crucial part of the experiment that lends itself easily to
being done in the U.S. and shipped pre-assembled to the Chooz site. 
\item At the end of the experiment, the PMTs can be recovered for re-use
in other experiments. As an example, the IMB PMTs are still in use after
almost twenty years! Thus purchase of these large, hemispherical, 
low-background  PMTs can be viewed as a long-term investment in the U.S. 
program.
\end{itemize}

Double-CHOOZ will have two PMT regions: (1) Inner Detector (ID)  PMTs
to view
the neutrino target and gamma-catcher region, and (2) Inner Veto (IV) PMTs
to view an active outer veto layer. In this proposal, we present plans to
construct the ID PMT array only. It is expected that construction of the 
IV array will be done by European collaborators. This is a natural separation,
since the ID array must detect relatively low light levels and be compatible
with the radiological standards of the target region, while the IV array
will be operating in the high light levels from cosmic-ray muons and can
have more relaxed radiological specifications. It is possible that
different PMTs might be used for these two arrays.\\

There are many factors and trade-offs involved in the design of
an ID PMT array for Double-CHOOZ. One important consideration is
PMT size. For a given collection area, larger  PMTs are favored over smaller
ones due to their somewhat lower cost per area and the reduced number
of electronics and high-voltage channels required. However, this
must be balanced with the difficulty in the handling and mounting
of large  PMTs in a confined space, the increased background due to
higher glass/area ratio, and the potential ``hole'' in coverage from the
loss of a single PMT. Experiments requiring large-area coverage 
(e.g. Super-Kamiokande) have used the largest  PMTs available
(twenty inches) while experiments requiring less coverage (e.g. CHOOZ) have
selected smaller  PMTs (eight inches). 
In this proposal we
consider the use of eight-inch  PMTs similar to the original CHOOZ 
experiment, although the possibility for larger (thirteen-inch)  PMTs is 
still being explored. The PMT budget of this proposal is conservative in
the sense that selection of the larger  PMTs would be less expensive.\\

\subsubsection{Required Coverage}

The required PMT coverage is based on three factors: (1) adequate light 
collection to allow a hardware threshold well below 1 MeV, (2) sufficient
energy resolution to reject backgrounds above 
the neutron capture gamma energy window at 8 MeV, and (3) adequate
energy resolution to allow shape comparison between the visible energy
spectrum in the near and far detectors. For Double-CHOOZ the first two
reasons dominate, as our simulations show very little change in sensitivity
over reasonable values of energy resolution at 1 MeV.\\

Simulations of the required coverage are notoriously difficult since they
must fold together light production efficiency, transport through the
waveshifter-laced scintillator (including re-emission of previously
shifted light), optical effects of the acrylic vessels, transport
through oil, and reflections from the walls and other structures. 
As an example of the difficulty, pre-experiment simulations for KamLAND
predicted roughly 100-150 p.e./MeV (using only 17-inch PMTs), 
whereas about 240 p.e./MeV was realized in the actual experiment.

\subsubsection{Requirements on Radiopurity}

PMTs are typically one of the most radioactive materials used in the
construction of neutrino detectors. This is due to the fact that they may
contain ~1 kg of glass per PMT - and glass typically has a rather high
concentration of the long-lived elements uranium (U), Thorium (Th),
and potassium (K). For example, Table~\ref{IPMT:T:assay} shows the radioactive
assay of the Electron Tubes, Inc. (ETI) model 9354 developed for use in
the Borexino experiment.

\begin{figure}
\centerline{\rotatebox[]{270}
{\scalebox{0.4}{\includegraphics[angle=90, width=\textwidth]{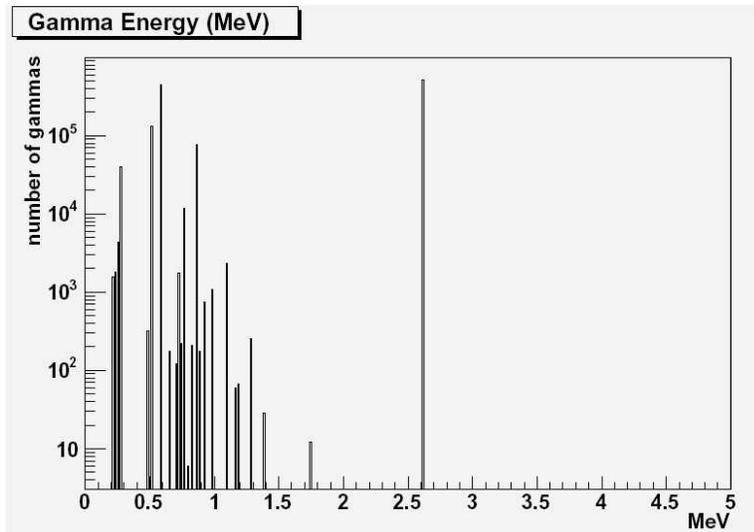}}}}
\caption{Simulation of the energy spectrum of $^{208}Th$ gamma rays
from beta decay to $^{208}Pb$.}
\label{IPMT:F:Tl_gammas}
\end{figure}

The most important backgrounds come from the decay of $^{40}K$ and
$^{208}Tl$ (part of the thorium chain).
$^{40}K$ has a single 1.46 MeV gamma from electron capture to the
first excited state of $^{40}K$. $^{208}Tl$ has many possible gammas from
beta decay to excited states of $^{208}Pb$, but all of them go through
the first excited state at 2.61 MeV. Figure~\ref{IPMT:F:Tl_gammas} shows the
gamma energy spectrum from a GEANT3 simulation of the Double-CHOOZ
detector backgrounds.\\

\begin{table}
\caption{Assay of the U, Th, K content of an ETI9354 PMT}
\begin{center}
\begin{tabular} {llr|rr|rr|rr}
material & component & mass & K (mg) & (+/-) & Th ($\mu g$) & (+/-) & U ($\mu g$) & (+/-) \\ \hline
bialkali & photocathode & & 2 & & & & & \\
 & & & & & & & & \\
glass & envelope & 635 & 38 & 10 & 19 & 6 & 19 & 13 \\
 & base (no pins) & 10 & 1 & 0 & 0 & 0 & 0 & 0 \\
 & & & & & & & & \\
ceramics & sideplates & 10 & 0 & 0 & 1 & 0 & 0 & 0 \\
 & rods/spacers & 13 & 0 & 0 & 1 & 0 & 0 & 0 \\
 & & & & & & & & \\
metals & dynodes & 86.6 & 0 & 1 & 3 & 1 & 0 & 1 \\
 & generator & 0.4 & 10 & 1 & 0 & 0 & 0 & 0 \\
 & & & & & & & & \\
plastics & overcap & 32 & 5 & 2 & 1 & 0 & 1 & 0 \\ \hline
\end{tabular}
\end{center}
\label{IPMT:T:assay}
\end{table}

Based on these data, each PMT has a total $^{40}K$ activity of 1.77 Hz. Taking
into account a branching ratio of 0.1067 for EC capture to the first excited
state gives a gamma activity of 0.19 Hz. Assuming secular equilibrium with
$^{232}Th$, the $^{208}Tl$ total activity is 0.097 Hz, with a gamma activity
of 0.23 Hz per PMT.\\

\begin{figure}
\centerline{\rotatebox[]{270}{\scalebox{0.4}{\includegraphics
[angle=90, width=\textwidth]{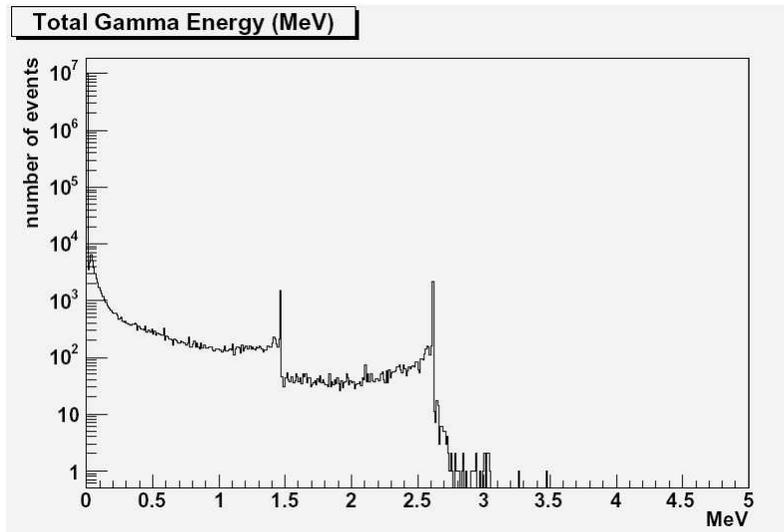}}}}
\caption{Simulation of the energy spectrum of gamma rays from $^{40}K$
and $^{208}Tl$ which enter the scintillator volumes.}
\label{IPMT:F:PMT_gammas}
\end{figure}

Using the detector simulation to propagate these gammas from the PMTs
to the scintillator-filled volumes results in the energy spectrum of
Figure~\ref{IPMT:F:PMT_gammas}. In this spectrum, the energy used
is the summed energy of all the gammas as they enter the active volumes
so that the effect of correlated gammas is included. In addition, the electrons
from the beta decay are also included, as they may produce gammas from 
bremsstrahlung. Knowing the absolute
activity then allows the number of events above a given energy threshold to
be extracted from this figure. Above 0.5 MeV we therefore expect 2.7 Hz of
gammas and above 1.0 MeV we expect 1.7 Hz.\\

To estimate the number of background events that come from this activity 
requires an estimate of the background from other sources in the 8 MeV
range (the energy of neutron capture on gadolinium). There are two 
major sources: (1) neutrons produced from muon spallation or muon capture,
and (2) external gammas in the range of 8 MeV produced
by nearby untagged muons. The rate of (1) is given in 
Table~\ref{t2} of Section 1 and is about 6 (65)  $h^{-1}$ for the 
far (near) detector. The rate of (2) can be estimated from the original CHOOZ
experiment, about 180~$h^{-1}$ at 300~m.w.e. 
Conservatively scaling from the
muon rate gives ~4300~$h^{-1}$ in the near detector (assuming 60~m.w.e.).
With a 200~$\mu sec$ window, the expected background rate from the accidental
coincidence of a neutron-like event with
a PMT-generated gamma is about 1.5 (18) $d^{-1}$ for the far (near) 
detector.  This is to be compared with an expected signal rate of about 
65 (3500) $d^{-1}$ (including efficiency). In practice, the accidental rate
can be measured to high precision using non-coincidental singles rates and
spectrum.  If a 1\% measurement is made, then using ETI9354 PMTs would 
contribute an uncertainty of only 0.02\% to the experimental result. Since
uncertainties below 0.05\% have a negligible effect of the results, we adopt
the assay requirements of Table~\ref{IPMT:T:specs} as specifications for
total PMT activity. These are essentially the summed values of 
Table~\ref{IPMT:T:assay} multiplied by two.\\

\begin{table}
\caption{Specifications for the allowed U, Th, K content per PMT}
\begin{center}
\begin{tabular} {ccc}
chain & specification \\ \hline
uranium & 40 $\mu g$\\
thorium & 50 $\mu g$\\
potassium & 110 $mg$\\ \hline
\end{tabular}
\end{center}
\label{IPMT:T:specs}
\end{table}

In addition to the ETI PMT, Hamamatsu Corporation also offers low-radioactivity
8-inch tubes. They have agreed to supply us with a sample of such a PMT
and we intend to assay it using the low background counting
facility at Alabama.\\

\subsubsection{Base Electronics}

We intend to use traditional passive divider chains similar to those
used in the Super-Kamiokande outer detector (designed at Louisiana State University) and KamLAND. Since
they consist only of resistors and capacitors, the reliability is 
very high. This is important since we plan to pot the electronics
in epoxy (see below) so a failure would be impossible to repair.\\

We plan to operate the  PMTs at a gain such that a single photoelectron
pulse into 50 ohms produces roughly a 20 mV pulse. At this value, simple,
commercially-available discriminators work well, pulse heights are large
compared with typical noise levels and HV ripple and dark noise values
are typically negligible compared to rates from low-energy singles. The
exact gain will depend on the model of PMT selected and the details of
the front-end electronics design. 
Louisiana State University will work with Drexel in the 
specification and design of the divider chain electronics.\\

We also plan to back-terminate the divider chain in order to reduce
the multiple reflections that can result from the large pulses generated
when a muon passes through the scintillator. This requires operating at
an increased gain to make up for signal loss, but the reduced dead time
loss more than makes up for this. A typical muon pulse is expected to
result in 200-300 p.e. per PMT, on average. This means a pulse height
in the several volt range as compared to around 100 mV or so for typical
neutrino interactions.\\

Currently, we plan to have the PMT supplier also build and attach the
divider chains according to our specifications. This is to allow them
to pot the electronics at the factory, which is very desirable from the
standpoint of catching defective  PMTs before delivery and ensuring
quality control. This was done by both the Super-Kamiokande and KamLAND
experiments and proved to be very satisfactory.\\

\subsubsection{Potting and Cabling}

In order to reduce the amount of cabling in the Double-CHOOZ detector,
we plan to have a single 50-ohm coaxial cable carrying both high voltage
and signal from the PMT to a patch panel above the detector. This is 
essentially the scheme used in the Super-Kamiokande outer detector as
well as LSND and MiniBooNE.
The patch panel will not only allow picking off of the signal for routing
to the front-end electronics, but also allow a transition from an expensive
Teflon-coated jacket (made for long-term oil immersion) to a cheaper PVC
jacket.\\

We plan to purchase  PMTs with the divider electronics already potted
for immersion in oil. The 17-inch  PMTs for KamLAND and the 8-inch PMTs
for the Super-Kamiokande veto were purchased in a similar fashion and the results were very
satisfactory, with less than 1\% failure after immersion. We have also
determined from comparing quotes from one supplier with our experience
in potting the KamLAND 20-inch  PMTs ourselves that the costs are very 
similar.\\

The potting is done by first attaching a small circuit board containing the
divider electronics to the pins of the PMT via floating leads instead of
sockets. This prevents any residual stress from unseating the PMT. Then the
electronics and PMT are aligned in an injection mold and a high-viscosity,
oil-resistant
epoxy is bonded to the PMT, cable, and circuit board in a single operation.
For KamLAND, this was done by Hamamatsu using A Nippon Pelnox two-component 
epoxy MG-151/HY-306. This epoxy is not only oil-resistant but also has a high
electrical resistance of more than $10^{10}$ Ohm-cm and a low density of
about 1.1 g/cm.\\

While this specific epoxy has worked well in the KamLAND oil, the 
Double-CHOOZ oil will be slightly different and thus we will test this material
for compatibility before use. Louisiana State University has experience in this area.\\

\subsubsection{The PMT Selection Process}

We plan to select the PMT type for Double-CHOOZ on the basis of several
criteria:\\

\begin{itemize}
\item proven compliance with our low activity specifications
\item proven ability to attach and pot diver chain electronics to our specifications
\item satisfactory peak-to-valley ratio (typically 2 or more) to
allow us to balance PMT gains and aid in energy calibration
\item proven quantum efficiency ($>20$\%)and reasonable transit time
 jitter (few ns) with full-face illumination
\item price
\end{itemize}

Test setups for PMT performance at Louisiana State University and PMT radioactivity at Alabama
will ensure  PMTs meet basic specifications and will be used for quality
control during the testing, cleaning, and assembly of the PMTs.\\

An important consideration is whether the supplier can deliver PMTs
at a rate needed for the Double-CHOOZ construction schedule. Based
on our experience, it typically takes about four months for a factory to
tool up an assembly line, and normally they can make about 200 PMTs
per month. Two potential suppliers that were contacted 
have confirmed this estimate. Thus, to meet the construction schedule
that calls for far detector operation beginning in summer 2007, we
must order the  PMTs by November 2005.\\

\subsubsection{PMT Testing}

We plan to have the  PMTs shipped first to Louisiana State University, where a high-throughput
testing facility will be constructed similar to those used in IMB, 
Super-Kamiokande, and KamLAND.  PMTs will be tested for gain and
dark noise versus high voltage, and an initial operating voltage
selected that will balance the PMT gains. Defective  PMTs will also
be identified and returned to the factory.\\

To meet the construction schedule, we must test 60 PMTs/week. Since
each test requires a preceding dark period of roughly 24 hours, we
need to make a facility capable of testing about twelve  PMTs at a time
(allowing the weekends to be used to make-up any failed tests).\\

After electronics testing, the  PMTs will be driven in batches of
~100 to Alabama for cleaning, mounting, and radioactivity spot-checking.\\

\subsubsection{PMT Mounting, Cleaning, and Assembly}

According to the present mechanical design, the ID PMTs will be mounted 
looking straight ahead on
rails with the spacing between PMTs optimized from simulation studies.
The rails will be provided as part of the PMT support structure built by the French
collaborators.
The design of the PMT mounts will be adapted from the designs successfully
implemented by other experiments, e.g. MiniBooNE, using the same basic type of PMT
in a similar environment.  The mount materials and the fabrication process will be
chosen to meet chemical compatibility requirements and radiological standards.

Prototype mounts will be built in the University of
Alabama Physics machine shop and test assembled with
PMTs.  These assemblies will be placed in a dark 
box, powered,  and counted over a long period
(at least one month) to check that addition of the mounts does not damage the 
PMTs.
 In addition, a mount will be shipped to France for checking that it can be 
mounted to the the rail system.
Once the design and fabrication process is
finalized, the job of making 1040 mounts from materials provided by the 
University of Alabama group will
be handed over to an outside shop.  

The PMTs received from Louisiana State University and the completed PMT mounts  will be cleaned and assembled
by the University of Alabama group.
A clean area with forced ventilation through 
high efficiency filters will be set up in one of the labs.  A 
team of undergraduate and
graduate students will degrease the PMT mounts with acetone, pass them through an ultrasound
bath in weak acid, rinse in water, and then thoroughly wipe down the mounts with alcohol.
The PMTs will also be wiped down with alcohol.    Within one hour of each being wiped
down with alcohol, the PMT and its mount will be assembled, wiped down again with alcohol,
allowed to dry, 
and then sealed in plastic.   The wipe tissues from alcohol cleaning of the first  assemblies
will be counted in the University of Alabama low background 
Ge detector to check that no significant surface radio-contamination survives to final cleaning.  Wipe tissues will be spot--checked throughout 
the cleaning and assembly stage.   Manpower will be assigned to cleaning and assembly
to keep  pace with PMT shipments from Louisiana State University.

\subsubsection{PMT Shipping, and Installation}

The PMT--mount assemblies will be individually packed into boxes and shipped to 
Argonne, which will be the staging point for shipment of Double-CHOOZ detector 
components to France.   Installation of the PMTs onto the support system rails will
be the responsibility of the Louisiana State University and the
University of Alabama groups.

%% file: highvoltage.tex
\subsection{High Voltage System}
\par 
A single vendor will be chosen to provide all of the major high voltage
components for the Double-CHOOZ PMT system.  Two very similar HV 
systems will be installed, one each at the near and far detector 
sites. One mainframe and one module type will be used throughout 
the system.  Common software will be written to meet the controls, 
monitoring and safety requirements as described elsewhere in this 
document.  This will simplify the system in general, which should 
result in a lower initial cost as well as reduced maintenance costs.  
We will consider all reasonable vendors but the two primary candidates
 are Connecticut-based Universal Voltronics (UV) and CAEN from Italy. 
 We have requested budgetary pricing information from these two vendors
 and the information we have received has been included in the WBS 
section 3.4.  Since the exact number of channels is unknown at this
 time, we requested pricing for 1070 channels of high voltage, rated
 at $<$3000~VDC $@$ 2.5mA/ch. This section and it corresponding WBS 
elements include information for the inner detector PMTs only. 
The PMT vendor will
 attach and pot the special 20-meter HV/signal cables to the PMT
 assemblies.

%% file: outerveto.tex
\subsection{Outer Veto}

Due to the shallow depth at the near detector, a high rate of muons is 
expected.  Since the primary background signal for this measurement will
be initiated by cosmic muons, an additional outer veto system is required.
This will be used to help identify muons which could cause neutrons or
other cosmogenic backgrounds and allow them to be eliminated from
the data set.
In some sense, the outer veto provides
redundancy for the inner veto in tagging background associated coincidences,
but such a redundancy is crucial to making a confident measurement of the
background.  Comparison of a single measurement with a full simulation would
not provide such confidence because the cross sections for muon spallation
products are not accurately measured.  In addition, the outer veto will provide
two other benefits:  1) it will achieve a tracking resolution not possible 
with the inner veto alone, reducing the volume of the inner detector
that is ``deadened" by a muon passing through, and 2) it will well measure
those muons which only clip the corners of the inner veto.  Such muons
are especially dangerous because the inner veto efficiency will be low for
these.  
The goal is to provide a system with 4$\pi$ coverage and
greater than 98\% efficiency.

\subsubsection{Expected Backgrounds}

Many sources of backgrounds have been investigated in the Double-CHOOZ LOI
\cite{bib:choozloi}.  The most significant were identified as coincident backgrounds 
arising from cosmic muons.  The primary, and highest rate, are fast neutrons.
These fast neutrons are spallation products from through-going muons in the
surrounding rock.  The neutrons can penetrate the detector and interact
with a nucleus in the target region, providing a prompt signal, and then
become captured by Gadolinium, providing the delayed signal.  

A secondary background, which will also be difficult to identify, results
from high energy muons which pass directly through the target region and 
interact with $^{12}$C nuclei in the target.  This can produce radioactive
isotopes of He and Li which undergo beta decay with a subsequent neutron
emission.  The beta-neutron signal will provide an identical signature
to the positron-neutron signal from a neutrino event.  Unfortunately, 
simply tagging the muon will not allow the elimination of these radioactive
events since the half-lives of these isotopes are between 0.1 - 1 second.
Thus any veto system will require some form of timing and pointing or 
tracking to be able to correlate a coincident signal with a previously 
through-going muon.

\subsubsection{Mechanical Constraints}

The near detector laboratory has not yet been constructed.  Therefore the
mechanical possibilities for a veto system are relatively open.  Given 
the shallow depth, we expect a significant muon rate from a wide angular
region of space.  It has been estimated that using a full 4$\pi$ coverage
will maximize the angular coverage while minimizing the total surface area
of the detector.  In addition, the ability to use muon detection on opposite
sides of the detector will allow rather accurate tracking through the central
region without requiring high spatial resolution of the individual
components.

At the far detector, the flexibility to construct a complete 4$\pi$ system
is not available.  The previously constructed laboratory is only big enough
to allow the inner detector system.  However, given that there is a 
significantly higher overburden at the far laboratory, the needs for large
angular coverage and precise tracking are not as high.  It is relevant,
however, to be able to compare the effects of the veto system on data 
selection.  We therefore envision placing a veto system in the access space
directly above the far detector.  This veto will be identical to the 
top component of the 4$\pi$ system at the near laboratory, thus providing
comparable results.  There will be an additional requirement that this
veto at the far laboratory be movable to allow operation of the ceiling 
crane during periods of maintenance.

\subsubsection{Detector Design}

\begin{figure}[htb]
\begin{center}
\includegraphics[width=.5\textwidth]{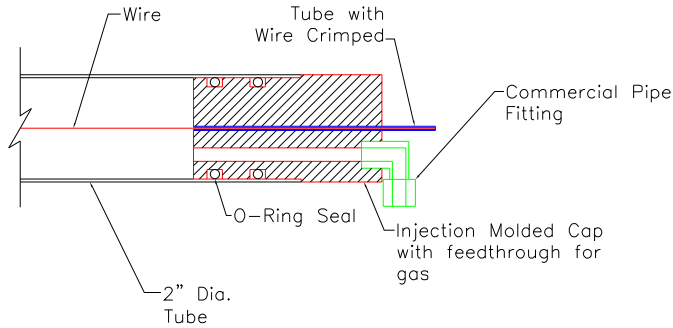}
\end{center}
\caption{Schematic design of the veto chamber end-cap.  The end-gap will be
constructed of injection molded plastic with penetrations for the gas and
wire feed-throughs.  The end-cap will be double O-ring sealed against 
gas leakage and the resistive wire will be crimped in place under 
tension. }
\label{Fig:vetoEndCap}
\end{figure}

The muon veto will use gas-filled proportional chambers with a resistive
wire for charge division.  This will allow very efficient coverage of
a maximum surface area while minimizing costs.  The design is loosely
based on the structures used in the Atlas muon chambers.  It will be
composed of a collection of 2 inch diameter aluminum extruded pipes
with O-ring sealed end-plugs.  A resistive wire will be strung
down the middle, as shown in Fig~\ref{Fig:vetoEndCap}.  

Since we expect to use charge division to identify the longitudinal position 
along the wire, full two dimensional tracking can be supplied with only a 
single orientation of the chambers.  This greatly simplifies the mechanical
construction.  The Double-CHOOZ detector is cylindrical in shape so
we have designed a system which is composed of a vertical barrel region
capped by identical top and bottom planes.

\begin{figure}[htb]
\begin{center}
\includegraphics[width=.75\textwidth]{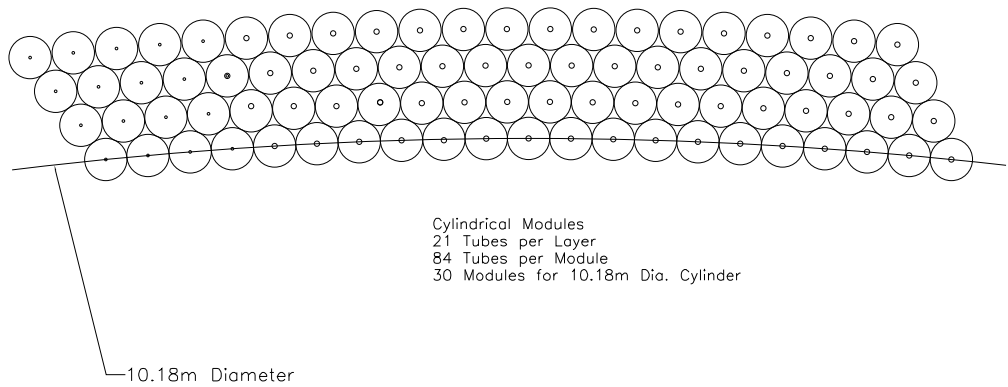}
\end{center}
\caption{Schematic of a single barrel region module for the muon veto system.
The 84 individual chambers will be epoxied together in the structure shown
to exactly encircle the inner detector.  The upper and lower planes will be 
similarly constructed out of 80 chambers and will be flat.}
\label{Fig:vetoModule}
\end{figure}

The top and bottom planes will cover a 10m x 10m area.  Each chamber will
be 10m long and 80 of these will be epoxied together to create a module.
The barrel region will be populated with vertical chambers in which a
basic module will consist of 84 chambers (shown in Figure~\ref{Fig:vetoModule}).
As designed, the entire detector will be surrounded by 4 layers of 
proportional chambers.   This should allow a high efficiency of muon 
detection while minimizing any impact from noise.

\subsubsection{Electronics}

\begin{figure}[htb]
\begin{center}
\includegraphics[width=.9\textwidth]{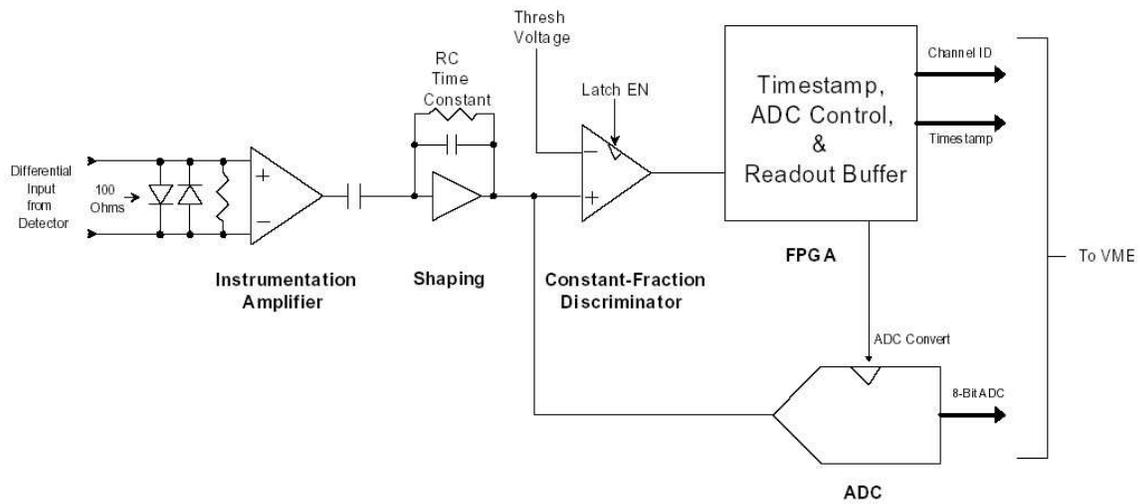}
\end{center}
\caption{A schematic drawing of the expected electronics readout.  The analog
signals from the detector will be the sum of 3 neighboring chambers and will
be amplified directly at the detector.  The signals will then be transported
differentially ~10m to the VME racks where this system will be used to 
digitize any self-triggered signal.}
\label{Fig:vetoElec}
\end{figure}

The signals at each end of the proportional chambers will be summed together
with 2 neighboring chambers to reduce the overall channel count.  This
signal will be immediately amplified and then transported to a standard 
VME crate.  There the analog signal will be processed as shown in 
Figure~\ref{Fig:vetoElec}: 
the analog signal is discriminated in a constant fraction discriminator which
then latches a timestamp in a TDC.  Then the charge is passed through a 
sample-and-hold integrator which is digitized by an 8-fold multiplexed 
ADC.  

This design has the advantage of simplicity and self triggering, and
should provide for sufficiently accurate time and
position resolution.  We expect that this system will have a maximum channel
deadtime of 100 $\mu$s.  Previous experience with such chambers at Soudan
\cite{bib:soudanNim} suggest that even above ground, a single tube rate 
of 100 Hz, dominated by local radioactivity, should be expected.  The 
expected deadtime would not have a noticeable impact on this kind of 
signal rate.

%% file: electronics.tex
\subsection{Electronics}
\begin{figure}
\includegraphics[width=\textwidth]{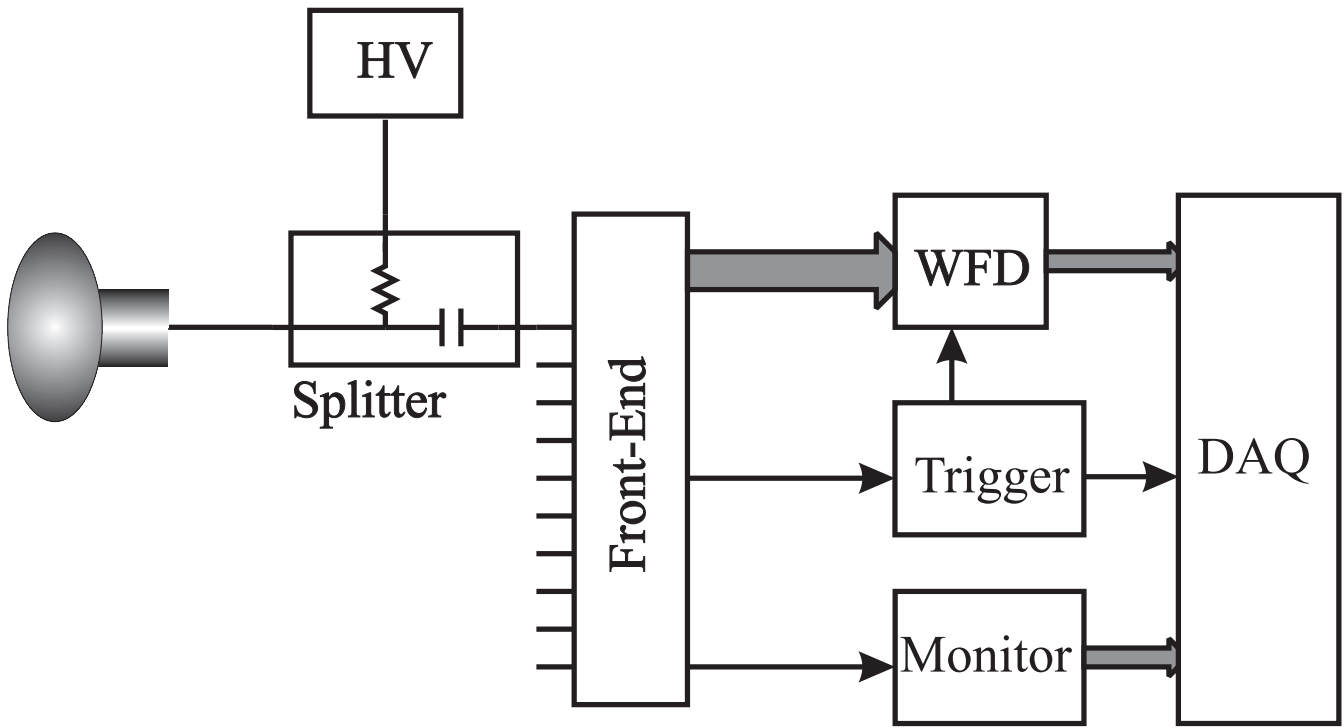}
\caption{Block diagram of Double-CHOOZ electronics. The PMT signal and HV
use the same cable within the detector, separated at the HV splitter. 
The analog front-end distributes the signal after (possible) amplification
to waveform digitizers (WFD), trigger, and monitoring subsystems.
}
\label{fig:electronics_overview}
\end{figure}

Figure~\ref{fig:electronics_overview}
shows an schematic of the overall electronics organization for the
Double-CHOOZ detectors.  The main signal path is from the PMTs, through
HV-splitters and front-end to a fast waveform digitization system where
the pulses are recorded. 

There will also be separate subsystems for providing high-voltage to
the PMTs, a trigger, and extensive monitoring of detector operation.

\subsubsection{HV-Splitters}

The cable to the PMTs will be a single, high-quality cable suitable
for oil-immersion.  Such cable is expensive, and it is economically
favorable to minimize the amount of such cable by using a single cable
for HV and signal inside the detector, with a splitter to separate
HV from signal at a point close to where the cable emerges from the
detector. 

In addition, the use of a single cable to the PMT greatly reduces ground-loop
pickup which, since it occurs on all signal channels simultaneously,
can be problematic for a detector where all PMTs are viewing the
same event.

Figure~\ref{fig:electronics_overview}
shows a schematic for a simple HV splitter. Although the HV splitter
is in the signal path, it will be implemented as part of the HV
system.

\subsubsection{Front-End Electronics}

The front-end electronics for the Double-CHOOZ experiment
handles preamplification, producing analog sums of PMT signals
for trigger and monitoring subsystems, pulse shaping and bandwidth
limitation, baseline restoration and pulse distribution
to other electronics subsystems.

\begin{figure}
\includegraphics[width=\textwidth]{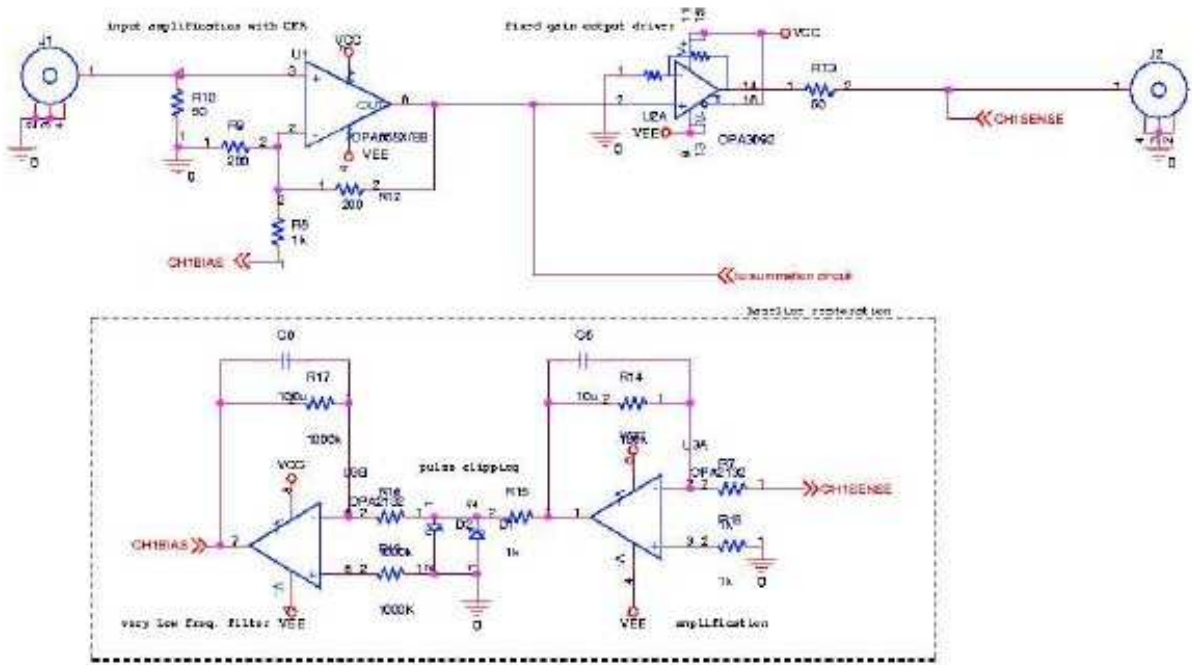}
\caption{A partial schematic diagram for the analog front-end electronics.
The use of DC-coupling in the signal path with baseline restoration
avoids AC-coupling over-shoot for large (muon) signals, while minimizing
DC offsets.}
\label{fig:electronics_front_end}
\end{figure}

Figure~\ref{fig:electronics_front_end}
shows a partial schematic for the front-end electronics. The
front-end modules will be purely analog, to prevent any digital
switching- or clock-noise pickup within the modules.

\subsubsection{Trigger}

\begin{figure}
\includegraphics[width=\textwidth]{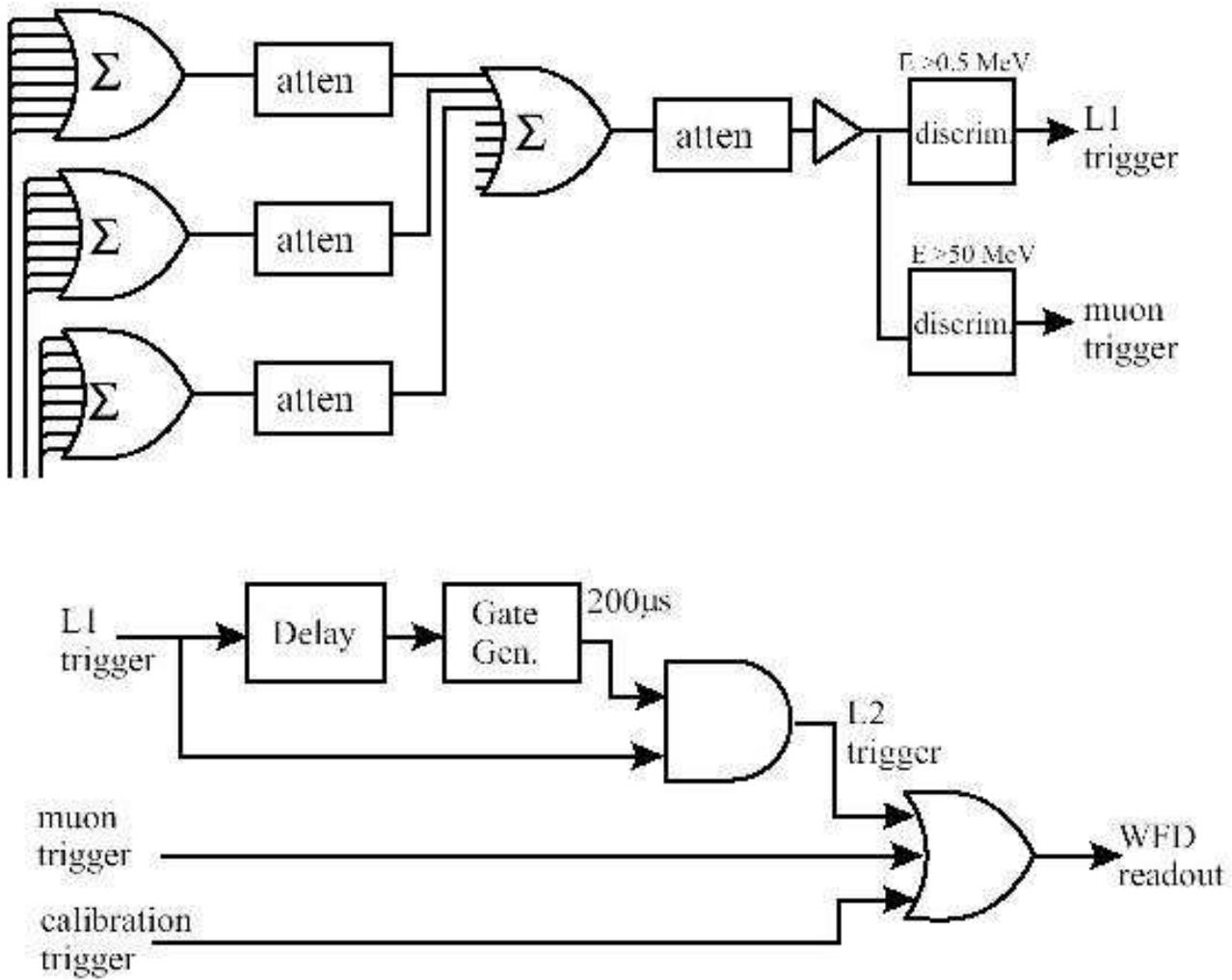}
\caption{Block diagram for the Double-CHOOZ trigger system. The
Level-1 (L1) trigger occurs when energy deposition is above $0.5\,$MeV. The
Level-2 (L2) trigger occurs with a two L1 triggers within 200$\,\mu$s. 
Other triggers, such as a muon trigger ($E>50\,$MeV) and a trigger
for calibration events, also cause WFD readout.}
\label{fig:electronics_triggers}
\end{figure}

Figure~\ref{fig:electronics_triggers}
shows a simplified schematic of the Double-CHOOZ trigger
system.

The basic (`Level 1', or `L1') trigger is a simple energy
deposition trigger.  In a detector of the Double-CHOOZ
geometry with long attenuation length scintillator, the
total light collected by the PMTs is a good measure of the
energy deposition.  While there are many ways to form
an energy trigger from the PMT signals, a simple analog
sum has the distinct advantage of being easily tested,
calibrated and modeled.   Other trigger options are also
being explored as possibilities for the experiment.

Signals from the PMTs are fanned in,
first in the front-end electronics, and further in the
trigger system, with calibrated attenuation between fan-in
stages to prevent electronic saturation. The final
signal is then passed through a linear low-pass filter
to minimize the effects of differing photon arrival-times
at the PMTs. 

Two discriminators are then used to define
an energy threshold for L1 triggers and for `muon' triggers.

The L1 trigger threshold will be set well below the minimum
positron energy deposition, to obtain full efficiency for positrons
produced in antineutrino interactions.  The `muon' trigger
threshold will probably be
at an energy deposition near 50$\,$MeV. The L1 trigger
will be sent to the waveform-recording electronics to cause the
capture of information from all PMT channels.  

The Level-2 (`L2') trigger is generated from having two
L1 triggers occur within a coincidence time of 200$\,\mu$s. This
coincidence can be done either in hardware or in software, and
a final implementation decision has not yet been made.  An L2
trigger causes the readout of waveform information stored from
the L1 triggers, possibly with extra L1 pulses that were prior
to the L2 trigger, as an aid to background rejection. 

Muon triggers, along with triggers from the veto system and
calibration triggers, will also cause pulses to be stored and read out
by the waveform system. 
 
Since one of the major goals of the Double-CHOOZ experiment is to 
make the two detectors as identical as possible, the trigger
thresholds must be set at the same energy.  One way for this to 
be accomplished is to use a particular low-energy gamma source
(such as ${}^{137}$Cs) as a calibration standard for the energy
threshold; this should allow the thresholds to be set at the
same energy rather precisely.

\subsubsection{PMT rate monitor}  

\begin{figure}
\includegraphics[width=\textwidth]{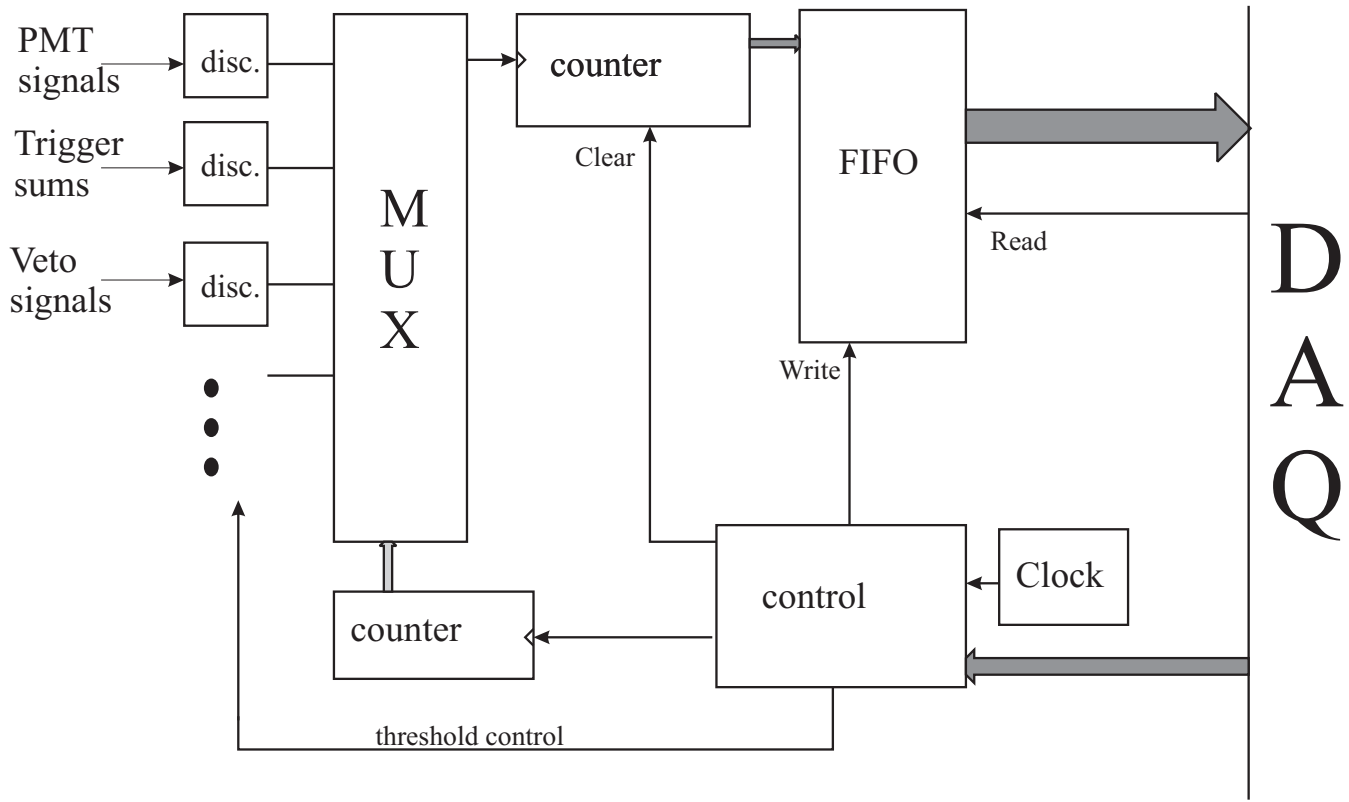}
\caption{
Block diagram of the PMT rate monitoring system. The monitor scans 
through PMT channels, counting single photoelectron pulses for
a fixed time interval, and storing the rates for later readout. This
subsystem can also monitor rates from the trigger system, providing
a continuous check on the operation of the trigger. 
}
\label{fig:electronics_ratemon}
\end{figure}

The PMT rate monitor subsystem 
shown schematically in Figure~\ref{fig:electronics_ratemon}
will be used 
to monitor the stability of the PMTs and associated electronics. 

One of the most sensitive indicators of PMT gain stability is
the single-photoelectron rate, so if that rate is stable it gives
one confidence that the PMTs are indeed stable.   

The PMT rate monitor will take fanned-in signals from the
front-end electronics, discriminate the signals at a low (sub-photoelectron)
level, then count pulses for fixed times to get a snapshot of the
PMT rates. 

The rate monitor could be implemented either with standard commercial
discriminator and scaler modules, or with a simple custom design. 
Because the number of channels is moderate (approx. 600 per detector),
it is not yet clear which approach is most cost-effective, but there
should be no significant difference in performance.

In addition to the fanned-in PMT signals from the front-end, 
the rate monitor should also look at rates from the further stages
of analog fan-in in the trigger, as well as the overall sums that
are used for the triggers and vetoes.  This will provide an end-to-end
stability check for the analog electronics in the experiment.

%
%


%% file: slow.tex
\subsection{Slow Monitoring}

\newcommand{\onewire}{1-Wire\textsuperscript{\textregistered}}
\newcommand{\ibutton}{iButton\textsuperscript{\textregistered}}

A slow monitoring and control system is required to control systematic
effects that could impact the experiment, to allow automated scans of
parameters such as thresholds and high voltages, and to provide
alarms, warnings, and diagnostic information to the experiment
operators.  The quantities to be monitored and controlled include
temperatures and voltages in electronics, experimental hall
environmental conditions, line voltages, liquid levels and
temperatures, gas system pressures, radon concentrations, photo-tube
high voltages, and discriminator settings and rates.  Most of these
functions can be accomplished using ``\onewire'' devices from Dallas
Semiconductor \cite{one-wire-webpage}.  The high voltage and
discriminator subsystems will have their own control and readback
hardware.  All slow monitoring and control systems will use the same
database and history log software.  A computer in each experimental
hall will monitor the local 1-wire bus and a local radon monitor,
acquire any data provided by other subsystems, record the data, and
make the data available via the local internet connection.

\subsubsection{Monitoring via 1-Wire interface}

\begin{figure}
\begin{center}
\includegraphics[width=\textwidth]{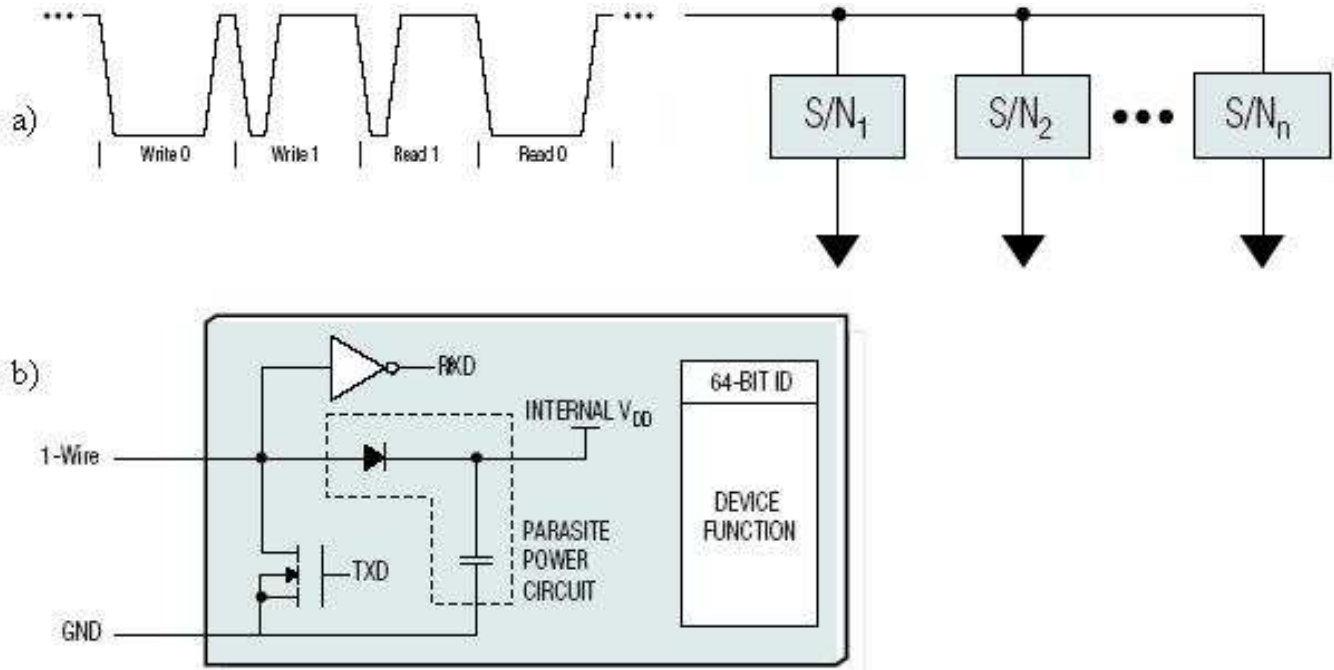}
\end{center}
\caption{The \onewire\ bus: (a) control, readback, and power provided
to multiple devices over a ``single'' wire; (b) parasite power circuit
captures power during high period of \onewire\ waveform.  (Adapted from
figures by Maxim IC / Dallas Semiconductor.)}
\label{Fig:onewirefigs}
\end{figure}

\begin{figure}
\begin{center}
\begin{tabular}{cc}
a) \includegraphics{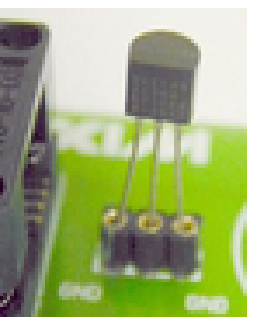} &
b) \includegraphics[width=.2\textwidth]{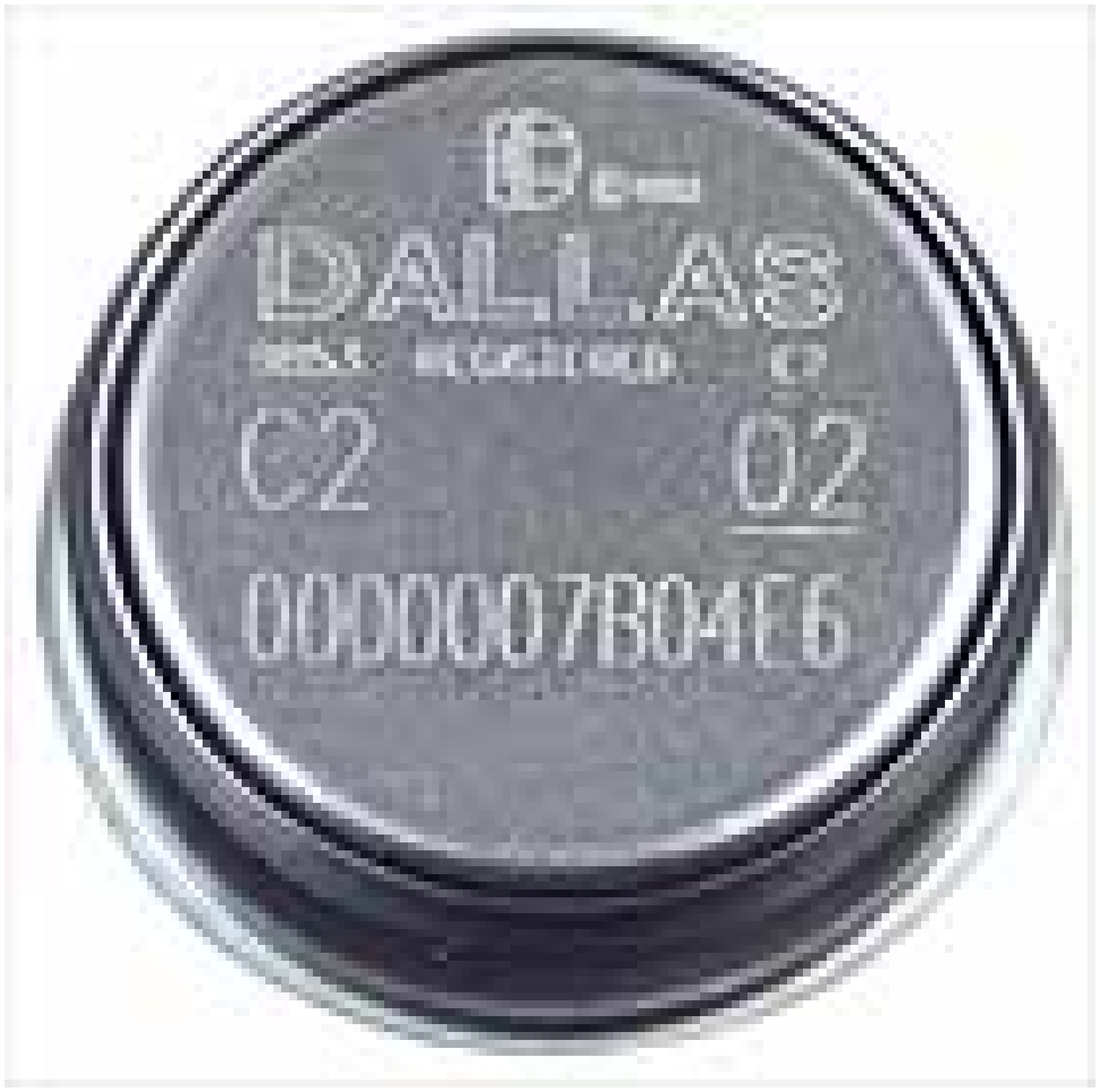}
\end{tabular}
\end{center}
\caption{Examples of \onewire\ devices: digital thermometer in (a) TO-92
package and (b) \ibutton\ case, 17.35~mm in diameter.  Figures are not
shown at the same scale.  (From Maxim IC / Dallas Semiconductor web
site \cite{one-wire-webpage}.)}
\label{Fig:packagephotos}
\end{figure}

The ``\onewire'' line of semiconductors from Maxim IC / Dallas
Semiconductor use a simple interface bus that supplies control,
readback, and power to an arbitrary number of devices over a single
twisted-pair connection \cite{one-wire-webpage}.
(Figure~\ref{Fig:onewirefigs}.)  A variety of sensor and control
functions are available in traditional IC packages and
stainless-steel-clad ``\ibutton s''.  (Figure~\ref{Fig:packagephotos}.)
Each device has a unique, factory-lasered and tested 64-bit
registration number used to provide device identification on the bus
and to assure device traceability.  Some devices are available in
individually calibrated NIST-traceable packages.  The features of low
cost, multidrop capability, unmistakable device ID, and versatility
make this an attractive choice for implementing the slow
instrumentation and control system.

Implementation details and expected performance for several subsystems
are given here.

 {\bf Crate and card temperatures and voltages:} Temperature and
 voltage monitoring can be included in any custom-built electronics at
 a component cost of only a few dollars per device using DS18S20 and
 DS2450 chips and low-cost modular connectors to connect to the
 \onewire\ bus.  Note in addition to the temperature and voltage
 functions, the unique ID on each chip provides automatic tracking of
 any card swaps.  Trivial custom boards containing only these
 components can be used to monitor temperature and bus voltages on
 crates which otherwise contain no custom-built electronics.

\begin{figure}
\begin{center}
\includegraphics[width=\textwidth]{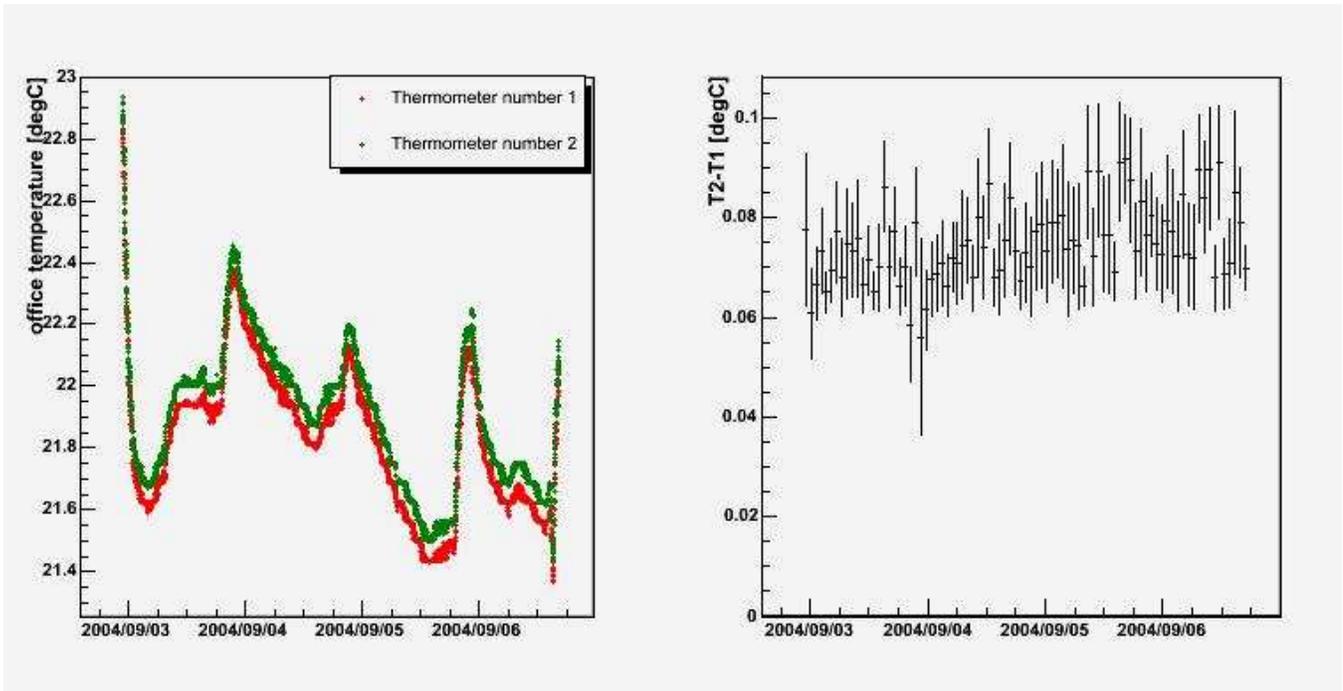}
\end{center}
\caption{Data from three day test of several \onewire\ samples: (a)
 temperature in an office at Kansas State University vs.\ time as read
 by two adjacent DS18S20 thermometer chips; (b) difference
 in reading of the two chips.}
\label{Fig:threedaytest}
\end{figure}

 In these chips, digitization of temperature and voltage is initiated
 by an explicit ``convert'' command from the bus master.  Temperature
 conversion takes 900~ms, and digitization of the four 12-bit channels
 of the DS2450 takes less than 4~ms total.  During conversion, the bus
 master may communicate with other devices if the chips have an
 external source of power; if a chip is powered parasitically from the
 bus, then the master must maintain the bus level high throughout the
 conversion.  Testing of samples provided by Maxim IC / Dallas
 Semiconductor confirm that multiple devices can maintain their
 internal state while all are powered parasitically from the same bus.
 Figure \ref{Fig:threedaytest} shows data from a three day period
 during which two DS18S20 thermometers were sampled once a second by a
 program written in Java.  The two thermometer chips were mounted in
 direct contact with each other, and recorded the same temperature to
 within a small fraction of a degree.  In this test, the thermometers
 were located about two meters from the bus master, and another
 \onewire\ device was connected on the same bus about three meters
 further downstream.  No failures or interruptions occurred during
 this period.

 {\bf Experimental hall environment:} The DS1923 ``Hygrochron''
 \ibutton\ looks ideal for monitoring temperature and humidity.  In
 addition to the humidity and temperature functions, this device
 features on-board battery backup and automatic logging to internal
 memory independent of external control.  Power or computer failures
 will not interrupt the temperature and humidity record.  This
 \ibutton\ is designed for tracking sensitive products during shipment
 or other handling.  Each DS1923 is individually calibrated and NIST
 traceable.  Another \ibutton, the DS1922, provides similar
 functionality without the humidity function.

 Monitoring of barometric pressure and other environmental factors is
 easily achieved using the DS2450 ADC and one or more external
 transducers.  (Note: A complete \onewire\ weather station is even
 available \cite{one-wire-weather}.)  Another interesting
 ``environmental'' condition to monitor is ambient light level in the
 experimental area: many systematic effects in past and present
 neutrino experiments have been attributed, correctly or incorrectly,
 to electrical or optical noise introduced by lighting, and a simple
 phototransistor addresses the issue handily.  The phototransistor
 technique also can be used to monitor status LEDs on devices which
 lack electrical status outputs.  AC line voltage is also easily
 monitored using a \onewire\ ADC and a trivial circuit.

 {\bf Liquid levels and temperatures, and gas pressures:} Transducers
 should be provided for monitoring important aspects of the detector
 such as scintillator and buffer oil levels, temperatures, and
 pressures in any gas systems used.  Transducers should produce
 voltages in the 0 to 5 V range for maximum compatibility with the
 DS2450 ADC.  The further specification, purchase, and installation of
 such transducers 
are the responsibility of the respective subsystems.

 {\bf Simple controls:} The DS2890 is a \onewire\ digitally controlled
 potentiometer.  It can be used to provide slow control for simple
 servos, power supplies, or other devices controllable by an external
 analog signal.  At present, there is no definite plan to use this
 capability, although the discriminator levels could possibly be
 controlled in this way.  Support for slow control as well as
 monitoring should be provided in the software for maximum
 flexibility.

\subsubsection{Radon monitoring}

Professional continuous radon monitors have become readily available
and relatively inexpensive.  An example is Sun Nuclear's Model 1027
\cite{SunNuc1027}.  Each experimental hall will have at least one
radon monitor read out by the slow control PC.  The data will be
stored and made available via the same interface used for all slow
monitor data.



\subsubsection{Interface to other subsystems}


Some hardware subsystems may have important slow monitor that cannot
be made available on the \onewire\ interface or the serial ports of
the slow monitor computers.  Examples may include the clean room
particle counters, the high voltage power supplies, and the
discriminator circuitry in the trigger system.  In such a case, either
the hardware itself or a computer which monitors and controls it
should make the data available via network TCP connection.
``Virtual'' monitor data, such as capture time, event rates, or other
quantities determined by online analysis, could also be recorded by
this mechanism.  The software on the master slow monitor computers
will poll these external servers and make all slow monitor data
available in a common framework.  This is preferable to each subsystem
providing a separate data interface.  In the common framework,
systematic correlations may be studied among any variables.  Support
for control functions and synchronization with externally controlled
devices should be provided in the software to allow for scans of
controlled parameters such as high voltages and threshold levels.  The
common framework will allow dependent variables observed in one
subsystem, {\it e.g.}, discriminator rates, to be easily correlated
with parameters monitored or controlled by some other subsystem, {\it
e.g.}, high voltage.


%% file: dchooz_laser.tex
\subsection{Laser System}
The laser calibration system proposed here is based on the experience with the
similar systems that have been employed successfully within the LSND,
MiniBooNE, CHOOZ, and KamLAND experiments.
The primary purpose of this system is to quantify and monitor pertinent
properties of each individual PMT: PMT gain, relative quantum efficiency,
pulse-height versus photoelectron linearity, and timing.
Other functions of the system include the measurement and monitoring of the
attenuation length of the scintillator over the lifetime of the experiment, and
the reconstruction of the light source locations.
In addition, this will be a valuable tool for commissioning the detector and
data acquisition testing.

The system will consist of a short-pulsed laser and a light distribution
module which transmits the light through quartz optical fibers to a
diffuser ball which can be positioned anywhere 
throughout the fiducial volume with the calibration deployment system.
The possibility of permanently deploying laser balls for monitoring purposes
is being considered.
Using a laser of wavelength within the absorption band of the scintillator the
light emitted by the scintillator-filled dispersion laser balls
will have the same
characteristics as the scintillator light.
The intensity of the laser light can be modulated via a system of
computer-controlled attenuator wheels.

The laser system will be operated by a stand-alone control system running
a real-time control program on a dedicated PC and could operate at a rate of up
to several tens of Hz.
A reference photo-diode would provide a tag signal for the data acquisition
system for each laser firing.
The single PE response, relative quantum efficiencies, and gain calibrations
can be obtained from low intensity, low frequency background runs using the
laser ball positioned at the center of the detector.
Special calibration runs with high intensity light levels 
determine uniquely the time offsets of the PMTs.
Time slewing corrections are determined from laser calibration runs covering
all intensities.

%% file: calibration.tex
\subsection{Calibration Deployment}
\subsubsection{Introduction}
The purpose of the calibration deployment system is to deploy calibration sources into the
target and gamma catcher regions.    The calibration sources and the motivation for using
them have already been described above.
 The deployment systems utilized by the near and 
far detectors will be identical.

The U.S. Double-CHOOZ collaboration has experience in the design and
operation of calibration source deployment systems, most notably for KamLAND
but including also Super-Kamiokande, K2K, and Palo Verde.

Calibration sources which the deployment system must
be designed to accommodate include point gamma sources,
terminated fibers illuminated by external lasers, and neutron sources.  These characteristic
dimensions of these source will range from a few mm to a few cm, and the masses of the
sources will range from a few tens of grams up to a few hundreds of grams.    
The calibration system
must be capable of positioning sources within every representation region of the target and
gamma catcher with an uncertainty which is small compared to the intrinsic vertex 
resolution of the detector in the energy range of 1--8 MeV; a deployment system with a 
positioning accuracy better than 2.5 cm will satisfy this requirement.

The materials and geometry of the deployment system must be chosen to minimize
uncertainties in the corrections for shadowing and absorption.   Neither the
materials of the deployment system itself, nor the process by which deployment
system introduces calibration sources into the detector can measurably increase
detector backgrounds or affect detector performance.   Setup of the deployment system
for inserting a particular source into the detector cannot be awkward or time consuming
and calibrations which are carried out frequently should be largely automated.

In what follows, we describe in the deployment system in 3 parts: (1) deployment 
methods, (2) detector interface, and (3) control systems.

\subsubsection{Deployment Methods}
 
 The methods of source deployment for the target region will be different from that of
 the gamma catcher region because of the different geometries and different calibration
 requirements, therefore, they are discussed separately.
 \vskip 0.2in
 \noindent{\underbar{{Target}}}
 
The complete set of calibration sources must be deployable throughout the 
representative  target  regions
because the experiment intends to use the complete target volume as the fiducial
volume.   (Recall from Section 1 that the  target  is cylindrical with a height of 2.8 m and a
diameter of 2.4~m; the center of the target is about 3.5~m below the top of the detector.)
Two design options are being considered to realize this 
capability, a cable--and--pulley 
system in two orthogonal planes and an articulated arm.    
The deployment system will access the target through a vertical
tube from the detector interface.  Actuators, as well as sensors for determining the 
deployment system position, would be located on the deployment system 
and in the detector interface.  
For the option of an articulated arm, given that the distance from the top of the inner veto to
the bottom of the target is about 5~m, mechanical members comprised of thin
metallic tubes or transparent plastics should suffice to provide adequate stiffness while
minimizing absorption and shadowing.
During a 
calibration, an operator would attach a calibration source to the source holder, deploy
the source into the target at the desired positions, and then retract the source to the
detector interface.    The design and operation of the deployment system must take into
account permanently mounted calibration sources, if any, in the target 
region.

In addition to a cable--and--pulley system or an articulated arm, a simple winch system
will be used to deploy sources along the symmetry axis of the target.    With this
system, a small subset of calibration sources can be quickly deployed on a frequent
basis for the purpose of monitoring detector stability.  The results of source 
calibrations on the symmetry axis will be used to 
determine when it is necessary to
repeat a full calibration of the target and to interpolate the detector calibration between
the full calibrations.

Those components of the deployment system which come in direct contact with the
scintillator will be checked for compatibility.  Compatibility tests will include soaking 
in liquid scintillator and then checking soak scintillator samples for changes in 
light yield, transparency, and radio--contamination.   Components that will not come
into direct contact with the scintillator but which will be exposed to its vapors  must 
also be tested.   After the first round of design is complete for the 
cable--and--pulley system or the articulated arm, a system prototype will
be built and tested in air or water.   All components of the deployment system will be
carefully cleaned before installation into the experiment.
\vskip 0.2in
\noindent{\underbar{{Gamma Catcher}}}

The gamma catcher requires its own calibration because its light yield and properties of
light and neutron transport will likely differ from that of the target.    These differences 
can be measured using gamma and neutron sources.  

To deploy sources in the gamma catcher, a set of guide tubes is being considered.  These
tubes would be transparent and small to avoid shadowing of scintillator light and to 
minimize dead material and absorption.  Most tubes would run from a manifold 
in the detector 
interface to insertion points at the 
 top of the gamma catcher.  The calibration sources would be attached to
a cable and inserted into a tube.  The tube would then guide the source to the top of
the gamma catcher, from which the source would lower vertically into the gamma catcher
region as additional cable is pushed into the tube.    Whether or not additional guide
tubes running through the gamma catcher to position sources in its top or bottom regions
are needed is being studied.  The position of the source can be determined from the length
of cable inserted into the tube and an accurate survey of the guide tube geometry.   A guide
tube system was used in the Palo Verde experiment to deploy sources with a precision of
2 cm over distances ranging up to 14 m.
Following the same practice as for the target, a small subset of sources would be 
deployed frequently to monitor stability.

A minimum of 6 guide tubes at three azimuthal positions and two radii would be installed.
The connection of the cable to the sources should be such that the same sources can
be deployed in the gamma catcher as in the target.  As with the deployment system for
the target, all components would have to be checked for compatibility before final selection and
carefully cleaned before installation into the experiment.

\subsubsection{Detector Interface}

The detector interface is the region which has access to the target and gamma catcher and
which can be accessed from the outside to introduce and remove calibration sources.  
The volume of the detector interface has to be large enough so that sources can be
easily manipulated and the deployment system can be assembled and disassembled as
needed safely and easily.
It must be connected to the experiment gas system and purged so that, when it is opened
to the detector, it has the same atmosphere to avoid introducing backgrounds.    The
glovebox will be equipped with radon and oxygen monitors for flagging leaks and monitoring
the progress of purging.
While the
detector interface is open to the detector, it must also be light tight, which may
mean that the operator must view the interior of the glovebox using infrared illumination
and cameras.   Feedthroughs for
laser fibers and control and power cables must be hermetic.  To bring sources into
the detector interface, a transfer box is needed: sources are placed into this box through
an external door, then the the transfer box is purged, after which the operator opens an
internal door and brings the source inside the detector interface.  

Most likely, the detector interface can be built by modifying a commercially available
glovebox.    The mechanical design of the detector will provide a defined set of flanges and
supports for the detector interface so that the design of the interface and deployment
systems can be largely decoupled from the design of the rest of the detector.

\subsubsection{Control Systems}
 
 All control and sensor channels for the deployment systems should be interfaced to
 a computer system for the purpose of automation and ready monitoring.  Although 
 it should be possible to monitor the state of the deployment system remotely, for
 safety and convenience, the computer system providing control as well as monitoring
 of the deployment system should be installed near the deployment interface.
 Interlocks,
 both hardware and software, must be implemented to ensure that the detector is
 not opened to light, that a valve between the detector interface and the detector
 is not closed while sources are being deployed, etc.  
 Full exercise of the control program
 and interlock system will be part of the testing of the deployment systems.

%% file: clean.tex
\subsection{Radiopurity maintenance}

A system is needed to exclude dust, radon, and radioactive krypton-85
present in the air from entering the active volume during detector
operation.  Given the anticipated rate of untaggable external neutron
captures in the central volume \cite{DCHOOZ_LOI_BKGRND_CHAPTER}, we
require a rate under 3~Hz of singles exceeding the prompt energy
analysis threshold in order to keep accidental backgrounds under
control.  This requirement is relatively weak in comparison to the
requirements met successfully by KamLAND and Borexino's Counting Test
Facility, but is still strict in comparison to what can be expected in
scintillator casually exposed to ordinary air, for reasons explained
below.

The US groups will work closely with European collaborators to develop
suitable systems to establish and maintain scintillator radiopurity.
These systems should include a clean room around the calibration
access ports, a clean calibration source preparation facility, and a
radon- and krypton-free nitrogen blanket to isolate the scintillator
from the air during normal operation and source deployment.

 \subsubsection{Potential for contamination from ${}^{222}$Rn,
 ${}^{85}$Kr, and dust}

 The average concentration of radon (${}^{222}$Rn) in fresh outdoor air
 is about 0.4 pCi/L (15~Bq/m${}^3$).  Indoors, particularly in
 underground structures, the radon concentration can be from one to
 several orders of magnitude higher.  The relative solubility factor
 for radon in mineral oil relative to radon in air is about 10
 \cite{InternationalCriticalTables}, so exposing a large surface area
 of the scintillator to air would result in thousands of Becquerels of
 radon activity in the active volume.

 In actuality, all but a small area of the scintillator is protected
 from the air.  Because the liquid in the chimney is usually fairly
 static, and the diffusion constant of radon in organic liquids is
 generally a little less than $3\times 10^{-5}$~cm${}^2$s${}^{-1}$
 \cite{InternationalCriticalTables}, a layer as thin as 10~cm is
 sufficient to limit the radon diffusion into the central volumes to
 acceptable levels.  However, during source deployment, the liquid in
 the chimneys is disturbed, and some scintillator from the top of the
 chimney may be brought into the fiducial volume.  In the event the
 top of a calibration chimney were in direct contact with lab air with
 a modest radon concentration of 100~Bq/m${}^3$ (below EPA's
 recommended action level), as little as a liter of scintillator from
 the top of the chimney dragged into a central volume would cause a
 significant and unacceptable increase in the singles rate.

 A similar situation exists with respect to the radioactive noble gas
 ${}^{85}$Kr, produced by fission reactions and emitted especially
 during reprocessing activities, such as those performed at the nearby
 La Hague reprocessing plant.  Atmospheric concentrations of
 ${}^{85}$Kr are typically at the 1~Bq/m${}^{3}$ level
 \cite{KryptonRefs}, but measurements made at Gent, Belgium, show that
 sudden jumps of many orders of magnitude often occur
 \cite{KryptonRefsGent}.  The relative solubility of krypton in
 organic liquids is about 1.  When atmospheric ${}^{85}$Kr levels are
 at their lowest, their potential for contributing activity to the
 scintillator is therefore two or three orders of magnitude less than
 that of radon; however, when atmospheric ${}^{85}$Kr levels are high,
 they could dominate over radon.  It should also be noted that the
 ${}^{85}$Kr spectrum contributes primarily to the lowest energy bins,
 thus concentrating the activity in a part of the spectrum we would
 like to use for measuring background.
 
 The activity and concentration of dust may vary from one laboratory
 to another.  To get an idea of the risk posed by dust, we consider
 the experience of the SNO experiment \cite{SNO_NIM_2000}.  They
 maintained ``class $2500\pm 500$'' cleanroom conditions throughout
 the experimental area, and attained deposition rates of
 300~$\mu$g/cm${}^{2}$/month.  An open 20-cm-diameter calibration port
 would thus collect a few micro-Becquerels per month of uranium,
 thorium, or potassium under these conditions.  However, particulate
 levels in ordinary, non-cleanroom areas may reach an equivalent of
 ``class 400 million'' \cite{Class400MillionRefs}, over one hundred
 thousand times as high as those seen in SNO.  Scaling from the SNO
 experience, we see the potential for a few tenths of a Becquerel per
 month to be deposited in the scintillator if no precautions were to
 be taken to avoid it.  Trivial measures such as keeping the ports
 covered when not in use may not be enough to prevent contamination on
 the order of a few Becquerels over the life of the experiment.
  

 \subsubsection{Control measures}
 

 Several measures should be taken to avoid contamination by dust,
radon, and radioactive krypton-85.  The calibration ports should be
kept closed when not in use.  The space between the top of the liquid
and the chimney covers should be flooded with pure nitrogen.  The
nitrogen should have less than 0.1 Bq/m${}^3$ of radon and ${}^{85}$Kr.
The air around the calibration access ports should be kept at class
10,000 or better.  Calibration sources will be cleaned and prepared as
described elsewhere in this document; after cleaning, it should be
possible to bring them to the clean deployment area without exposing
them to unclean air.  Radon levels in the area around the calibration
chimneys will be monitored as described in the slow monitoring section
of this proposal.  We will investigate the possibility of obtaining
estimates or forecasts ${}^{85}$Kr levels from European authorities or
monitoring it ourselves.
With these precautions in place, the cleanliness
requirements for this experiment should be easily met.


%% file: costs.tex
\section{Cost and Schedule}
\subsection{Overview of Costs}
\par At the Chooz site, the laboratory previously used by the CHOOZ experiment is
vacant and available as a far site for use with minimal preparation.
Electricite de France (EdF) made a major contribution to the CHOOZ experiment
in constructing the laboratory, and has been
asked to contribute
to Double-CHOOZ by constructing the near detector
laboratory. The relationship between the CHOOZ experiment and EdF was
very cooperative and cordial; the success of a \thc experiment such as 
Double-CHOOZ requires such close cooperation. 
We are optimistic that EdF will again
be a willing partner in cutting-edge neutrino science.
\par The current estimate for the cost of both detectors, (not including the
near detector lab) is 9.0-9.5 million euros. In France, IN2P3 and CEA/Saclay have
approved a contribution of 2.2-2.5 million euros. proposed contributions from
Germany and the other collaborators are in the early stages of the funding
process. The CHOOZ-US collaborators are requesting
\$4.858M as detailed in Section~\ref{sec:costdet}, primarily for phototubes,
electronics and an outer veto system for the near detector.
These crucial
parts of the experiment coincide with the expertise of the U.S. collaborators,
who have extensive experience not only in phototubes, but also in high
voltage and front-end electronics, laser calibrations, deployment
systems, and muon tracking systems. 

\subsection{Overview of Schedule}
The anticipated schedule calls for
the completion of an R\&D phase in 2004 followed by project definition.
This will include a prototype to evaluate technical solutions. Production
phase would be February 2005-October 2007, with the far detector
completed in October 2007. The
construction of the near detector is scheduled to be completed
in March 2008.
Detector operations would be for three years, 2008-2011.
\par First results are in principle possible with just the far detector because the
luminosity of the important original CHOOZ experiment will be matched in
just a few months.  Using both detectors, Double-CHOOZ will reach a
sensitivity $\mxang$ of 0.05 in 2009 and 0.03 in 2011.  Whether running
any longer at that time makes sense will depend on an evaluation of systematic
errors and backgrounds achieved to date, as well as the world situation
regarding $\quq$.
\par Photomultipliers and the high
voltage system
would be returned to the U.S. after the experiment is over. 

\subsection{Work Breakdown Structure (WBS) }
\par The CHOOZ-US scope, consisting of cost, schedule and an 
understanding of US participation, is driven by the European
project completion date of April 2008.  Provided here is a
description of US scope delineated by work elements, a description
of those elements, the cost structure for those elements, escalation
considerations, and a schedule to meet Double-CHOOZ goals.
\subsubsection{Work Breakdown Structure Description}
The following describes US contributions to the Double-CHOOZ project
These contributions provide essential components, functionality,
and assured integration of US and European systems.   They form the
foundations of work packages used to track schedule, cost and
deliverables.  The WBS is organized in six major work packages.
Work is further resolved into ensuing levels until reasonable details 
are addressed.
This structure becomes the cost and schedule template for the
conceptual design cost estimate.  The major Double-CHOOZ WBS follows:\\
1. Inner Detector \\
2. Outer Detector \\
3. Signals and Processing \\
4. Calibration \\
5. Management and Common Projects\\
A terse description is provided for second level WBS elements which are part
of the US project scope.
\begin{description}
\item[1.3] PMTs
\par Provides for the System of phototubes for the near detector target, 
and the the far detector target.
Provides for the purchase of 1050 photomultipliers.  16 spares
are included.  The tube is the 8 inch ETI 9354, Hamamatsu R5912,
or similar low-background PMT.  We have used a quote from Hamamatsu
as the basis for costing the PMT, PMT base electronics, potting,
and under-oil cable. 
Quality Control Checkout, DOE tagging and inventory will be performed at Louisiana State University.
Phototube purchasing will be done by Louisiana State
University  and task progress will be monitored
by Argonne National Laboratory.
A structure inside the detector(s) for mounting and installing
the phototubes will be developed at the University of Alabama.
\item[2.2] Outer Veto
\par Provides a system of 5000 gas proportional chambers
for the identification of muons in the near detector and to
veto associated background events.
This requires the design of modular assemblies of the
proportional chambers as well as the associated electronics,
gas and support systems.  After the completion of technology testing and
design reviews,  
procurement and production will be shared between
Argonne National Lab and the University of Tennessee.  Argonne National Lab
will be responsible for progress monitoring and reporting as well as
supervision of final shipment and installation at the experimental 
detector laboratories.
\item[3.1] Analog Front End Electronics
\par Based on requirements for 1024 phototubes, 
provides for the electronics for readout of
all phototubes in the near and far Double-CHOOZ detectors plus
20\% spares.
Includes design and engineering of preliminary and final designs,
Preliminary Design Reviews (PDR) and Final Design Reviews (FDR), component
monitoring, coordination of the generation of software with the project
scientific staff, all testing and supporting documentation and
operating manuals.  It will include assembly and testing at Drexel
and installation at the near and far detector laboratories for Double-CHOOZ
in France.
The front-end electronics task at Drexel University will
use existing graduate student RA and travel support at approximately 2/3
FTE level, as the Drexel group makes
the transition from KamLAND to Double-CHOOZ over the 
next two years, however some additional operations funding
is included here as necessary for the Double-CHOOZ effort.

\item[3.4] High Voltage
\par 
Based on requirements, provides for a single system for
high voltage supply to all photomultiplier tubes.
Two primary candidates for HV systems are Connecticut-based Universal
Voltronics and CAEN from Italy. Cost quotations from both are in hands.
This item provides for 1024 channels.
DOE tagging and inventory will be performed at the University of
Tennessee.
This contract will be monitored by Argonne National Laboratory.
Quality assurance, cabling, testing, installation, and maintenance of the
high voltage system will be provided by the University of Tennessee.
\item[3.5] Slow Control and Monitoring
\par Based on requirements, provides a system to control and scan items such 
as thresholds, high voltage settings and temperatures, and to provide alarms,
warnings and diagnostic information to the experiment operators.
This provides for the purchase from Maxim IC/Dallas Semiconductor
1-Wire interface chips and other components needed to control and readback
hardware.  Assembly will take place at Kansas State University and Installation
at the Double-CHOOZ laboratories in the Ardennes region of France.
The PMT monitoring is to provide periodic 
samples of PMT singles rates at a sub-photoelectron threshold, with
readout and alarm software integrated
into the main DAQ.   
Assembly and testing will be performed
at Drexel University, and installation at the 
Double-CHOOZ laboratories.
\item[4.2] Mechanical Deployment of Sources in 3 dimensions
\par 
Provides for the accurate deployment
of calibration sources within
the target and gamma catcher regions
of the detectors.  The development and assembly will be done by the
University of Alabama, with assistance from the Argonne National Lab for
mechanical engineering.
\item[4.3] Laser System
\par Provides for a laser system to quantify and monitor pertinent properties
of each photomultiplier.  
The system will be developed and tested at
the University of Alabama and installed in the Double-CHOOZ laboratories.
\item[5.1 \& 5.3] Project Coordination, Engineering Coordination
\par Consists of tasks which are the responsibility of
CHOOZ-US project engineer at Argonne
National Laboratory.  Includes project engineering.
Systems engineering and
integration will be responsible for defining interfaces, pre-installation logistics and
installation logistics and day-to-day communications with European Collaboration
counterparts.  
\item[5.2 \& 5.4] Technical Direction and Cost and Effort Monitoring
\par Consists of CHOOZ-US management at Argonne
National Laboratory.  Includes administrative tasks in accordance
with the management plan.  
Documentation and performance tracking will be the responsibility
of the Project Manager.  
\item[5.5] Progress Monitoring and Reporting
\par Provides for administrative expenses involved with quarterly reporting
of progress on costs and schedules to the Department of Energy.
\item[5.6] Shipping
\par Shipping involves transportation expenses for final components of
US systems from Argonne to the Double-CHOOZ site in the Ardennes region of France.
\end{description}
\subsubsection{Full WBS}
The Full Work Breakdown Structure is as follows:\\
1.	Inner Detector\\
1.1.	Vessel Mechanics \\
1.2.	Liquid Scintillator \\
1.3.	PMTs\\
1.3.1.	PMT Procurement\\
1.3.2.	Dark Box Construction\\
1.3.3.	Checkout and testing of PMTs\\
1.3.4.	Mount and Assembly of PMTs\\
2.	Outer Detector\\
2.1.	Inner Veto \\
2.2.	Outer Veto\\
2.2.1.	Design of Modules and Fixtures\\
2.2.2.	Module Factory Setup and Tooling\\
2.2.3.  Component Procurement \\
2.2.4.	Module Assembly and Repair\\
2.2.5.	Wire Stringing\\
2.2.6.	Module Testing\\
2.2.7.	Design of Support System\\
2.2.8.	Support System Construction\\
2.2.9.	Design of Gas Systems and Prototype Testing\\
2.2.10.	Gas System Construction\\
2.2.11.	Electronics Design and Prototype\\
2.2.12.	Electronics Small System Test\\
2.2.13.	Electronics Production\\
2.2.14.	Supervision of Installation on site.\\
3.	Signals and Processing\\
3.1.	Analog Front End Electronics\\
3.1.1.	Electronics and Trigger Design and Prototype\\
3.1.2.	Components\\
3.1.3.	Assembly and Testing\\
3.2.	Digital Electronics \\
3.3.	Data Acquisition System \\
3.4.	High Voltage\\
3.4.1.	High Voltage Power Supplies\\
3.4.2.	High Voltage System Components\\
3.4.3.	Design, Assembly and Installation\\
3.5.	Slow Control/Monitoring\\
3.5.1.	Slow Control Components\\
3.5.2.	Slow Control Design, Assembly, Installation and Travel\\
3.5.3.	PMT monitoring Design and Prototype\\
3.5.4.  PMT Monitoring Assembly and Installation\\
4.	Calibration\\
4.1.	Radioactive Sources \\
4.2.	Mechanical deployment of sources in 3-dimensions\\
4.2.1.	Engineering Design\\
4.2.2.	Target System Fabrication/Assembly/Testing\\
4.2.3.	Gamma Catcher System Fabrication/Assembly/Testing\\
4.2.4.	Shipping Travel and Installation\\
4.3.	Laser System\\
4.3.1.	Laser Head, Driver and Switch Box\\
4.3.2.	Installation and checkout\\
5.	Management and Common Projects\\
5.1.	Project Coordination\\
5.2.	Technical Direction\\
5.3.	Engineering Coordination\\
5.4.    Cost and Effort Monitoring\\
5.5.	Progress Monitoring and Reporting\\
5.6.    Shipping to Experiment Location 
\subsection{Cost \& Schedule Details}
\label{sec:costdet}
The cost estimation for the CHOOZ-US project was made on the basis
of the specific requirements for each WBS system.  The estimates
were generated by specific engineers or supervising physicists with
previous experience from other similar projects.  Where possible,
commercial vendors and catalog information was used for costs related
to materials and sub-contracted services.  The following assumptions
were made:
\begin{enumerate}
\item US systems will not be subject to any taxes or tariffs for shipping and
installation at the experimental site in France.
\item The French laboratory at Chooz will provide all suitable skilled 
technicians labor necessary for installation.
\item All M\&S estimates are free of overhead costs, consistent with the 
regulations of the participating universities.
\item Estimated effort rates are fully burdened with respect to the
listed institution.
\item No overtime rates are assumed for the duration of the project.
\item The effort cost listed for each task refers to the incurred costs
which are over and above the base operating funding supplied by the DOE.
Argonne National Laboratory has agreed to support 25\% of the cost of
all Argonne effort from base funding.
\item All costs are listed in FY05 dollars.  No escalation has been applied.
\end{enumerate}

Detailed cost estimates are attached in a separate document.
They are organized according to the WBS and are listed
for each third level task.  Additional details such as specific component
costs, written quotes, catalog pricing, and detailed drawings 
are available.  
Each estimate is listed with a notation
showing its source:
\begin{description}
\item[Vendor Quote:] a written or emailed estimate for an item or system 
supplied within the last 3 months.
\item[Catalog Price:] Prices for commercially available products or services
obtained from currently valid catalogs or web-based quotations.
\item[Shop Quote:] Supplied by a university or laboratory based shop within the
last 3 months.  Includes all effort, tooling and equipment.  
\item[Engineer Estimate:] Consists of a detailed design and cost estimate
provided by an engineer with previous experience with the specified system.
Drawings, design assumptions and calculations are supplied where appropriate.
\item[Physicist Estimate:] Consists of a design and cost provided by a 
physicist with previous experience with the specified or similar system. 
These estimates are usually scaled from previously completed successful 
projects. 
\end{description}

Contingency costs have been included to cover uncertainties which may
result from unforeseen and unpredictable conditions or from uncertainties 
within the defined scope.  Each level 3 element of the WBS has been assigned
a contingency based on the engineer's judgment as to the solidity of the
estimate and scope definition.  Generally, solid quotes were given 10\% 
contingency.  It is understood that many of the catalog prices are 
more likely to afford cost savings as the un-explored economies of scale
become available at time of purchase.  Most labor and travel estimations
were given a 50\% contingency except in circumstances where an experienced
engineer had total control of the definition of the work and scope.

\subsubsection{Project Construction Schedule}
The CHOOZ-US project schedule has been developed after consultation with
the European collaborators.  It is intended to mesh with the activities
and schedules described in the Double-CHOOZ-LOI\cite{bib:choozloi}.  The 
specified schedule has data taking beginning at the far detector in 
early 2007 with the near detector coming on line at the beginning of 2008.
We therefore assume that the CHOOZ-US project will continue uninterrupted
over the 3 years from July 2005 to July 2008 and that sufficient funds will
be provided.  With that in mind, a short description of the schedule for
each task follows:
\begin{description}
\item[1.3 PMTs -]The production of PMTs has the longest timescale for the
project.  The order will therefore be placed immediately on funding of the
project.  We expect to receive the order in 2 batches one in each of the
first two years of the project.  Checkout and testing of the PMTs will
proceed with availability.  The operating voltage will be determined
during these tests at Louisiana State University.
Installation of the PMTs will
occur around the end of 2006 for the far detector and in the fall of
2007 for the near detector.
\item[2.2 Outer Veto -]The outer veto will not need to be available until
the near detector installation.  As a result, the first year will primarily
be spent in design and prototype testing.  Production is expected to begin
in spring 2006 and last until the fall of 2007.  Final installation at
the experimental location is expected at the end of 2007.
\item[3.1 Analog Front End Electronics -]The electronics designs and 
prototype testing will be completed by early 2006 when final production can 
begin.  The electronics system will be shipped to the experiment and tested
by early 2007 for availability during the far detector startup.
The front-end electronics task at Drexel University will
use existing graduate student RA and travel support at approximately 2/3
FTE level, as the Drexel group makes
the transition from KamLAND to Double-CHOOZ over the 
next two years, however some additional operations funding
is included here as necessary for the Double-CHOOZ effort.
\item[3.4 High Voltage -]
The high voltage system design is almost
complete
and uses mostly commercially available components.  Procurement of the
components and checkout will happen by the middle of 2006 and
installation
at the experiment will be completed by the end of that year.
\item[3.5 Slow Control and Monitoring -]The slow control and PMT rate 
monitoring systems will both go through design and production by July
2006.  They will then be shipped to the experiment and be available for
the testing and use by early 2007.
\item[4.2 Mechanical Deployment of Sources in 3-dimensions -]
The
calibration
deployment system will complete its design phase by the end of 2005.
This
will allow production to begin rapidly so that it can be assembled in
conjunction with the main inner detector in France.  This will be
completed
for the far detector by fall of 2006 and for the near detector by the
middle of 2007.
\item[4.3 Laser System -]The laser calibration system requires only 
commercially available components which will be shipped directly to the
experimental location.  The purchase order will be submitted in 2006
such that delivery, installation and checkout can be performed by the 
middle of 2007. 
\item[Management and Common Projects -]Management of individual projects
and the integration and coordination with the European collaborators 
will be a continuous process during the life of the project.  The 
transportation of US components to the experimental location is expected to
occur in 3 shipments.  These shipments are loosely expected to occur
in fall 2006, spring 2007 and fall 2007.
\end{description}
On the basis of the preceding schedule considerations, the required funding,
including contingency, is shown in Table~\ref{T:CostProfile}.
The plan to start this project
in the summer of 2005 will utilize forward funding from
two of the Universities.

\begin{table}[htb]
\caption{CHOOZ-US Cost Profile.  The costs are shown for each major
task over the expected 3 years of the project (defined from July 1 2005
to June 30 2008).  The listed costs are in FY05 dollars and contain
the contingencies described in the detailed budget description.}
\begin{center}
\begin{tabular}{ll|r|r|r|r} \hline
WBS & Description & Year 1 & Year 2 & Year 3 & Totals \\
\hline
1.3 & PMTs & 
\$1,050,553 & \$1,056,705 & \$7,801 &	\$2,115,060 \\ 
2.2 & Outer Veto &
\$509,414 &	\$595,755 &	\$213,909 &	\$1,319,079 \\ 
3.1 & Front End Electronics &
\$4,500 &	\$264,525 &	\$0 &	\$269,025 \\
3.4 & High Voltage &
\$402,689 &	\$16,065 &	\$0 &	\$418,755 \\
3.5 & Slow Control/Monitoring &
\$74,135 &	\$18,824 &	\$0 &	\$92,959 \\
4.2 & Calibration Deployment &
\$151,096 &	\$88,082 &	\$0 &	\$239,178 \\
4.3 & Laser System &
\$0 &	\$55,031 &	\$0 &	\$55,031 \\
5.  & Management/Common Projects &
\$110,025 &	\$122,818 &	\$116,421 &	\$349,265 \\
\hline
& Totals &
\$2,302,413 &	\$2,217,809 &	\$338,133 &	\$4,858,356 \\
\hline
\end{tabular}
\end{center}
\label{T:CostProfile}
\end{table}

\subsection{Project Management}

\par The CHOOZ-US management plan is driven by several priorities:\\
\begin{itemize}
\item The project is a partnership between several Universities and Argonne
National Laboratory. The latter should provide Project
Management and Engineering Support to the CHOOZ-US group.
It is crucial that the group management be aware of the needs
and capabilities of both types of institution to ensure project success.
\item The detector systems in the U.S. project scope must be interfaced with
other detector systems being designed and constructed by groups in Europe.
To avoid confusion and potential technical conflicts, there must be clear
and consistent coordination 
among all collaborators.
\item A quick measurement of $\theta_{13}$ is top priority of Double-CHOOZ.
This necessitates beginning immediately on the design and specification of
detector systems, especially for long lead time items such as PMTs. To make
this feasible we must forward-fund these efforts through the universities. It
is crucial that these funds are used in the most effective way for all
U.S. detector subsystems and not just those in the scope of the universities
providing the forward funding.
\item Each institution has a stake in providing operating funds for detector
construction and it is crucial that a coherent plan is developed for
coordination of the operating fund requests from the various U.S. groups.
\end{itemize}

\par
\begin{figure}[ptbh]
\vspace*{2.0cm} 
\begin{center}
\includegraphics[width=7.0in]{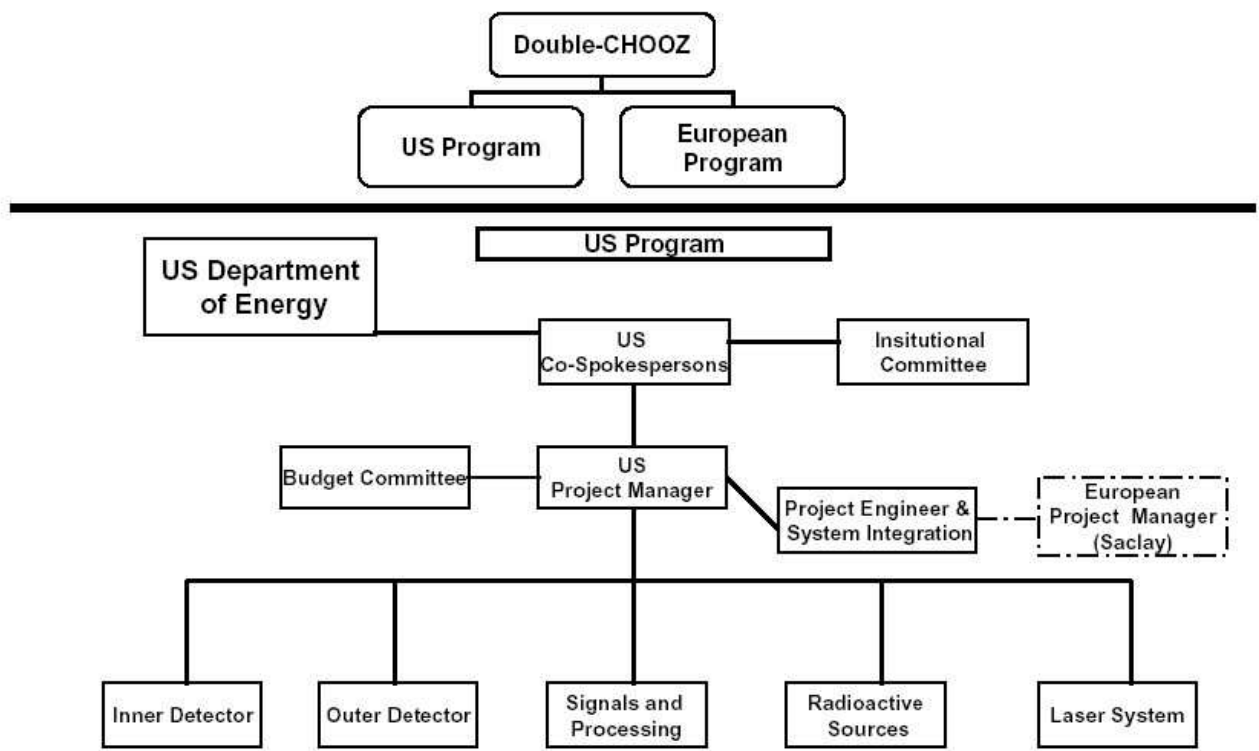} 
 \caption{ Proposed management structure for CHOOZ-US.}
\label{fig:management}
\end{center}
 \end{figure}

 The organization structure is shown in Figure~\ref{fig:management}.
The Double-CHOOZ Spokesperson is Herve de Kerret from
APC and PCC College de France who has overall responsibility for
Double-CHOOZ.  The CHOOZ-US collaboration will operate as an
integral part of the Double-CHOOZ, but for reporting and management
purposes, a CHOOZ-US structure will be maintained.  
The principal investigator from each institution in the U.S.
will be a member of the
Institutional Committee, which has the responsibility of electing
the US co-spokespersons.
To coordinate laboratory and university efforts the
Institutional Committee is initially electing two US Co-Spokespersons.
Maury Goodman is a physicist at
Argonne National Laboratory
and a recognized leader on the Soudan 2 experiment. He was also
spokesman for P-822, the original proposal for a long-baseline
neutrino beam at Fermilab.  Professor Robert Svoboda is a senior 
professor at Louisiana State University and was
Co-Convener for Solar Neutrinos on the Super-Kamiokande experiment and
University Coordinator for the construction of the KamLAND detector.
They will carry overall responsibility for all technical and budgetary
aspects of CHOOZ-US contributions.  They will
also be responsible for coordinating the overall project with the European
collaborators and will report to the host country spokesperson on the
overall progress of the U.S. construction effort.  In the remainder of this
section, the co-spokespersons will refer only to CHOOZ-US.

Reporting to the Co-Spokespersons on the technical coordination of
the detector construction project is the Project Manager. Project management
will be the responsibility of Argonne National Laboratory, with
Physicist Dr. David Reyna designated as the initial Project
Manager.  He will work with the Project Engineer to
\begin{itemize}
\item Develop Memoranda of Understanding (MOUs) with each
institution which would be used
to coordinate all spending and to spell out budget and schedule expectations.
\item Track project costs and milestones toward completion of the
project.
\item Organize periodic reviews of the work of the university and
laboratory groups in cooperation with the Department of Energy.
\item Monitor the work breakdown structure and submit quarterly
reports to the Department of Energy.
\end{itemize}
The two CHOOZ-US co-spokespersons together with the Project
Manager have primary responsibility for communicating with the
Department of Energy.  The Co-spokesperson from Argonne and the Project 
Manager are responsible for assuring the DOE 
that the CHOOZ-US project is proceeding appropriately.  

\par The Project Engineer for the project is Victor Guarino, an engineer
at Argonne National Laboratory. He will serve as the point-of-contact for technical issues
in coordinating with the European groups and will oversee the 
engineering plan for the U.S. portion of the detector, including projects
in the university scope. He will work in close concert
with Florence Ardellier, the designated European Project Manager from Saclay.

\par The project budget will be reviewed on a regular basis by a Budget
Committee, which consists of the two Co-Spokespersons, US Project Manager,
Project Engineer, and the PI's from the University of Tennessee 
(Y.Kamyshkov) and Drexel University (Charles Lane). They will make
recommendations to the co-spokespersons on funding priorities and on 
the allocation of contingency based
on the reports from the US Project Manager and inputs from all the 
university Principle Investigators (PI's). 

\par At every stage of this project, including design, assembly and
installation, safety will be a paramount concern.  This will be done by
incorporating the following principles at every stage of the project:
clear management responsibilities, documentation, a working atmosphere 
of safety, Integrated Safety Management.  A Quality Assurance plan for
the project will be developed in accordance with policies and recommendations 
from the Department of Energy and any other appropriate agencies.

\par Once Double-CHOOZ
is commissioned and fully operational, the US Project Manager will
be replaced with an analysis structure selected by the Institutional
Committee in coordination with the entire collaboration.
All collaborating groups are free to participate in any of the 
scientific goals of Double-CHOOZ.

%% file: appendix.tex
\newpage
\appendix
\section{The U.S. Collaboration}
\par The signers of this proposal have extensive experience in reactor neutrino
experiments: CHOOZ (Lane), Palo Verde (Busenitz) and
KamLAND (Bugg, Busenitz,
Dazeley, Efremenko, Horton-Smith,
Kamyshkov, Lane, and Svoboda). There is also experience
in a wide variety of other neutrino experiments including IMB (LoSecco
and Svoboda),
Soudan 2 (Goodman), MACRO (Lane), SNO (Kutter), MINOS (Goodman and Reyna),
Brookhaven 704 (LoSecco), 
LSND (Stancu and Metcalf), MiniBooNE (Stancu and Metcalf), 
Fermilab E1A (LoSecco), Fermilab E594 (Goodman) 
and
Super-Kamiokande (Dazeley and Svoboda).
\section{The Full Collaboration}
\thispagestyle{empty}
\vspace*{5mm}
\noindent{ \bf 
F.~Ardellier~$^5$,
I.~Barabanov~$^{10}$,
J.C.~Barri\`ere~$^5$,
M.~Bauer~$^{7}$,
S.~Berridge$^{20}$,
L.~Bezrukov~$^{10}$,
Ch.~Buck~$^{13}$,
W.~Bugg$^{20}$, 
J.~Busenitz$^1$, 
C.~Cattadori~$^{8,9}$,
B.~Courty~$^{2,15}$,
M.~Cribier~$^{2,5}$,
F.~Dalnoki-Veress~$^{13}$,
N.~Danilov~$^4$,
S.~Dazeley$^{12}$, 
H.~de Kerret~$^{2,15}$,
A.~Di~Vacri~$^{8,19}$,
G.~Drake$^3$,
Y.~Efremenko$^{20}$, 
A.~Etenko~$^{16}$,
M.~Fallot~$^{17}$,
Ch.~Grieb~$^{18}$,
M.~Goeger~$^{18}$,
M.~Goodman$^{3}$,
J.~Grudzinski$^3$,
V.~Guarino$^3$,
A.~Guertin~$^{17}$,
G. Horton-Smith$^{11}$,
C.~Hagner~$^{21}$,
W.~Hampel~$^{13}$,
F.X.~Hartmann~$^{13}$,
P.~Huber~$^{18}$,
J.~Jochum~$^{7}$,
Y.~Kamyshkov$^{20}$, 
T.~Kirchner~$^{17}$,
Y.S.~Krylov~$^{4}$,
D.~Kryn~$^{2,15}$,
T.~Kutter$^{12}$
T.~Lachenmaier~$^{18}$,
C.~Lane$^c$,   
Th.~Lasserre~$^{2,5}$, 
Ch.~Lendvai~$^{18}$,
A.~Letourneau~$^{5}$,
M.~Lindner~$^{18}$,
J.~LoSecco$^{14}$,
F.~Marie~$^{5}$,
J.~Martino~$^{17}$,
R.~McNeil$^{12}$,
G.~Mention~$^{2,15}$,
W.~Metcalf$^{12}$, 
A.~Milsztajn~$^{5}$,
J.P.~Meyer~$^{5}$,
D.~Motta~$^{13}$,
L.~Oberauer~$^{18}$,
M.~Obolensky~$^{2,15}$,
L.~Pandola~$^{8,19}$,
W.~Potzel~$^{18}$,
D.~Reyna$^3$,
S.~Sch\"onert~$^{13}$,
U.~Schwan~$^{13}$,
T.~Schwetz~$^{18}$,
S.~Scholl~$^{7}$,
L.~Scola~$^{5}$,
M.~Skorokhvatov~$^{16}$,
I.~Stancu$^1$, 
S.~Sukhotin~$^{15,16}$,
R.~Svoboda$^{12}$, 
R.~Talaga$^3$,
D.~Vignaud~$^{2,15}$,
F.~von~Feilitzsch~$^{18}$,
W.~Winter~$^{18}$,
E.~Yanovich$^{10}$}\\
\\

\noindent{$^1$ \rm University of Alabama, Tuscaloosa, Alabamba 35487-0324, USA}\\
{$^2$ \rm APC, 11 place Marcelin Berthelot, 75005 Paris, France} \\
{$^3$ \rm Argonne National Lab, Argonne, IL 60439-4815, USA}\\
{$^4$ \rm IPC of RAS, 31, Leninsky prospect, Moscow 117312, Russia}\\
{$^5$ \rm DAPNIA (SEDI, SIS, SPhN, SPP), CEA/Saclay, 91191 Gif-sur-Yvette, France}\\
{$^6$ \rm Drexel University, Philadelphia, Pennsylvania 19104, USA}\\
{$^7$ \rm Eberhard Karls Universit\"at, Wilhelmstr. D-72074  T\"ubingen, Germany}\\
{$^8$ \rm INFN, LGNS, I-67010  Assergi (AQ), Italy}\\
{$^9$ \rm INFN Milano, Via Celoria 16, 20133 Milano, Italy}\\
{$^{10}$ \rm INR of RAS, 7a, 60th October Anniversary prospect, Moscow 117312, Russia}\\
{$^{11}$ \rm Kansas State Univeristy, Manhattan 
Kansas 66506-26031, USA}\\
{$^{12}$ \rm Louisiana State University, Baton 
Rouge, Louisiana 70803-4001, USA}\\
{$^{13}$ \rm MPI f\"ur Kernphysik, Saupfercheckweg 1, D-69117 Heidelberg, Germany}\\
{$^{14}$ \rm University of Notre Dame, South Bend, Indiana 46556, USA}\\
{$^{15}$ \rm PCC Coll\`ege de France, 11 place Marcelin Berthelot, 75005 Paris, France}\\
{$^{16}$ \rm RRC Kurchatov Institute, 123182 Moscow, Kurchatov sq. 1, Russia}\\
{$^{17}$ \rm Subatech (Ecole des Mines), 4, rue
  Alfred~Kastler, 44307 Nantes, France \\
{$^{18}$ \rm TU M\"unchen. James-Franck-Str., D-85748 Garching, Germany}\\
{$^{19}$ \rm University of L'Aquila, Via Vetoio 1, I-67010 Coppito,
  L'Aquila, Italy \\
{$^{20}$ \rm University of Tennessee, Knoxville, Tennessee 37996-1200, USA}\\
{$^{21}$ \rm Universit\"at Hamburg, Luruper Chaussee 149, D-22761 Hamburg, Germany}\\